\documentclass[prb,aps,superscriptaddress,dvips,11pt]{revtex4}
\usepackage{feynmp}
\usepackage{amsmath}
\usepackage{amsfonts}
\usepackage{amssymb}
\usepackage{epsfig}
\usepackage{graphicx}
\usepackage{subfigure}
\usepackage{bbm}
\usepackage{color}
\usepackage{longtable}
\usepackage{booktabs}
\usepackage{multirow}
\usepackage{graphicx}
\usepackage{epsfig}

\newcommand{\Rmnum}[1]{\expandafter\@slowromancap\romannumeral #1@}
\allowdisplaybreaks[3]
\begin{document}

\title{Nonperturbative Dyson-Schwinger equation approach to strongly
interacting Dirac fermion systems}

\author{Xiao-Yin Pan}
\affiliation{Department of Physics, Ningbo University, Ningbo,
Zhejiang 315211, China}
\author{Zhao-Kun Yang}
\affiliation{Department of Modern Physics, University of Science and
Technology of China, Hefei, Anhui 230026, China}
\author{Xin Li}
\affiliation{Department of Modern Physics, University of Science and
Technology of China, Hefei, Anhui 230026, China}
\author{Guo-Zhu Liu}
\altaffiliation{Corresponding author: gzliu@ustc.edu.cn}
\affiliation{Department of Modern Physics, University of Science and
Technology of China, Hefei, Anhui 230026, China}

\begin{abstract}
Studying the strong correlation effects in interacting Dirac fermion
systems is one of the most challenging problems in modern condensed
matter physics. The long-range Coulomb interaction and the
fermion-phonon interaction can lead to a variety of intriguing
properties. In the strong-coupling regime, weak-coupling
perturbation theory breaks down. The validity of $1/N$ expansion
with $N$ being the fermion flavor is also in doubt since $N$ equals
to $2$ or $4$ in realistic systems. Here, we investigate the
interaction between (1+2)- and (1+3)-dimensional massless Dirac
fermions and a generic scalar boson, and develop an efficient
non-perturbative approach to access the strong-coupling regime. We
first derive a number of self-consistently coupled Ward-Takahashi
identities based on a careful symmetry analysis and then use these
identities to show that the full fermion-boson vertex function is
solely determined by the full fermion propagator. Making use of this
result, we rigorously prove that the full fermion propagator
satisfies an exact and self-closed Dyson-Schwinger integral
equation, which can be solved by employing numerical methods. A
major advantage of our non-perturbative approach is that there is no
need to employ any small expansion parameter. Our approach provides
a unified theoretical framework for studying strong Coulomb and fermion-phonon interactions. It may also be used to approximately handle the Yukawa coupling between fermions and order-parameter fluctuations around continuous quantum critical points. Our approach
is applied to treat the Coulomb interaction in undoped graphene. We
find that the renormalized fermion velocity exhibits a logarithmic
momentum-dependence but is nearly energy independent, and that no
excitonic gap is generated by the Coulomb interaction. These
theoretical results are consistent with experiments in graphene.
\end{abstract}

\maketitle

%%%%%%%%%%%%%%%%%%%%%%%%%%%%%Main Body%%%%%%%%%%%%%%%%%%%%%%%%%

\section{Introduction \label{Sec:Introdunction}}

Developing efficient theoretical and numerical methods to handle the
strong interactions of quantum many-body systems is absolutely one
of the most challenging problems of condensed matter physics. In
ordinary Fermi liquid systems, weak repulsive interaction is known
to be irrelevant at low energies. This ensures that the conventional
method of weak-coupling perturbative expansion is applicable
\cite{AGD, ColemanBook}. Using perturbation theory, one can expand a
physical quantity as the sum of an infinite number of terms, each of
which is proportional to certain power of a small coupling constant
$\lambda$. Usually one only needs to compute the leading one or two
terms since the contributions of all the sub-leading terms are
supposed to be negligible. Apparently, the perturbation theory is
valid only when $\lambda$ is sufficiently small. It is broadly
recognized that the inter-particle interaction is strong in many
condensed matter systems, such as cuprate superconductors
\cite{Lee06}, heavy fermion compounds \cite{Heavy}, and certain
types of Dirac/Weyl semimetals \cite{CastroNeto, Kotov12, Vafek14,
Wehling14, Armitage17, Tang18}. In these materials, strong
interactions may lead to a variety of non-Fermi liquid (NFL)
behaviors and quantum phase transitions. When the coupling parameter
$\lambda$ is at the order of unity or much larger than unity, the
traditional method of perturbative expansion breaks down and can no
longer be trusted.

In order to study strong inter-particle interactions, it is
necessary to go beyond the framework of weak-coupling perturbative
expansion. A frequently used method is to generalize the fermion
flavor $N$ to a large number and expand physical quantities in
powers of $1/N$. As $N\rightarrow \infty$, one might be able to
consider only the leading one or two terms, based on the expectation
that all the higher order contributions are suppressed. This
expansion scheme has been previously applied to investigate strongly
correlated electronic systems \cite{Polchinski94, Altshuler94,
Kim94, Kim97, Rantner01, Rech06, Kim08, Metlitski20101,
Metlitski20102, Raghu19}. However, the main problem of this approach
is that in most realistic systems the physical value of fermion
flavor is $N=2$, corresponding to spin degeneracy. It is unclear
whether the results obtained in the $N\rightarrow \infty$ limit are
still reliable as $N$ is reduced down to its physical value.
Actually, the $1/N$ expansion scheme may be invalid even in the
$N\rightarrow \infty$ limit. As argued by Lee \cite{Lee09}, the
leading contribution of $1/N$ expansion contains an infinite number
of Feynman diagrams as $N\rightarrow \infty$ in the U(1) gauge model
of spin liquids.

Over the last 15 years, Dirac semimetal materials
\cite{CastroNeto, Kotov12, Vafek14, Wehling14, Armitage17, Tang18}
have been extensively studied. Such materials do not have a finite
Fermi surface, and the conduction and valence bands touch at
discrete points, around which relativistic Dirac fermions emerge as
low-lying elementary excitations. Graphene \cite{Geim05, Geim07} and
surface state of three-dimensional topological insulator
\cite{Vafek14, Wehling14, Armitage17, Hasan10, Qi11} are two typical
$(1+2)$-dimensional Dirac semimetals. $(1+3)$-dimensional Dirac
semimetal may be realized in TiBiSe$_{2-x}$S$_{x}$ \cite{XuSuYang11,
Sato11}, Bi$_{2-x}$In$_{x}$Se$_{3}$ \cite{Brahlek12, Wu13}, and also
Na$_{3}$Bi and Cd$_{3}$As$_{2}$ \cite{Wangzj12, Wangzj13,
LiuZK14A, Neupane14, LiuZK14B, Borisenko14, HeLP14,
ZhangFaXianXiu19}. Dirac fermions exhibit different properties from
the Schrodinger electrons excited around the finite Fermi surface of
a normal metal. The unique electronic structures of Dirac
semimetals lead to prominent new features. The first new feature is
that, Dirac fermions have more degrees of freedom than Schrodinger
electrons. The latter only have two spin components, thus the unity
matrix (in spin-independent cases) and the Pauli matrices (in
spin-dependent cases) suffice to describe the action. In contrast,
Dirac fermions have additional quantum numbers, such as sublattice
and valley. In the case of graphene, one usually needs to introduce
a number of $4\times 4$ gamma matrices to define the action
\cite{CastroNeto, Kotov12}. This makes the structure of correlation
functions more complicated. Another new feature is that, while the
Coulomb interaction is always short-ranged due to static screening
and thus is irrelevant in the low-energy regime in metals with a
finite Fermi surface, it remains long ranged in undoped Dirac
semimetals as a result of vanishing density of states (DOS) at
band-touching points. The long-range Coulomb interaction produces
unconventional FL behaviors in some semimetals \cite{Kotov12,
Gonzalez99} and NFL behaviors in some other semimetals \cite{Moon13,
Herbut14, Savary14, Janssen17, Isobe16, Wangmit17, Wangnfl18,
WangLiuZhang19, Han19}. It can result in strong renormalization of
fermion velocity and other many-body effects \cite{Gonzalez94,
Son07, DasSarma07, Polini07, Vafek07, Mishchenko07, Foster08, Son08,
Vafek08, Kotov08, Kotov09, Vozmediano10, Fogler12, Barnes14,
Sharma16, Elias11, Hofmann14, Throckmorton15, Chae12, Lanzara11}.
When the Coulomb interaction becomes sufficiently strong, it might
lead to an excitonic semimetal-insulator quantum phase transition
\cite{Khveshchenko01, Gorbar02, Khveshchenko04, Khveshchenko09,
Liu09, Gamayun10, WangLiu12, WangLiu14, Gonzalez12, Gonzalez12jhep,
Carrington16, Carrington18, Drut09A, Armour10, Buividovich12,
Ulybyshev13, Tupitsyn17, Xiao18}. Apart from the Coulomb
interaction, the interaction between Dirac fermion and phonon might
be important, and has been investigated using various techniques
\cite{Louie07, Roy14, Meng19, Scalettar19}. In particular, recent
quantum Monte Carlo (QMC) simulations \cite{Meng19, Scalettar19}
have claimed to reveal a charge density wave (CDW) order caused by
fermion-phonon interaction.

When the Coulomb interaction or the fermion-phonon interaction is
strong, the weak-coupling perturbation theory becomes invalid. The
validity of $1/N$ expansion is also questionable since the physical
flavor is usually $N=2$ in realistic Dirac semimetals. Although
large-scale QMC simulation \cite{Armour10, Buividovich12,
Ulybyshev13, Tupitsyn17, Tang18, Drut09A} and other numerical
methods, such as dynamical mean field theory (DMFT) \cite{DMFT}, can
be applied to investigate on-site interactions, their capability of
accessing the strong-coupling regime of long-range interactions is
in doubt. It is urgent to seek a more powerful non-perturbative
method to handle strong couplings.

In a recent publication \cite{Liu19}, the authors have developed a
non-perturbative Dyson-Schwinger (DS) equation approach to
investigate the superconductivity mediated by electron-phonon
interaction in metals with finite Fermi surface. This approach goes
beyond the conventional Migdal-Eliashberg (ME) theory \cite{Migdal,
Eliashberg}. A significant advance achieved in Ref.~\cite{Liu19} is
that, the full electron-phonon vertex function can be completely
determined by solving two coupled Ward-Takahashi identities (WTIs)
derived rigorously from global U(1) symmetries. Making use of this
result, it is shown in Ref.~\cite{Liu19} that the DS equation of
fully renormalized fermion propagator is self-closed and can be
efficiently solved by numerical tools. In distinction to the method
of weak-coupling expansion, the DS equation approach does not
involve any small expansion parameter and is reliable even in the
strong coupling regime. The widely used QMC simulations suffer from
the fermion-sign problem and become inadequate at low temperatures.
DMFT \cite{DMFT} ignores long-range correlations and breaks down in
low-dimensional systems. By comparison, our DS equation approach is
applicable to all temperatures and all (physically meaningful)
spatial dimensions, and works well for both short- and long-range
interactions.

The approach developed in Ref.~\cite{Liu19} is of broad
applicability, not restricted to electron-phonon systems. In this
paper, we will show that this approach can be generalized to study
the strong correlation effects in Dirac fermion systems. In order
not to lose generality, we consider a model that describes the
interaction between massless Dirac fermion, represented by $\psi$,
and a scalar boson, represented by $\phi$. The dispersion of Dirac
fermion may be isotropic or anisotropic. The scalar boson could be
the phonon induced by lattice vibrations, or the scalar potential
that effectively represents the long-range Coulomb interaction. The
scalar boson could also be identified as the quantum fluctuation of
certain (say nematic or CDW) order parameter, but the situation
becomes more complex in this case. We will make a unified,
model-independent analysis and prove that the DS equation of Dirac
fermion propagator $G(p)$ is self-closed as long as the boson field
does not have self-interactions. The exact fermion-boson vertex
function appearing in such a self-closed equation is obtained from a
number of coupled WTIs that are derived rigorously from special
global U(1) transformations of the effective action of the system.
By using this approach, the quasiparticle damping, the Fermi
velocity renormalization, the possible formation of excitonic
pairing, and the interplay of these many-body effects can be
simultaneously extracted from the numerical solutions of the DS
equation. All the results are valid for any value of fermion flavor
and any value of fermion-boson interaction strength parameter.

There is an important difference between conventional
electron-phonon systems and Dirac fermion systems. In the former
case, the vertex function is calculated from two WTIs induced by two
symmetries and two symmetry-induced conserved currents \cite{Liu19}.
In the latter case, however, there are no sufficient
symmetry-induced WTIs. To completely determine the vertex function,
we need to employ both symmetry-induced conserved currents and
asymmetry-related non-conserved currents to derive a sufficient
number of generalized WTIs. Not all non-conserved currents are
directly useful. We will demonstrate how to construct useful
non-conserved currents and how to obtain the corresponding
generalized WTIs from such non-conserved currents.

To illustrate how our approach works in realistic systems, we take
undoped graphene as an example. In particular, we restrict our
interest to the impact of long-range interaction, leaving the
fermion-phonon interaction for future research. The effective fine
structure constant of undoped graphene is of the order of unity,
i.e., $\alpha \sim 1$, implying that Dirac fermions experience a
strong Coulomb interaction. In addition, the physical flavor is
$N=2$ if four-component spinor is adopted. Thus, this system
actually does not have a suitable small parameter. Previous
field-theoretical analysis carried out by means of small-$\alpha$
expansion and $1/N$ expansion have not provided conclusive results
about the behavior of fermion velocity renormalization and the fate
of excitonic insulating transition. Actually, it was revealed in
Refs.\cite{Barnes14, Kolomeisky15, Hofmann14} that the perturbation
series expressed in powers of $\alpha$ diverges already at the
leading two or three orders, implying that conventional perturbation
theory is unreliable. The validity of results obtained by using
$1/N$ expansion is also far from clear. To circumvent ambiguities
induced by perturbative expansion, in this work we apply our
non-perturbative approach to revisit the strong Coulomb interaction
between Dirac fermions. We obtain the exact solutions of the self-consistent DS equation of the full Dirac fermion propagator. Our results show that, the renormalized fermion velocity exhibits a logarithmic momentum dependence at a fixed
energy, but is nearly energy independent at a fixed momentum.
Moreover, after carrying out extensive calculations, we confirm that
the Coulomb interaction cannot dynamically open an excitonic gap in
realistic graphene materials. These theoretical results are
qualitatively in good agreement with experiments.

The rest of the paper is organized as follows. In
Sec.~\ref{Sec:Model}, we define the effective action describing the
interaction between Dirac fermions and scalar bosons. In
Sec.~\ref{Sec:DSEs}, we present the coupled DS integral equations of
full fermion propagator, full boson propagator, and full
fermion-boson interaction vertex function. In Sec.~\ref{Sec:WTIs},
we derive a number of coupled WTIs satisfied by various current
vertex functions by performing a rigorous functional analysis. In
Secs.~\ref{Sec:unitymatrix} and \ref{Sec:gamma0}, we provide the
explicit expressions of the corresponding WTIs for two different
sorts of fermion-boson interaction terms, respectively. The exact
relations between current vertex functions and fermion-boson
interaction vertex functions are derived and analyzed in
Sec.~\ref{Sec:D0D}. In Sec.~\ref{Sec:graphene} we present a
systematic investigation of the quantum many-body effects induced by
the Coulomb interaction in graphene by solving the exact DS equation
of fermion propagator without making any approximation. In
Sec.~\ref{Sec:Summary}, we briefly summarize the main results of this paper. We define all the used gamma matrices in Appendix \ref{App:gammamatrices}, and provide the detailed derivation of the DS equations of fermion and boson propagators in Appendix \ref{App:DSEs}.

\section{Model \label{Sec:Model}}

The model considered in this work describes the interaction between
massless Dirac fermions and some sort of scalar boson. We will first
present the generic form of the action and then discuss three
different physical systems described by the action.

Our starting point is the following partition function
\begin{eqnarray}
\mathcal{Z} = \int \mathcal{D}\phi \mathcal{D}\psi \mathcal{D}{\bar
\psi}e^{iS[\phi,\psi,{\bar\psi}]},
\end{eqnarray}
which is defined as a functional integration over all possible field
configurations weighted by the total action
\begin{eqnarray}
S[\phi,\psi,{\bar\psi}] = S_f[\psi,{\bar\psi}] + S_b[\phi] +
S_{fb}[\phi,\psi,{\bar\psi}],
\end{eqnarray}
where $S_f[\psi,{\bar\psi}]$ is the action for the free Dirac
fermion field $\psi$, $S_b[\phi]$ for the scalar boson field $\phi$,
and $S_{fb}[\phi,\psi,{\bar\psi}]$ for the fermion-boson coupling.

For free Dirac fermions, its action $S_f[\psi,{\bar\psi}]$ is defined via the Lagrangian density $\mathcal{L}_{f}[\psi,{\bar\psi}]$ as follows
\begin{eqnarray}
S_f[\psi,{\bar\psi}] &=& \int dx \mathcal{L}_{f}[\psi,{\bar\psi}]
\nonumber \\
&=& -i\sum_{\sigma=1}^N \int dx{\bar\psi}_\sigma(x)
({i{\partial_t}\gamma^{0}-\mathcal{H}_f}){\psi_\sigma}(x).
\end{eqnarray}
Here, $x=(t,{\bf x})$ denotes the $(1+d)$-dimensional coordinate
vector with $d=2$ or $d=3$, and $dx = dtd{\bf x}$. The conjugate of
spinor field $\psi$ is ${\bar\psi} = \psi^{\dag}\gamma^0$. The
flavor index is denoted by $\sigma$, which sums from $1$ to $N$. In
the case of $d=3$, $\psi$ naturally has four components within the
standard Dirac theory of relativistic fermions. Accordingly, we
should use four standard $4\times 4$ matrices $\gamma^{\mu}$, which
satisfy Clifford algebra $\{\gamma^{\mu},
\gamma^{\nu}\}=2g^{\mu\nu}$, to define
$\mathcal{L}_{f}[\psi,{\bar\psi}]$. Definitions of $\gamma^{\mu}$
are presented in Appendix \ref{App:gammamatrices}. In the case of
$d=2$, there are two possible representations of $\psi$
\cite{Appelquist86}. One may still use the four-component spinor
representation, just like in the case of $d=3$. Another option is to
introduce two-component representation of $\psi$ and to define
$\mathcal{L}_{f}[\psi,{\bar\psi}]$ in terms of $2\times 2$ Pauli
matrices along with unit matrix $I$. There is an important
difference between these two options: one could define and discuss
chiral symmetry, defined via $\gamma^{5}$ that satisfies the
relation $\{\gamma^{5},\gamma^{\mu}\}=0$, only when four-component
representation is adopted. As illustrated in
Ref.~\cite{Appelquist86}, it is not possible to define chiral
symmetry in terms of two-component spinor. Later we wish to study
the phenomenon of dynamical chiral symmetry breaking induced due to
excitonic pairing. Therefore, throughout this paper we always adopt
four-component spinor. All the results can be directly applied to
the case of two-component spinor, except those regarding chiral
symmetry (breaking). The Hamiltonian density $\mathcal{H}_f$ is
\begin{eqnarray}
\mathcal{H}_f &=& -i\sum_{i=1}^d{\gamma^i}({{v_i}{\partial_i}})
\rightarrow -i\sum_{i=1}^d \gamma^{i}\partial_i,
\end{eqnarray}
where $\gamma^{i}$ is the spatial component of $\gamma^{\mu}$ and
$v_{i}$ is the fermion velocity along the $i$-direction. For
notational simplicity, we absorb velocities $v_i$ into $\partial_i$,
which is equivalent to taking $v_i=1$. It is easy to recover $v_i$
whenever necessary.

The free action of boson field $\phi$ is formally written as
\begin{eqnarray}
S_b[\phi] &=& \int dx \mathcal{L}_{b}[\phi] \nonumber \\
&=& -i\int dx \phi^{\dag}(x) \frac{\mathbb{D}}{2}\phi(x),
\end{eqnarray}
where the operator $\mathbb{D}$ defines the equation of the free
motion of boson, i.e., $\mathbb{D}\phi = 0$. The expression of
$\mathbb{D}(x)$ is system dependent and will be given later.

The fermion-boson interaction is described by a Yukawa-type coupling
term
\begin{eqnarray}
S_{fb}[\phi,\psi,{\bar\psi}] &=& \int dx
\mathcal{L}_{fb}[\phi,\psi,{\bar\psi}] \nonumber \\
&=& -ig \sum^N_{\sigma=1}\int dx \phi(x){\bar\psi}_\sigma(x)
\gamma^{m}\psi_\sigma(x), \label{Eq:yukawacoupling}
\end{eqnarray}
where $g$ is the coupling constant and $\gamma^{m}$ is an arbitrary
gamma matrix. This term describes a certain sort of interaction for
any given expression of $\gamma^{m}$. For instance, if the scalar
boson couples to the fermion density operator $\psi^{\dag}\psi =
\bar{\psi}\gamma^{0}\psi$, one should choose $\gamma^{m} =
\gamma^{0}$.

The scalar field $\phi$ might describe any type of scalar bosonic mode.
Here we consider three frequently encountered cases.

\subsection{Coulomb interaction}

The pure Coulomb interaction is modeled by a direct density-density
coupling term
\begin{eqnarray}
H_{C} = \frac{1}{4\pi}\frac{e^{2}}{v\epsilon}\sum_{\sigma,\sigma'}
\int d^2\mathbf{x} d^2 \mathbf{x}'\rho_{\sigma}(\mathbf{x})
\frac{1}{\left|\mathbf{x} - \mathbf{x}'\right|}
\rho_{\sigma'}^{\dag}(\mathbf{x}'),
\end{eqnarray}
where the fermion density operator is $\rho_{\sigma}(\mathbf{x})
\equiv \psi_{\sigma}^{\dag}(\mathbf{x})\psi_{\sigma}(\mathbf{x}) =
{\bar \psi}_{\sigma}(\mathbf{x})\gamma^{0}
\psi_{\sigma}(\mathbf{x})$. In order to use our approach, it is
convenient to introduce an auxiliary scalar field $a_{0}$ and then
to re-express the Coulomb interaction by the following Lagrangian
density \cite{Son07, Barnes14}
\begin{eqnarray}
\mathcal{L}_{b}[a_{0}] &=& a_{0} \frac{\mathbb{D}}{2}a_{0}, \\
\mathcal{L}_{fb}[a_{0},\psi,{\bar\psi}] &=& -i g
\sum^N_{\sigma=1}a_{0}{\bar \psi}_{\sigma} \gamma^{0}\psi_{\sigma}.
\end{eqnarray}
After making Fourier transformations, the inverse of operator
$\mathbb{D}$ is converted into the free boson propagator, which is
$D_{0}(\mathbf{q}) = \frac{2\pi e^{2}}{v\epsilon|\mathbf{q}|}$ in
$(1+2)$ dimensions and $D_{0}(\mathbf{q}) = \frac{4\pi
e^{2}}{v\epsilon|\mathbf{q}|^{2}}$ in $(1+3)$ dimensions. Notice there are no self-coupling terms of the boson field $a_{0}$. This is because the Coulomb interaction originates from a U(1) gauge interaction.

\subsection{Fermion-phonon interaction}

Phonons are generated by the vibration of lattices, and exist in all
semimetals. The free motion of phonon field and its coupling to
Dirac fermions are described by
\begin{eqnarray}
\mathcal{L}_{b}[\varphi] &=&
\varphi^{\dag}\frac{\mathbb{D}}{2}\varphi,
\\
\mathcal{L}_{fb}[\varphi,\psi,{\bar\psi}] &=& -i g
\sum^N_{\sigma=1}\varphi{\bar\psi}_{\sigma} \gamma^{0}\psi_{\sigma},
\end{eqnarray}
where the operator $\mathbb{D} = -\frac{\partial_{t}^2 +
\Omega_{\mathbf{\nabla}}^2}{\Omega_{\mathbf{\nabla}}}$ with
$\Omega_{\mathbf{\nabla}}$ being the real-space correspondence of
phonon dispersion $\Omega_{\mathbf{q}}$. The coupling of massless
Dirac fermions to phonons has attracted considerable interest,
especially in the context of graphene. But most theoretical studies
are based on either first-principle calculations or weak-coupling ME
theory. The strong fermion-phonon coupling regime is rarely
considered. While the Migdal theorem is valid in ordinary metals
with a large Fermi surface, it turns out to break down in Dirac
semimetals whose Fermi surface shrinks to isolated points.

Our approach is applicable to electron-phonon interaction as long as
the free motion of phonons is described by harmonic oscillation,
namely, the action does not contain self-coupling between $\varphi$
fields. The harmonic oscillation approximation works well in most
realistic crystals, and such self-coupling terms as
$(\varphi^{\dag}\varphi)^{2}$ are usually irrelevant in the
low-energy region.

\subsection{Yukawa interaction near quantum critical point}

When a Dirac fermion system undergoes a continuous quantum phase
transition, the originally gapless semimetal is turned into a
distinct ordered phase, which might exhibit superconductivity, CDW,
antiferromagnetism, or electronic nematicity. Near the quantum
critical point, the quantum fluctuation of the corresponding order
parameter could be very strong and result in a variety of remarkable
quantum critical phenomena \cite{Kim08, Wangmit17, Pan18,
Liunematic12, Wang19, Xiao19, Lang19, LeeSS07, Grover14, Jian15,
Liu19npj}.

The quantum fluctuation of an order parameter is described by a
scalar boson field $\phi$, whose free Lagrangian density is
\begin{eqnarray}
\mathcal{L}_b = \frac{1}{2}\left[(\partial_t\phi)^2 - (\nabla
\phi)^2 - r\phi^2\right],
\end{eqnarray}
in which the operator $\mathbb{D} = -(\partial_{t}^{2} - \nabla^{2}
- r)$. Here, the effective boson mass $r$ measures the distance of
the system to quantum critical point, with $r=0$ at the transition.
In momenta space, the free boson propagator is known to be
\begin{eqnarray}
D_{0}(q) = \frac{1}{q^{2}+r}.
\end{eqnarray}
The fermion-boson coupling term is already given by
Eq.~(\ref{Eq:yukawacoupling}). The expression of $\gamma^{m}$
appearing in Eq.~(\ref{Eq:yukawacoupling}) is determined by the
definition of order parameter. For an order parameter defined by
$\langle \bar{\psi}M_{\mathrm{OP}}\psi \rangle$, one should identity
$\gamma^{m} = M_{\mathrm{OP}}$. If the boson represents the quantum
fluctuation of an excitonic order parameter \cite{Pan18}, which is
of the form $\bar{\psi}\psi$, one should choose $\gamma^{m} = I$.
When $(1+2)$-dimensional Dirac fermions couple to nematic quantum
fluctuations \cite{Kim08, Liunematic12}, $\gamma^{m} = \gamma^{1}$
or $\gamma^{m}=\gamma^{2}$.

Different from the two cases of Coulomb interaction and
fermion-phonon interaction, there is an additional self-coupling
term for order-parameter fluctuation:
\begin{eqnarray}
\mathcal{L}_{\phi^4} = u\phi^{4}(x).
\end{eqnarray}
The existence of this additional term makes the DS equations much
more complicated. Only when such a $\phi^4$ term is absent, could
our approach be exact. We will discuss this issue in greater details
in Sec.~\ref{Sec:D0D}.

\section{Dyson-Schwinger equations of correlation functions \label{Sec:DSEs}}

In this section we do not specify the physical origin of the boson
field $\phi$, and most of our results are independent of what the
boson field stands for.

In quantum field theory and quantum many-body theory, all the
physical quantities are defined in terms of various $n$-point
correlation functions
\begin{eqnarray}
\langle \mathcal{O}_{1}\mathcal{O}_{2}...\mathcal{O}_{n}\rangle,
\end{eqnarray}
where $\mathcal{O}$'s are Heisenberg operators and
$\langle...\rangle$ indicates that the statistical average is
carried out over all the possible configurations. The full fermion
and boson propagators are two two-point correlation functions
defined as
\begin{eqnarray}
G(x) &=& -i\langle\psi \bar{\psi}\rangle, \\
D(x) &=& -i\langle \phi \phi^{\dag}\rangle.
\end{eqnarray}
In the non-interacting limit, they are reduced to free propagators
\begin{eqnarray}
G_{0}(x) &=& -i\langle\psi\bar{\psi}\rangle_{0}, \\
D_{0}(x) &=& -i\langle\phi\phi^{\dag}\rangle_{0}.
\end{eqnarray}
In the momentum space, the free fermion propagator has the form
$G_{0}(p)=\frac{1}{\gamma^{\mu}p_{\mu}}$. The expression of free
boson propagator is model dependent, as already discussed in
Sec.~\ref{Sec:Model}.

As shown in Appendix \ref{App:DSEs}, the free and full propagators
are related by the following self-consistent DS integral equations
\begin{eqnarray}
G^{-1}(p) &=& G^{-1}_{0}(p) + ig^2 \int \frac{dk}{(2
\pi)^{(1+d)}}\gamma^{m}G(k)D(k-p)\Gamma_{\mathrm{int}}(k,p),
\label{eq:DSEG} \\
D^{-1}(q) &=& D^{-1}_{0}(q) - ig^2 N\int\frac{dk}{(2
\pi)^{(1+d)}}{\mathrm{Tr}}\left[\gamma^{m} G(k+q)
\Gamma_{\mathrm{int}}(k+q,k)G(k)\right], \label{eq:DSED}
\end{eqnarray}
where $dk\equiv dk_{0}d^{d}\mathbf{k}$. For simplicity, the DS
equations are expressed in the momentum space. These two DS
equations can be derived rigorously by performing field-theoretic
analysis within the framework of functional integral (calculational
details are presented in Appendix \ref{App:DSEs}). Here,
$\Gamma_{\mathrm{int}}(k,p)$ stands for the proper (external-legs
truncated) fermion-boson vertex function defined via the following
three-point correlation function
\begin{eqnarray}
D(k-p)G(k)\Gamma_{\mathrm{int}}(k,p)G(p) = \langle \phi \psi {\bar
\psi}\rangle.\label{eq:Gammaint}
\end{eqnarray}
To determine propagators $G(p)$ and $D(q)$, one needs to first
specify the vertex function $\Gamma_{\mathrm{int}}(k,p)$. By
carrying out functional calculations, one can show that
$\Gamma_{\mathrm{int}}$ satisfies its own DS equation
\begin{eqnarray}
\Gamma_{\mathrm{int}}(k,p) = \gamma^{m} - \int \frac{dp'}{(2
\pi)^{(1+d)}} G(p'+k) \Gamma_{\mathrm{int}}(k,p')G(p')K_{4}(p,p',k),
\end{eqnarray}
where $K_{4}(p,p',q)$ denotes the kernel function defined via a
four-point correlation function $\langle \psi{\bar \psi}\psi {\bar
\psi}\rangle$, namely
\begin{eqnarray}
G(p+p'+k)G(p')K_{4}(p,p',k)G(p)G(k) = \langle \psi {\bar \psi}\psi
{\bar \psi}\rangle.
\end{eqnarray}
$K_{4}(p,p',q)$ also satisfies its own DS integral equation that in
turn is associated with five-, six-, and higher-point correlation
functions. Repeating the same manipulations, one would derive an
infinite hierarchy of coupled integral equations \cite{Itzykson}.
The full set of DS integral equations are exact and contain all the
interaction-induced effects. Unfortunately, they seem not to be
closed and thus are intractable. This seriously hinders the
application of DS equations to realistic physical systems.

To make the DS equations closed, a frequently used strategy is to
introduce hard truncations. For instance, one might argue that all
the four- and higher-point correlation functions are unimportant so
that the fermion-boson vertex function can be replaced by its bare
expression, i.e.,
\begin{eqnarray}
\Gamma_{\mathrm{int}}(k,p) \to \gamma^{m}. \nonumber
\end{eqnarray}
This approximation is known as the Migdal's theorem \cite{Migdal}.
As long as the Migdal's theorem is valid, one can ignore all the
vertex corrections and simplify the DS equations (\ref{eq:DSEG}) and
(\ref{eq:DSED}) to
\begin{eqnarray}
G^{-1}(p) &=& G^{-1}_{0}(p) + ig^2 \int \frac{dk}{(2
\pi)^{(1+d)}}\gamma^{m}G(k)D(k-p)\gamma^{m}, \nonumber \\
D^{-1}(q) &=& D^{-1}_{0}(q)- ig^2 N\int\frac{dk}{(2\pi)^{(1+d)}}
{\mathrm{Tr}}\left[\gamma^{m}G(k+q)\gamma^{m} G(k)\right].\nonumber
\end{eqnarray}
These two coupled equations are often called ME equations, since
they are formally similar to the ME equations originally derived to
describe phonon-mediated superconductivity \cite{AGD, Migdal,
Eliashberg}. In practical studies of ME equations, one often uses
the free boson propagator $D_{0}(q)$ to approximate the full
propagator $D(q)$, or employs random phase approximation (RPA) to
express the boson propagator as $D(q) =
\frac{1}{D_{0}^{-1}(q)-\Pi_{\mathrm{RPA}}(q)}$, where the
polarization function $\Pi_{\mathrm{RPA}}(q)$ is approximately
computed by using the free fermion propagator $G_{0}(p)$ and the
bare vertex. However, the Migdal's theorem is not always valid, and
it breaks down in a large number of strongly correlated systems
\cite{Kivelson18, Liu19}. In systems where Migdal's theorem becomes
invalid, we need to carefully incorporate the contributions of
fermion-boson vertex corrections into both $G(p)$ and $D(q)$. This
is extremely difficult because the full vertex function
$\Gamma_{\mathrm{int}}(k,p)$ contains an infinite number of Feynman
diagrams. Computing the simplest triangle diagram of vertex
corrections is already very difficult, let alone the more
complicated multi-loop diagrams. When the fermion-boson interaction
becomes strong, there is no reason to expect that lower-order
diagrams make more significant contributions than higher-order
diagrams. As discussed in Sec.~\ref{Sec:Introdunction}, generalizing
the fermion flavor $N$ to an unphysically large value does not help
solve the problem. Another possible strategy is to assume (in most
cases without a convincing reason) some kind of \emph{Ansatz} for
the vertex function, and then to insert it into the DS equations of
$G(p)$ and $D(q)$. Nevertheless, this kind of \emph{Ansatz} usually
comes from unjustified experience and hence is \emph{ad hoc}.

In Ref.~\cite{Liu19}, we have developed an efficient
non-perturbative approach to determine the electron-phonon vertex
corrections. It is not necessary to compute any specific Feynman
diagram of vertex corrections nor to introduce any \emph{Ansatz}.
The core idea of our approach \cite{Liu19} is to incorporate the
full vertex function into DS equations of $G(p)$ and $D(q)$ by
utilizing two coupled WTIs derived from two global U(1) symmetries.
However, different from the electron-phonon system considered in
Ref.~\cite{Liu19}, the Dirac fermion systems do not have
sufficiently many symmetries to entirely determine the vertex
function. To obtain the exact vertex function, we will generalize
the approach proposed in Ref.~\cite{Liu19} and use both symmetric
and asymmetric global U(1) transformations to derive all the related
WTIs.

\section{Generalized Ward-Takahashi identities \label{Sec:WTIs}}

The fermion propagator and vertex function are connected via a
number of generalized WTIs. The aim of this section is to derive all
the involved WTIs. The basic strategy adopted here was originally
proposed by Takahashi \cite{Takahashi86} in the context of quantum
gauge theories, and later re-formulated by Kondo \cite{Kondo97} and
He \emph{et al.} \cite{He01} in the context of quantum
electrodynamics (QED). The application of this method in
(1+3)-dimensional QED was not successful, and the WTIs seem not to
be closed due to the complexity of the model. Indeed, QED exhibits
both Lorentz invariance and local gauge invariance. Due to the Lorentz
invariance, a large number of WTIs are coupled to each other and
thus intractable. It is very difficult to compute physical quantities, because one usually needs to introduce a Wilson line to maintain local gauge invariance. Moreover, there might be anomalies in gauge theories. For the idea of Takahashi to work, it would be more suitable to consider condensed matter systems that do not respect Lorentz symmetry nor local gauge symmetry. In Ref.~\cite{Liu19}, we have shown that the full electron-phonon vertex function can be
determined by two coupled WTIs in metals with a finite Fermi
surface. Now we generalize the approach to Dirac fermion systems.

It should be emphasized that there are two types of vertex
functions: one is the interaction vertex function
$\Gamma_{\mathrm{int}}$ defined by Eq.~(\ref{eq:Gammaint}); the
other is called current vertex function $\Gamma_{M}^{\mu}$ because
it is defined by $\langle j^{\mu}_{M} \psi\bar{\psi}\rangle \sim
G\Gamma_{M}^{\mu} G$ with $j^{\mu}_{M}$ being a composite current
operator. The interaction vertex function $\Gamma_{\mathrm{int}}$
enters into the DS equations of fermion and boson propagators, as
shown by Eq.~(\ref{eq:DSEG}) and Eq.~(\ref{eq:DSED}), and therefore
is the quantity that we really need. It should be noted that
$\Gamma_{\mathrm{int}}$ does not necessarily satisfy any WTI. It is
the current vertex function $\Gamma_{M}^{\mu}$ that enters into
various WTIs, since $\Gamma_{M}^{\mu}$ is related to some type of symmetry-induced current. The exact relation between interaction
and current vertex functions will be derived in Sec.~\ref{Sec:D0D}.
The aim of this section is to demonstrate how to determine current
vertex functions. We will first define a number of generalized
current operators and then use them to derive the corresponding current vertex
functions. All the current vertex functions can be unambiguously
obtained if we could find a sufficient number of coupled WTIs.

It is known that the action of the system respects a global U(1)
symmetry, defined by a global change of the phase of fermion field,
i.e.,
\begin{eqnarray}
\psi_{\sigma}(x) \to e^{i\theta}\psi_{\sigma}(x), \nonumber
\end{eqnarray}
where $\theta$ is supposed to be an infinitesimal constant.
According to Noether theorem, this symmetry leads to the
conservation of current $j^{\mu}(x)={\bar
\psi}_{\sigma}(x){\gamma}^{\mu}\psi_{\sigma}(x)$, namely
$\partial_{\mu}j^{\mu}(x)=0$. The relation between symmetry and
conserved current is always valid at the classical level. When the
fields are quantized, such a symmetry is converted into a universal
relation between two- and three-point correlation functions. In
particular, the current vertex function and the fermion
propagator satisfy a WTI. But the current vertex function
$\Gamma_{M}^{\mu}$ defined via this current has three components in
$(1+2)$ dimensions and four components in $(1+3)$ dimensions, and
thus cannot be determined by one single WTI. $\Gamma_{M}^{\mu}$
could be unambiguously determined only when there are a sufficient
number of WTIs. Remarkably, there do exist several additional WTIs
that couple to the ordinary WTI. Nevertheless, the additional WTIs
are hidden and should be found out very carefully.

We now demonstrate how to derive all the related WTIs. It turns out
the functional integral formulation of quantum field theory provides
the most compact and elegant framework for the derivation of
intrinsic relations between correlation functions. Using functional
integral techniques \cite{Itzykson}, the mean value of operator
$\mathcal{O}(x)$, which might be the product of an arbitrary number
of field operators, is defined as
\begin{eqnarray}
\langle \mathcal{O}(x)\rangle_J =
\frac{[[\mathcal{O}(x)]]_J}{[[1]]_J},
\end{eqnarray}
where the numerator is given by
\begin{eqnarray}
[[\mathcal{O}(x)]]_J = \int \mathcal{D}\phi \mathcal{D}\psi_\sigma
\mathcal{D}{\bar\psi}_{\sigma} \mathcal{O}(x)
\exp\left(i\int{dx}[\mathcal{L} + J\phi+ {\bar
\eta}_{\sigma}\psi_\sigma + {\bar\psi}_{\sigma}\eta_\sigma]\right),
\end{eqnarray}
and the denominator is just the partition function
\begin{eqnarray}
[[1]]_J \equiv \mathcal{Z}[J,{\bar \eta},\eta] = \int
\mathcal{D}\phi \mathcal{D}\psi_\sigma \mathcal{D}{\bar
\psi}_{\sigma} \exp\left(i\int {dx}[\mathcal{L} + J\phi+ {\bar
\eta}_{\sigma}\psi_\sigma + {\bar\psi}_{\sigma}\eta_\sigma]\right).
\end{eqnarray}
Here, $J$, $\eta$, and $\bar\eta$ are the external sources of
$\phi$, $\bar\psi$, and $\psi$, respectively. For notational
simplicity, we will use one single subscript $J$ to stand for all
the possible external sources, i.e., $\langle \mathcal{O}\rangle_{J}
\equiv \langle \mathcal{O} \rangle_{J,\eta,{\bar \eta}}$.

The partition function $\mathcal{Z}$, also known as the generating
functional of correlation functions \cite{Itzykson}, should be
invariant under an arbitrary infinitesimal variation of any field
operator. Based on the fact that $\delta \mathcal{Z} = 0$ for any $\delta {\bar \psi}$, we obtain the following average of the equation of motion (EOM) of field operator $\psi(x)$ in the presence of external sources
\begin{eqnarray}
\langle{i{\gamma}^\mu \partial_\mu \psi_\sigma(x) +
g\phi(x)\gamma^m\psi_\sigma(x)+\eta_\sigma(x)}\rangle_J = 0.
\label{Eq:EOMpsi}
\end{eqnarray}
Now we introduce a $4\times 4$ matrix $\Theta$, and require that it
satisfies either the condition
\begin{eqnarray}
{\widehat \Theta}\equiv\gamma^0 \Theta^{\dagger}\gamma^0 = \Theta,
\label{Eq:constrain1}
\end{eqnarray}
which henceforth is referred to as constraint I, or another
condition
\begin{eqnarray}
{\widehat \Theta}\equiv\gamma^0 \Theta^{\dagger}\gamma^0 = -\Theta,
\label{Eq:constrain2}
\end{eqnarray}
which henceforth is referred to as constraint II. We multiply
$\Theta$ to the average of EOM given by Eq.~(28) from the left side,
and then find that
\begin{eqnarray}
\langle{i\Theta {\gamma}^\mu \partial_\mu
\psi_\sigma(x)+g\phi(x)\Theta \gamma^m\psi_\sigma(x)+\Theta
\eta_\sigma(x)}\rangle_J = 0.
\end{eqnarray}
Performing functional derivative $\frac{\delta}{-i\delta\eta(y)}$ on
this equation leads us to
\begin{eqnarray}
\langle i{{\bar \psi}_{\sigma}(y)\Theta {\gamma}^\mu
\partial_\mu \psi_\sigma(x)+g\phi(x){\bar \psi}_{\sigma}
(y)\Theta \gamma^m \psi_\sigma(x) + {\bar\psi}_{\sigma}
(y)\Theta\eta_\sigma (x) + i \delta(x-y)\mathrm{Tr}\Theta} \rangle_J
= 0. \label{Eq:EOMofpsi}
\end{eqnarray}
Similarly, since $\delta \mathcal{Z} = 0$ for any $\delta \psi$, we get the average of the EOM of
field operator ${\bar \psi}$:
\begin{eqnarray}
\langle i({\partial_\mu}{\bar \psi}_{\sigma}(x))\gamma^{\mu} -
g\phi(x){\bar \psi}_{\sigma}(x)\gamma^{m} - {\bar
\eta}_\sigma(x)\rangle_J = 0.\label{Eq:EOMbarpsi}
\end{eqnarray}
This time, we multiply $\Theta$ from the right side and then obtain
\begin{eqnarray}
\langle i({\partial_\mu}{{\bar\psi}_{\sigma}}(x)){\gamma}^\mu
\Theta-g\phi(x){\bar\psi}_{\sigma}\gamma^m \Theta - {\bar
\eta}_{\sigma}(x)\Theta\rangle_J = 0.
\end{eqnarray}
Accordingly, we should carry out functional derivative
$\frac{\delta}{i\delta {\bar \eta}(y)}$, which gives rise to
\begin{eqnarray}
\langle i({\partial_\mu}{{\bar\psi}_{\sigma}}(x)){\gamma}^\mu \Theta
\psi_\sigma(y) - g\phi(x){\bar\psi}_{\sigma}(x) \gamma^{m}\Theta
\psi_\sigma(y) - {\bar\eta}_{\sigma}(x)\Theta \psi_\sigma(y) -i
\delta(x-y)\mathrm{Tr}\Theta \rangle_J = 0. \label{Eq:EOMofbarpsi}
\end{eqnarray}

Comparing Eq.~(\ref{Eq:EOMofpsi}) and Eq.~(\ref{Eq:EOMofbarpsi}), we
observe that the Yukawa-coupling term, described by coupling
constant $g$, can be eliminated by proper manipulations. Now suppose
that $\Theta$ satisfies constraint I and one more constraint
\begin{eqnarray}
[\Theta,\gamma^m] \equiv \Theta\gamma^{m}-\gamma^{m}\Theta=0,
\label{Eq:constraint3}
\end{eqnarray}
which henceforth is referred to as constraint III. After adding
Eq.~(\ref{Eq:EOMofpsi}) to Eq.~(\ref{Eq:EOMofbarpsi}) and taking the
limit $x\rightarrow y$, we find the following identity holds
\begin{eqnarray}
\langle {\bar\psi}_{\sigma}(x)i\Theta {\gamma}^\mu
(\partial_{\mu}{\psi}_{\sigma}(x)) + (\partial_\mu
{\bar\psi}_{\sigma})i{\gamma}^\mu \Theta \psi_\sigma(x)+{\bar
\psi}_{\sigma}(x) \Theta\eta_\sigma(x) - {\bar \eta}_\sigma(x)\Theta
\psi_\sigma(x)\rangle_J = 0. \label{Eq:generalizednoether1}
\end{eqnarray}
Then we suppose $\Theta$ satisfies both constraint II and an
additional condition
\begin{eqnarray}
\{\Theta,\gamma^m\} \equiv \Theta\gamma^{m}+\gamma^{m}\Theta=0,
\label{Eq:constraint4}
\end{eqnarray}
which henceforth is referred to as constraint IV. For $\Theta$
satisfying constraints II and IV, we subtract
Eq.~(\ref{Eq:EOMofpsi}) from Eq.~(\ref{Eq:EOMofbarpsi}) and then
take the limit $x\rightarrow y$, which leads to another identity
\begin{eqnarray}
\langle -{\bar \psi}_{\sigma}(x)i\Theta {\gamma}^\mu
(\partial_{\mu}\psi_{\sigma}(x)) + (\partial_\mu
{\bar\psi}_{\sigma})i{\gamma}^\mu \Theta
\psi_\sigma(x)-{\bar\psi}_{\sigma}(x) \Theta\eta_\sigma(x) - {\bar
\eta}_\sigma(x)\Theta \psi_\sigma(x)\rangle_J = 0.
\label{Eq:generalizednoether2}
\end{eqnarray}

The two identities given by Eq.~(\ref{Eq:generalizednoether1}) and
Eq.~(\ref{Eq:generalizednoether2}) play a crucial role in our
approach and thus warrants a deeper analysis. Below we would like to
prove that these two identities can alternatively be
derived from a number of generalized global U(1) transformations.
For this purpose, we extend the ordinary global U(1) transformation
$\psi_{\sigma}\rightarrow e^{i\theta}\psi_{\sigma}$ for a particular
flavor $\sigma$ to the following more generic U(1) transformation
\begin{eqnarray}
&&{\psi'_\sigma} = {e^{i\theta \Theta}}\psi_\sigma =
\psi_{\sigma}+\Delta\psi_\sigma, \label{Eq:genericvariation1}
\\
&&{{\bar\psi}_{\sigma}}' = {\bar \psi}_{\sigma}e^{-i\theta
\widehat{\Theta}}= {\bar \psi}_{\sigma} + \Delta{\bar\psi}_{\sigma}
\label{Eq:genericvariation2},
\end{eqnarray}
where $\Theta$ is an arbitrary $4\times 4$ Hermitian or
anti-Hermitian matrix satisfying either constraint I or constraint
II. The infinitesimal variations of field operators are
\begin{eqnarray}
\Delta\psi_{\sigma} = i\theta \Theta \psi_{\sigma}, \quad \Delta
{\bar\psi}_{\sigma} = -i\theta{\bar\psi}_{\sigma}\widehat{\Theta}.
\end{eqnarray}
Under the above generic transformations, the change of the total
action is
\begin{eqnarray}
\Delta S &=& S[{\psi'}_\sigma,{\bar\psi}_{\sigma}'] -
S[{\psi_\sigma},{\bar\psi}_{\sigma}]
\nonumber\\
&=& -i\theta \int dx \{{\bar\psi}_{\sigma}{\widehat \Theta} i
\gamma^{\mu} \partial_\mu{\psi_\sigma}+(\partial_\mu
{\bar\psi}_{\sigma}) i \gamma^\mu{\Theta}
{\psi_\sigma} \nonumber \\
&& + g \phi({\bar\psi}_{\sigma}{\widehat \Theta}\gamma^m
{\psi_\sigma}-{\bar\psi}_{\sigma}\gamma^m {\Theta}{\psi_\sigma})
+{\bar\psi}_{\sigma}{\widehat \Theta}{\eta_\sigma} -
{\bar\eta}_{\sigma}{\Theta}{\psi_\sigma}\}.
\end{eqnarray}
In this expression, ${\bar\psi}_{\sigma}{\widehat \Theta} i
\gamma^\mu \partial_\mu{\psi_\sigma}+(\partial_\mu
{\bar\psi}_{\sigma}) i \gamma^\mu{\Theta} {\psi_\sigma}$
comes from the infinitesimal variation of the free fermion term,
i.e., $\Delta\mathcal{L}_{f}$, and is bilinear in spinor field. In
comparison, $g \phi({\bar\psi}_{\sigma}{\widehat \Theta}\gamma^m
{\psi_\sigma}-{\bar\psi}_{\sigma}\gamma^m {\Theta}{\psi_\sigma})$
comes from the infinitesimal variation of the Yukawa coupling term,
i.e., $\Delta\mathcal{L}_{fb}$. The quantum many-body system under
consideration should be thermodynamically stable and robust against
an arbitrary infinitesimal variation of spinor field. This means
that the partition function $\mathcal{Z}$, which sums over all the
possible field configurations, must be invariant under the
transformations defined by
Eqs.~(\ref{Eq:genericvariation1}) and (\ref{Eq:genericvariation2}) for any
small parameter $\theta$. Therefore, the following equation should
be valid
\begin{eqnarray}
\langle {\bar\psi}_{\sigma}{\widehat \Theta} i \gamma^{\mu}
\partial_\mu {\psi_\sigma}+(\partial_{\mu}{\bar\psi}_{\sigma})
i \gamma^\mu{\Theta}{\psi_\sigma} + g
\phi({\bar\psi}_{\sigma}{\widehat \Theta}\gamma^m {\psi_\sigma} -
{\bar\psi}_{\sigma}\gamma^m {\Theta}{\psi_\sigma}) +
{\bar\psi}_{\sigma}{\widehat \Theta}{\eta_\sigma} -
{\bar\eta}_{\sigma}{\Theta}{\psi_\sigma} \rangle_J = 0.
\label{Eq:genericidentity}
\end{eqnarray}
We are particularly interested in two cases. Firstly, if the matrix
$\Theta$ satisfies constraints I and III simultaneously, the third
term in the l.h.s of this equation vanishes, which leads to
Eq.~(\ref{Eq:generalizednoether1}). Secondly, if $\Theta$ satisfies
constraints II and IV simultaneously, the third term in the l.h.s of
this equation also vanishes, which leads to
Eq.~(\ref{Eq:generalizednoether2}).

The two identities Eq.~(\ref{Eq:generalizednoether1}) and
Eq.~(\ref{Eq:generalizednoether2}) can be regarded as a generalized
version of the Noether theorem. To understand this, let us take a
further look at the generic U(1) transformations defined by
Eqs.~(\ref{Eq:genericvariation1}) and (\ref{Eq:genericvariation2}). In
principle, after performing such transformations, the total
Lagrangian $\mathcal{L} = \mathcal{L}_{f}+\mathcal{L}_{fb}+\mathcal{L}_{b}$ would be modified in three possible ways:

(1) For some special choices of $\Theta$, the total Lagrangian
$\mathcal{L}$ is invariant in the absence of external sources. In
this case, the transformation $\psi_{\sigma}\rightarrow
e^{i\theta\Theta}\psi_{\sigma}$ should be identified as a symmetry
transformation. The simplest choice of this type is $\Theta = I$. At
the level of classical field theory, Noether theorem tells us that
the electric current $j^{\mu}(x)={\bar \psi}\gamma^{\mu}\psi$ is
conserved and satisfies $\partial_{\mu}j^{\mu}=0$. In the framework
of quantum field theory, current conservation should be re-phrased
as the vanishing of the mean value of $\partial_{\mu}j^{\mu}$,
namely $\langle \partial_{\mu}j^{\mu}\rangle = 0$. In the presence
of external sources, which are introduced to generate correlation
functions, the mean value $\langle \partial_{\mu}j^{\mu}\rangle$ no
longer vanishes but instead satisfies a Slavnov-Taylor identity
(STI) \cite{Itzykson, Liu19}
\begin{eqnarray}
i\langle\partial_{\mu}j^{\mu}\rangle_J = \langle{\bar
\eta}\psi\rangle_J - \langle{\bar \psi}\eta\rangle_J,
\end{eqnarray}
which can be easily obtained from Eq.~(\ref{Eq:genericidentity}) by
taking $\Theta = I$. This STI is reduced to $\langle
\partial_{\mu} j^{\mu}\rangle = 0$ only in the zero-source
limit $J=\eta={\bar \eta}=0$. Apparently, the ordinary Noether
theorem is just the zero-source limit of one special ($\Theta$ being
unit matrix) form of the generalized identity given by
Eq.~(\ref{Eq:genericidentity}). After performing functional
derivatives of the STI with respective to external sources, one
would obtain (see Ref.~\cite{Liu19} for details) a WTI that relates
the vertex function defined via conversed current $j^{\mu}$ to the
full fermion propagator. If a system has two global U(1) symmetries,
there would be two STIs and, accordingly, two WTIs. For instance,
the interacting electron-phonon system investigated in
Ref.~\cite{Liu19} has two global U(1) symmetries, corresponding to
charge conservation and spin conservation, respectively, which then
leads to two WTIs. As shown in Ref.~\cite{Liu19}, the charge-related
WTI and the spin-related WTI are indeed coupled to each other.
Making use of such a crucial fact, the time- and spatial-components
of current vertex functions can be completely determined and
expressed purely in terms of full fermion propagator.

(2) The Dirac fermion systems are more complicated than the
electron-phonon system studied in Ref.~\cite{Liu19}. The spinor
field of Dirac fermion has four components, and the number of
current vertex functions are larger than that of global U(1)
symmetries. That means, symmetry-induced WTIs are not sufficient to
determine current vertex functions. In this paper, we develop a very
powerful method to obtain a sufficient number of generalized WTIs
based on both symmetric and asymmetric global U(1) transformations.
Below we demonstrate how to employ our method. Now suppose the
matrix $\Theta$ is carefully selected such that the global
transformations $\psi_{\sigma}\rightarrow
e^{i\theta\Theta}\psi_{\sigma}$ leave the fermion-boson coupling
term $\mathcal{L}_{fb}$ unchanged but alter the free fermion term
$\mathcal{L}_{f}$. The boson sector $\mathcal{L}_{b}$ is always
invariant under U(1) transformations of spinor field and thus will
not be discussed further. Now the generalized identity
Eq.~(\ref{Eq:genericidentity}) becomes
\begin{eqnarray}
\langle {\bar\psi}_{\sigma}{\widehat \Theta} i \gamma^{\mu}
\partial_\mu {\psi_\sigma}+(\partial_{\mu}{\bar\psi}_{\sigma})
i\gamma^{\mu}{\Theta}{\psi_\sigma} + {\bar\psi}_{\sigma}{\widehat
\Theta}{\eta_\sigma}-{\bar\eta}_{\sigma}{\Theta}{\psi_\sigma}
\rangle_J = 0, \label{Eq:asymmetricidentity}
\end{eqnarray}
which are consistent with Eq.~(\ref{Eq:generalizednoether1}) and
(\ref{Eq:generalizednoether2}). Notice that the transformations
$\psi_{\sigma}\rightarrow e^{i\theta\Theta}\psi_{\sigma}$ cannot be
identified as symmetries of the system since they do not keep
$\mathcal{L}_{f}$ invariant. Therefore, there is no conserved
current even in the zero-source limit and the first two terms
appearing in the mean value of Eq.~(\ref{Eq:asymmetricidentity})
cannot be expressed as the divergence of any current operator.
However, the identity given by Eq.~(\ref{Eq:asymmetricidentity}), or
equivalently by Eq.~(\ref{Eq:generalizednoether1}) and
Eq.~(\ref{Eq:generalizednoether2}), can still generate a number of
useful exact relations between two- and three-point correlation
functions.

(3) For all the other choices of $\Theta$, the interaction term
$\mathcal{L}_{fb}$ is changed by the transformations
$\psi_{\sigma}\rightarrow e^{i\theta\Theta}\psi_{\sigma}$. Although
the generic identity given by Eq.~(\ref{Eq:genericidentity}) is
still valid, it is rarely useful no matter whether $\mathcal{L}_{f}$
is invariant or not. The reason of this fact will become clear soon.

We deliberately choose the $\Theta$ matrices to satisfy constraints
I and III simultaneously or satisfy constraints II and IV
simultaneously. Then the first two possibilities can be unified. We
obtain Eq.~(\ref{Eq:generalizednoether1}) for $\Theta$ matrices
satisfying constraints I and III, and
Eq.~(\ref{Eq:generalizednoether2}) for $\Theta$ matrices satisfying
constraints II and IV. To illustrate the importance of these two
identities, we perform functional derivatives $\frac{\delta}{i\delta
{\bar \eta}_{\alpha}(y)}$ and
$\frac{\delta}{-i\delta{\eta}_\beta(z)}$ in order (here $\alpha$ and
$\beta$ denotes the $\alpha$ and $\beta$ components of $\sigma$) and
set $J = \eta = {\bar \eta} = 0$ at the end. For flavor $\sigma$,
such operations turn Eq.~(\ref{Eq:generalizednoether1}) into
\begin{eqnarray}
&&\partial_{\mu}\langle{\bar\psi}_{\sigma}(x)\frac{1}{2}\left
\{\Theta,\gamma^{\mu}\right\} \psi_{\sigma}(x)\psi_{\alpha}(y){\bar
\psi}_{\beta}(z)\rangle_{c} = -
\delta(x-y)\langle\Theta\psi_{\alpha}(y){\bar \psi}_{\beta}(z)
\rangle_{c} \nonumber \\
&&+\delta(x-z)\langle\psi_{\alpha}(y){\bar\psi}_{\beta}(z)
\Theta\rangle_{c} + \langle{\bar \psi}_{\sigma}(x)
\frac{1}{2}[\Theta,\gamma^{\mu}]({\overleftarrow \partial}_{\mu} -
{\partial}_{\mu})\psi_{\sigma}(x) \psi_{\alpha}(y){\bar
\psi}_{\beta}(z)\rangle_{c}. \label{Eq:STI1}
\end{eqnarray}
Here, the notation $\langle ...\rangle_c$ indicates that only
connected Feynman diagrams are taken into account. The
transformation $\psi_{\sigma}\rightarrow
e^{i\theta\Theta}\psi_{\sigma}$ may or may not be a symmetry of the
system. Below we discuss these two cases separately.

If $\psi_{\sigma}\rightarrow e^{i\theta\Theta}\psi_{\sigma}$ is a
symmetry of the system, $\Theta$ must commutate with all ${
\gamma}^{\mu}$'s, obeying $[\Theta,{\gamma}^{\mu}]=0$. Then the
above identity can be re-written as
\begin{eqnarray}
\langle \partial_{\mu} j^{\mu}_{\sigma}(x) \psi_{\alpha}(y){\bar
\psi}_{\beta}(z) \rangle_{c} = -\delta(x-y)
\langle\Theta\psi_{\alpha}(y){\bar \psi}_{\beta}(z) \rangle_{c} +
\delta(x-z)\langle\psi_{\alpha}(y){\bar
\psi}_{\beta}(z)\Theta\rangle_{c}, \label{Eq:symmetricwti1}
\end{eqnarray}
where $j^{\mu}_{\sigma}(x)={\bar\psi}_{\sigma}(x)\frac{1}{2}\left
\{\Theta,{\gamma}^{\mu}\right\} \psi_{\sigma}(x)$ is a
symmetry-induced conserved current. To proceed, we introduce a
generic current operator
\begin{eqnarray}
j^{\mu}_{M}(x) = {\bar \psi}_{\sigma}(x) M^{\mu}\psi_\sigma(x),
\label{Eq:vectorcurrent}
\end{eqnarray}
where $M^{\mu}$ is a matrix. Note that this current does not need to
be conserved. Although in principle $M^{\mu}$ could be any matrix,
here we are particularly interested in two sorts of expressions
\begin{eqnarray}
M^{\mu} = \frac{1}{2}\{\Theta,{\gamma}^{\mu}\} \quad \mathrm{and}
\quad M^{\mu} = \frac{1}{2}[\Theta,{\gamma}^{\mu}].
\end{eqnarray}
The above composite current operator can be used to define the
following correlation function \cite{Liu19, Takahashi57, Boyer67}
\begin{eqnarray}
\langle{j^{\mu}_M(x)\psi_\alpha(y){\bar\psi}}_\beta(z)\rangle_c =
\int d\xi_1 d\xi_2 \left(G(y-\xi_1) \Gamma_{M}^{\mu}(\xi_1-x,
x-\xi_2) G(\xi_2-z)\right)_{\alpha\beta}, \label{Eq:currentvertex}
\end{eqnarray}
where the current vertex function
$\Gamma_{M}^{\mu}(\xi_1-x,x-\xi_2)$ is obtained by truncating the
two external legs (i.e., external fermion propagators) of
$\langle{j^\mu_M(x)\psi_\alpha(y){\bar\psi}}_\beta(z)\rangle_c$. The
Fourier transformations of the Dirac fermion propagator and the
current vertex function are given by
\begin{eqnarray}
G(y-\xi_1)=\int \frac{dk}{(2\pi)^{(1+d)}} e^{-ik (y-\xi_1)}{ G(k)},
\quad G(\xi_2-z)=\int \frac{dp}{(2\pi)^{(1+d)}} e^{-ip (\xi_2-z)}{
G(p)},
\end{eqnarray}
and
\begin{eqnarray}
\Gamma_{M}^{\mu}(\xi_1-x,x-\xi_2)=\int \frac{dk dp}{(2\pi)^{2(1+d)}}
\Gamma_{M}^{\mu}(k,p) e^{-ik(\xi_1-x)-ip(x-\xi_2)}.
\end{eqnarray}
After carrying out Fourier transformations, we will obtain a number
of exact identities between the current vertex function
$\Gamma_{M}^{\mu}(k,p)$ and the full fermion propagator $G(k)$. In
the simplest case, $\Theta=I$, we would turn
Eq.~(\ref{Eq:symmetricwti1}) into
\begin{eqnarray}
(k_{\mu} - p_{\mu})\Gamma_{{\gamma}^{\mu}}(k,p) = -G^{-1}(k) +
G^{-1}(p),
\end{eqnarray}
which is precisely the ordinary, U(1)-symmetry induced WTI.

If $\psi_{\sigma}\rightarrow e^{i\theta\Theta}\psi_{\sigma}$ is not
a symmetry of the system, $\Theta$ does not commutate with all
${\gamma}^{\mu}$'s. In this case, the identity given by
Eq.~(\ref{Eq:STI1}) becomes
\begin{eqnarray}
\langle \partial_{\mu} j^{\mu}_{\sigma}(x) \psi_{\alpha}(y){\bar
\psi}_{\beta}(z) \rangle_{c} &=& -\delta(x-y)
\langle\Theta\psi_{\alpha}(y){\bar \psi}_{\beta}(z) \rangle_{c} +
\delta(x-z)\langle\psi_{\alpha}(y){\bar \psi}_{\beta}(z)
\Theta\rangle_{c} \nonumber \\
&& + \langle{\bar\psi}_{\sigma}(x)\frac{1}{2}[\Theta,{
\gamma}^{\mu}]({\overleftarrow \partial}_{\mu} -
{\partial}_{\mu})\psi_{\sigma}(x)\psi_{\alpha}(y){\bar
\psi}_{\beta}(z)\rangle_{c}. \label{Eq:asymmetricSTI1}
\end{eqnarray}
Since the last term of right-hand side (r.h.s.) does not identically
vanish, the current
$j^{\mu}_{\sigma}(x)={\bar\psi}_{\sigma}(x)\frac{1}{2}\left
\{\Theta,{\gamma}^{\mu}\right\} \psi_{\sigma}(x)$ is not conserved.
However, despite the absence of ordinary symmetry-induce WTI, we
emphasize that the identity given by Eq.~(\ref{Eq:asymmetricSTI1})
is still strictly valid and provides very useful information. The
key observation is that, one can identify
${\bar\psi}_{\sigma}(x)\frac{1}{2}[\Theta,\gamma^{\mu}]
\psi_{\sigma}(x)$ as a current operator and then use its divergence
to define another current vertex function $\Gamma_{M}^{\mu}$. In
fact, if we perform functional derivatives $\frac{\delta}{i\delta
{\bar \eta}_{\alpha}(y)}$ and
$\frac{\delta}{-i\delta{\eta}_\beta(z)}$ to
Eq.~(\ref{Eq:generalizednoether2}), we would obtain
\begin{eqnarray}
&&\partial_{\mu}\langle{\bar\psi}_{\sigma}(x)\frac{1}{2}
\left[\Theta,{\gamma}^{\mu}\right]\psi_{\sigma}(x)
\psi_{\alpha}(y){\bar \psi}_{\beta}(z)\rangle_{c} =
\delta(x-y)\langle\Theta\psi_{\alpha}(y){\bar \psi}_{\beta}(z)
\rangle_{c} \nonumber \\
&& +\delta(x-z)\langle\psi_{\alpha}(y){\bar \psi}_{\beta}(z)
\Theta\rangle_{c} -\langle{\bar \psi}_{\sigma}(x)\frac{1}{2}
\{\Theta,{\gamma}^{\mu}\}({\overleftarrow \partial}_{\mu} +
{\partial}_{\mu})\psi_{\sigma}(x)\psi_{\alpha}(y){\bar
\psi}_{\beta}(z)\rangle_{c}. \label{Eq:STI2}
\end{eqnarray}
It it important to notice that the divergence of the current
${\bar\psi}_{\sigma}(x)\frac{1}{2}[\Theta,\gamma^{\mu}]
\psi_{\sigma}(x)$ appears in the mean value of the left-hand side
(l.h.s.) of this identity. Since usually $\{\Theta,{\gamma}^{\mu}\}
\neq 0$, the bilinear operator ${\bar\psi}_{\sigma}(x)
\frac{1}{2}[\Theta,\gamma^{\mu}]\psi_{\sigma}(x)$ represents an
asymmetry-related, non-conserved current (its divergence does not
vanish). Although this current is not conserved, it is still very
useful. A remarkable fact is that, the two strictly valid identities
Eq.~(\ref{Eq:STI1}) and Eq.~(\ref{Eq:STI2}) are self-consistently
coupled. Now it is convenient to decompose the current vertex
functions $\Gamma_{M}^{\mu}(\xi_1-x,x-\xi_2)$ defined in terms of
$M^{\mu} = \frac{1}{2}\{\Theta,{\gamma}^{\mu}\} =
\frac{1}{2}\left(\Theta{\gamma}^{\mu} + {\gamma}^{\mu}\Theta\right)$
and $M^{\mu} = \frac{1}{2}[\Theta,{\gamma}^{\mu}] =
\frac{1}{2}\left(\Theta{\gamma}^{\mu} - {\gamma}^{\mu}\Theta\right)$
into two more elementary functions
$\Gamma_{\Theta{\gamma}^{\mu}}(\xi_1-x,x-\xi_2)$ and
$\Gamma_{{\gamma}^{\mu}\Theta}(\xi_1-x,x-\xi_2)$. The unknown
functions $\Gamma_{\Theta{\gamma}^{\mu}}(\xi_1-x,x-\xi_2)$ and
$\Gamma_{{\gamma}^{\mu}\Theta}(\xi_1-x,x-\xi_2)$ can be completely
determined by solving Eq.~(\ref{Eq:STI1}) and Eq.~(\ref{Eq:STI2}).

Next we Fourier transform Eq.~(\ref{Eq:STI1}) and
Eq.~(\ref{Eq:STI2}) from real space to momentum space. The functions
$\Gamma_{\Theta{\gamma}^{\mu}}$ and $\Gamma_{{ \gamma}^{\mu}\Theta}$
are related to the fermion propagators via the identity
\begin{eqnarray}
k_{\mu}{\Gamma_{{\gamma}^{\mu}\Theta}(k,p) -
p_{\mu}\Gamma_{\Theta{\gamma}^{\mu}}}(k,p) = -G^{-1}(k)\Theta +
\Theta G^{-1}(p) \label{Eq:WTIS1}
\end{eqnarray}
if $\Theta$ satisfies constraints I and III and via the identity
\begin{eqnarray}
k_{\mu}\Gamma_{\gamma^{\mu}\Theta}(k,p) +
p_{\mu}{\Gamma_{\Theta{\gamma}^{\mu}}}(k,p) = -G^{-1}(k)\Theta -
\Theta G^{-1}(p) \label{Eq:WTIS2}
\end{eqnarray}
if $\Theta$ satisfies constraint II and IV. Some of these identities
result from symmetric transformations and thus are just the ordinary
WTIs. The rest of the identities result from special asymmetric
transformations and are different from ordinary WTIs. However, for
simplicity, we will universally call them (generalized) WTIs. For a
given $\Theta$, there are a certain number of unknown functions
$\Gamma_{{\gamma}^{\mu}\Theta}$ and $\Gamma_{\Theta{
\gamma}^{\mu}}$. If we could find a sufficient number of WTIs, we
would able to completely determine these unknown functions and
express them purely in terms of fermion propagators.

Now we explain why we have deliberately chosen $\Theta$ to leave the
fermion-boson coupling term $\mathcal{L}_{fb}$ unchanged. In fact,
if $\mathcal{L}_{fb}$ is changed by the transformations
$\psi_{\sigma}\rightarrow e^{i\theta\Theta}\psi_{\sigma}$, the third
term of l.h.s. of Eq.~(\ref{Eq:genericidentity}) does not vanish.
Then an additional term
\begin{eqnarray}
\langle g \phi(x)\big({\bar\psi}_{\sigma}(x){\widehat
\Theta}\gamma^m {\psi_\sigma}(x) - {\bar\psi}_{\sigma}(x) \gamma^m
{\Theta}{\psi_\sigma}(x)\big) \psi_{\alpha}(y){\bar \psi}_{\beta}(z)
\rangle_J
\end{eqnarray}
would appear in both Eq.~(\ref{Eq:STI1}) and Eq.~(\ref{Eq:STI2}).
This is a five-point correlation function that is related to an
infinite number of higher-point correlation functions. Once such a five-point correlation function is incorporated, the generalized WTIs
given by Eqs.~(\ref{Eq:WTIS1}-\ref{Eq:WTIS2}) would not be
self-closed and the current vertex functions $\Gamma_{{
\gamma}^{\mu}\Theta}$ and $\Gamma_{\Theta{\gamma}^{\mu}}$ could
never be expressed purely in terms of fermion propagators. Different
from $\mathcal{L}_{fb}$, it does not matter if the free term
$\mathcal{L}_{f}$ is changed by asymmetric transformations
$\psi_{\sigma}\rightarrow e^{i\theta\Theta}\psi_{\sigma}$. This is
because $\mathcal{L}_{f}$ is bilinear in spinor field $\psi(x)$ and,
consequently, its variation $\Delta \mathcal{L}_{f}$ is also
bilinear in $\psi(x)$. As demonstrated in the above analysis, one
can always define a number of non-conserved currents on the basis of
$\Delta \mathcal{L}_{f}$ and then derive the same number of
asymmetry-induced WTIs, provided that the interaction term
$\mathcal{L}_{fb}$ is unchanged by these special asymmetric
transformations.

The formation of superconductivity induced by the electron-phonon
interaction in metals with a finite Fermi surface was previously
addressed in Ref.~\cite{Liu19}. In that case, the fermionic
excitations are described by two-component Nambu spinor and there
are only two unknown current vertex functions. Owing to the
relatively simple structure of free electron Lagrangian density
$\mathcal{L}_{f}$, the two current vertex functions can be
determined by solving two symmetry-induced WTIs (corresponding to
charge conservation and spin conservation, respectively). In Dirac
semimetals, the Dirac fermions have a more complicated kinetic term
$\mathcal{L}_{f}$. In order to determine all the involved current
vertex functions, we have to employ both symmetry-induced WTIs and
asymmetry-induced WTIs. Therefore, the results presented in this
section have significantly broadened the scope of application of the
approach originally developed in Ref.~\cite{Liu19}.

Our next step is to determine $\Gamma_{{\gamma}^{\mu}\Theta}$ and
$\Gamma_{\Theta{\gamma}^{\mu}}$. Most realistic semimetals are
theoretically defined and experimentally fabricated in (1+2)- or
(1+3)-dimensions, thus we study only these two cases.

\section{Fermion-boson coupling $\phi{\bar\psi}\psi$ \label{Sec:unitymatrix}}

In this section, we investigate the case in which the boson field
$\phi$ couples to ${\bar\psi}\psi$ defined via the unity matrix $I$.
The Yukawa coupling term $\phi{\bar\psi}\psi$ describes the
interaction between massless Dirac fermions and the quantum critical
fluctuation of the order parameter that is induced by dynamical
chiral symmetry breaking \cite{Pan18}. In this case the constraint
III is always satisfied, thus we only need to ensure that the
Constraint I is simultaneously satisfied.

\subsection{$(1+2)$ dimensions \label{Sec:unitymatrix012}}

We first consider $(1+2)$-dimensional Dirac semimetals. There are
four possible choices of $\Theta$. Two new variables $q=k-p$ and
$P=k+p$ are introduced to simplify notations.

(1) Choose $\Theta = \gamma^0$. We obtain
\begin{eqnarray}
&& q_0 \Gamma_I - P_1 \Gamma_{\gamma^0 \gamma^1} - P_2
\Gamma_{\gamma^0 \gamma^2} \nonumber \\
&& = -G^{-1}(k)\gamma^{0}+\gamma^{0}G^{-1}(p) = \mathcal{B}_0.
\end{eqnarray}

(2) Choose $\Theta=\gamma^1$. We obtain
\begin{eqnarray}
&&-P_0\Gamma_{\gamma^0 \gamma^1}+q_1\Gamma_I + P_2\Gamma_{\gamma^1
\gamma^2} \nonumber \\
&& =G^{-1}(k)\gamma^1 - \gamma^1 G^{-1}(p) = \mathcal{B}_1.
\end{eqnarray}

(3) Choose $\Theta = \gamma^2$. We obtain
\begin{eqnarray}
&& -P_0\Gamma_{\gamma^0 \gamma^2} - {P_1\Gamma_{\gamma^1
\gamma^2}} + q_2\Gamma_I \nonumber \\
&& = G^{-1}(k)\gamma^2 - \gamma^2 G^{-1}(p) = {\mathcal B}_2.
\end{eqnarray}

(4) Choose $\Theta=i\gamma^{012} = i\gamma^{0}\gamma^{1}\gamma^{2}$.
We obtain
\begin{eqnarray}
&&q_0\Gamma_{\gamma^1 \gamma^2}+q_1\Gamma_{\gamma^0 \gamma^2} -
q_2\Gamma_{\gamma^0\gamma^1} \nonumber
\\
&& = -G^{-1}(k)\gamma^{012} + \gamma^{012} G^{-1}(p) =
\mathcal{B}_3.
\end{eqnarray}
Note that $\gamma^{012} = -i\tau_3\otimes I$ if one uses $4\times 4$
matrices and $\gamma^{012} = -iI$ if one uses $2 \times 2$ matrices.

We now see that the four current vertex functions $\Gamma_I$,
$\Gamma_{\gamma^0 \gamma^1}$, and $\Gamma_{\gamma^0 \gamma^2}$, and
$\Gamma_{\gamma^1 \gamma^2}$ satisfy four different WTIs. In order
to obtain these four functions, it is now convenient to define a
matrix $M_{\mathcal{B}}$ defined as follows
\begin{eqnarray}
M_{\mathcal{B}} \begin{pmatrix}
\Gamma_I\\
\Gamma_{\gamma^0\gamma^1}\\
\Gamma_{\gamma^0\gamma^2}\\
\Gamma_{\gamma^1\gamma^2}
\end{pmatrix}
\equiv \begin{pmatrix}
q_0& -P_1& -P_2& 0\\
q_1& -P_0&  0& P_2\\
q_2&  0& -P_0& -P_1\\
0& -q_2& q_1& q_0
\end{pmatrix}\begin{pmatrix}
\Gamma_I\\
\Gamma_{\gamma^0\gamma^1}\\
\Gamma_{\gamma^0\gamma^2}\\
\Gamma_{\gamma^1\gamma^2}
\end{pmatrix} = \begin{pmatrix}
\mathcal{B}_0\\
\mathcal{B}_1\\
\mathcal{B}_2\\
\mathcal{B}_3\end{pmatrix}.
\end{eqnarray}
The inverse of $M_{\mathcal{B}}$ has the expression
\begin{eqnarray}
&& M_{\mathcal{B}}^{-1}=\frac{1}{q_0P_0 - q_1P_1 - q_2P_2}
\begin{pmatrix}
P_0&  -P_1& -P_2& 0\\
q_1&  -q_0&  0& P_2\\
q_2& 0&  -q_0& -P_1\\
0&  -q_2&  q_1& P_0
\end{pmatrix}.
\end{eqnarray}
The invertibility of this sort of matrix will be discussed in
Sec.~\ref{Sec:gamma012}. Then $\Gamma_I$,
$\Gamma_{\gamma^{0}\gamma^{1}}$, $\Gamma_{\gamma^{0} \gamma^{2}}$,
and $\Gamma_{\gamma^{1}\gamma^{2}}$ can be easily computed from the
following equations
\begin{eqnarray}
\begin{pmatrix}
\Gamma_I\\
\Gamma_{\gamma^0\gamma^1}\\
\Gamma_{\gamma^0\gamma^2}\\
\Gamma_{\gamma^1\gamma^2}
\end{pmatrix}
=\frac{1}{q_0P_0 - q_1P_1 - q_2P_2}\begin{pmatrix}
P_0&  -P_1& -P_2& 0\\
q_1&  -q_0&  0& P_2\\
q_2& 0&  -q_0& -P_1\\
0&  -q_2&  q_1& P_0
\end{pmatrix}\begin{pmatrix}
\mathcal{B}_0\\
\mathcal{B}_1\\
\mathcal{B}_2\\
\mathcal{B}_3\end{pmatrix}.
\end{eqnarray}
Since the Yukawa coupling is $\phi {\bar\psi}\psi$, we are only
interested in $\Gamma_{I}$, which depends on the Dirac fermion
propagator as follows
\begin{eqnarray}
\Gamma_I = \frac{P_0\mathcal{B}_0 - P_1\mathcal{B}_1 -
P_2\mathcal{B}_2}{q_0P_0 - q_1P_1 - q_2P_2}.
\end{eqnarray}

\subsection{$(1+3)$ dimensions \label{Sec:unitymatrix013}}

In this subsection we consider the case of $(1+3)$-dimensional Dirac
semimetal. The WTIs can be derived by utilizing the same
calculational procedure as $(1+2)$-dimensional system.

(1) Choose $\Theta = \gamma^0$. We obtain
\begin{eqnarray}
&&q_0\Gamma_I - P_1\Gamma_{\gamma^0 \gamma^1} - P_2 \Gamma_{\gamma^0
\gamma^2} - P_3\Gamma_{\gamma^0 \gamma^3}
\nonumber \\
&& = -G^{-1}(k)\gamma^0 + \gamma^0 G^{-1}(p) = \mathcal{D}_0.
\end{eqnarray}

(2) Choose $\Theta = \gamma^1$, we obtain
\begin{eqnarray}
&& -P_0\Gamma_{\gamma^0\gamma^1} + q_1\Gamma_I + P_2
\Gamma_{\gamma^1 \gamma^2} + P_3\Gamma_{\gamma^1\gamma^3} \nonumber
\\
&& = G^{-1}(k)\gamma^1 - \gamma^1 G^{-1}(p) = \mathcal{D}_1.
\end{eqnarray}

(3) Choose $\Theta = \gamma^2$, we obtain
\begin{eqnarray}
&& -P_0\Gamma_{\gamma^0\gamma^1} - P_1\Gamma_{\gamma^1\gamma^2} +
q_2 \Gamma_I +P_3 \Gamma _{\gamma^2\gamma^3}
\nonumber \\
&& = G^{-1}(k)\gamma^2 - \gamma^2G^{-1}(p) = {\mathcal{D}}_2.
\end{eqnarray}

(4) Choose $\Theta = \gamma^{012} = \gamma^{0}\gamma^{1}\gamma^{2}$.
We obtain
\begin{eqnarray}
&&q_0\Gamma_{\gamma^1\gamma^2}+q_1 \Gamma_{\gamma^0\gamma^2} - q_2
\Gamma_{\gamma^0\gamma^1} + P_3\Gamma_{\gamma^{0123}} \nonumber \\
&& =-G^{-1}(k)\gamma^{012} + \gamma^{012}G^{-1}(p)= \mathcal{D}_3.
\end{eqnarray}
Here $\gamma^{0123} = \gamma^{0}\gamma^{1}\gamma^{2}\gamma^{3} =
-i\gamma^5$.

(5) Choose $\Theta = \gamma^{3}$. We obtain
\begin{eqnarray}
&& -P_0\Gamma_{\gamma^0\gamma^3} - P_1\Gamma_{\gamma^1\gamma^3} -
P_2\Gamma_{\gamma^2\gamma^3}+q_3\Gamma_I \nonumber \\
&& = G^{-1}(k)\gamma^3 - \gamma^3G^{-1}(p) = \mathcal{D}_4.
\end{eqnarray}

(6) Choose $\Theta = \gamma^{013} = \gamma^{0}\gamma^{1}\gamma^{3}$.
We obtain
\begin{eqnarray}
&& q_0\Gamma_{\gamma^1\gamma^3} + P_1\Gamma_{\gamma^0\gamma^3} + P_2
\Gamma_{\gamma^{0123}} - q_3 \Gamma_{\gamma^0 \gamma^1} \nonumber \\
&& = -G^{-1}(k)\gamma^{013} + \gamma^{013}G^{-1}(p) = \mathcal{D}_5.
\end{eqnarray}

(7) Choose $\Theta=\gamma^{023}=\gamma^{0}\gamma^{2}\gamma^{3}$. We
obtain
\begin{eqnarray}
&&q_0\Gamma_{\gamma^2\gamma^3} - P_1\Gamma_{\gamma^{0123}}
+q_2\Gamma_{\gamma^0\gamma^3} - q_3\Gamma_{\gamma^0\gamma^2}
\nonumber \\
&& =-G^{-1}(k)\gamma^{023} + \gamma^{023}G^{-1}(p) = \mathcal{D}_6.
\end{eqnarray}

(8) Choose $\Theta = \gamma^{123} =
-\gamma^{1}\gamma^{2}\gamma^{3}$. We obtain
\begin{eqnarray}
&& P_0\Gamma_{\gamma^{0123}} + q_1\Gamma_{\gamma^{2}\gamma^{3}} -
q_2\Gamma_{\gamma^{1}\gamma^{3}} + q_3\Gamma_{\gamma^{1}\gamma^{2}}
\nonumber \\
&& = G^{-1}(k)\gamma^{123} + \gamma^{123}G^{-1}(p) = \mathcal{D}_7.
\end{eqnarray}

Combining the above eight equations, we obtain
\begin{eqnarray}
M_{\mathcal{D}} \begin{pmatrix}
\Gamma_I\\
\Gamma_{{\gamma^0}{\gamma^1}}\\
\Gamma_{{\gamma^0}{\gamma^2}}\\
\Gamma_{{\gamma^0}{\gamma^3}}\\
\Gamma_{{\gamma^1}{\gamma^2}}\\
\Gamma_{{\gamma^1}{\gamma^3}}\\
\Gamma_{{\gamma^2}{\gamma^3}}\\
\Gamma_{\gamma^{0123}}
\end{pmatrix} \equiv \begin{pmatrix}
q_0& -P_1 & -P_2 & -P_3 & 0 & 0 &0  & 0\\
q_1 & -P_0  &0  &0  &P_2  & P_3 &0  &0 \\
q_2 &0 & -P_0 & 0& -P_1 & 0 & P_3  &0 \\
0 & -q_2 &  q_1 & 0 & q_0 & 0 & 0 & P_3 \\
q_3 & 0 & 0 & -P_0 & 0 & -P_1 & -P_3 &0 \\
0 & -q_3& 0 & P_1 & 0 & q_0 & 0 & -P_2\\
0 &0 & -q_3 & -q_2 & 0 & 0 & q_0 & P_1 \\
0 &0  &0  &0  & q_3 & -q_2 & q_1 & P_0
\end{pmatrix}\begin{pmatrix}
\Gamma_I\\
\Gamma_{{\gamma^0}{\gamma^1}}\\
\Gamma_{{\gamma^0}{\gamma^2}}\\
\Gamma_{{\gamma^0}{\gamma^3}}\\
\Gamma_{{\gamma^1}{\gamma^2}}\\
\Gamma_{{\gamma^1}{\gamma^3}}\\
\Gamma_{{\gamma^2}{\gamma^3}}\\
\Gamma_{\gamma^{0123}}
\end{pmatrix}=\begin{pmatrix}
{\mathcal D}_0\\
{\mathcal D}_1\\
{\mathcal D}_2\\
{\mathcal D}_3\\
{\mathcal D}_4\\
{\mathcal D}_5\\
{\mathcal D}_6\\
{\mathcal D}_7
\end{pmatrix},
\end{eqnarray}
where $M_{\mathcal{D}}$ is an $8\times 8$ matrix. Using the inverse
of $M_{\mathcal{D}}$, which is complicated and will not be
explicitly given here, one can express $\Gamma_I$ purely in terms of
Dirac fermion propagators.
%\begin{eqnarray}
%M_{\mathcal{D}} = \begin{pmatrix}
%q_0& -P_1 & -P_2 & -P_3 & 0 & 0 &0  & 0\\
%q_1 & -P_0  &0  &0  &P_2  & P_3 &0  &0 \\
%q_2 &0 & -P_0 & 0& -P_1 & 0 & P_3  &0 \\
%0 & -q_2 &  q_1 & 0 & q_0 & 0 & 0 & P_3 \\
%q_3 & 0 & 0 & -P_0 & 0 & -P_1 & -P_3 &0 \\
%0 & -q_3& 0 & P_1 & 0 & q_0 & 0 & -P_2\\
%0 &0 & -q_3 & -q_2 & 0 & 0 & q_0 & P_1 \\
%0 &0  &0  &0  & q_3 & -q_2 & q_1 & P_0
%\end{pmatrix}.
%\end{eqnarray}

\section{Fermion-boson coupling $\phi{\bar\psi}\gamma^{0}\psi$ \label{Sec:gamma0}}

In this section we consider the model in which $\gamma^m = \gamma^0$
and calculate the corresponding current vertex function, which will
be denoted by the symbol $\Upsilon_{\gamma^{0}}$. The matrix
$\Theta$ to be used here should satisfy constraint III or Constraint
IV. We need to be careful and make sure that $\Theta$ also satisfies
Constraint I in the former case and satisfies constraint II in the
latter case. All the WTIs will be derived from either
Eq.~(\ref{Eq:WTIS1}) or Eq.~(\ref{Eq:WTIS2}), depending on the
concrete expression of each $\Theta$.

\subsection{$(1+2)$ dimensions \label{Sec:gamma012}}

When one is studying the effects of Coulomb interaction or
fermion-phonon interaction in graphene or other types of
two-dimensional Dirac semimetals, the Yukawa-coupling
$g\phi{\bar\psi}\gamma^{0}\psi$ is encountered. The WTIs to be
derived here will be very useful in such studies.

(1) Apparently, the simplest choice of matrix $\Theta$ is $\Theta =
I$. For this choice, it is easy to check that the constraints I and
III are satisfied. We have already mentioned that
$\psi_\sigma\rightarrow e^{i\theta}\psi_\sigma$ is a symmetry of the
total Lagrangian density $\mathcal{L}$. Thus we could use
Eq.~(\ref{Eq:WTIS1}) and obtain the following identity
\begin{eqnarray}
&&q_{0}{\Upsilon_{\gamma^{0}}}(k,p) + q_{1}{\Upsilon_{\gamma^{1}}}
+ q_{2}{\Upsilon_{\gamma^{2}}}(k,p) \nonumber \\
&& = -G^{-1}(k)+G^{-1}(p) = {\mathcal{A}}_{0}.
\label{Eq:ordinaryWTI12}
\end{eqnarray}
This is the ordinary symmetry-induced WTI. This WTI by itself is of
little practical usage since one single identity cannot determine
three unknown current vertex functions $\Upsilon_{\gamma^{0}}$,
$\Upsilon_{\gamma^{1}}$, and $\Upsilon_{\gamma^{2}}$. Fortunately,
there are more WTIs.

(2) Choose $\Theta = \gamma^{01} = \gamma^{0}\gamma^{1}$. This
matrix satisfies the constraints II and IV, i.e.,
$\hat{\Theta}=-\Theta$ and $\{\gamma^0, \Theta\} = 0$. Using
Eq.~(\ref{Eq:WTIS2}) and the following relations
\begin{eqnarray}
\gamma^0 {\gamma^{01}} =-\gamma^{01}{\gamma ^0}, \quad
\gamma^1\gamma^{01}= -\gamma^{01}{\gamma^1}, \quad
\gamma^2{\gamma^{01}}={\gamma^{01}}\gamma^2,
\end{eqnarray}
we obtain
\begin{eqnarray}
&& -q_0 {\Upsilon_{\gamma^1}} -q_1{\Upsilon_{\gamma^0}}
-{P_2}\Upsilon_{\gamma^{012}} \nonumber \\
&& =G^{-1}(k)\gamma^{01} + \gamma^{01}G^{-1}(p) = \mathcal{A}_1.
\end{eqnarray}
Apart from $\Upsilon_{\gamma^{0}}$ and $\Upsilon_{\gamma^{1}}$,
there appears a fourth unknown function $\Upsilon_{\gamma^{012}}$.

(3) Choose $\Theta = \gamma^{02} = \gamma^{0}\gamma^{2}$. This
matrix also satisfies the constraints II and IV simultaneously.
Based on Eq.~(\ref{Eq:WTIS2}) and the following relations
\begin{eqnarray}
\gamma^{0}\gamma^{02}=  -\gamma^{02}\gamma^{0}, \quad
\gamma^{2}\gamma^{02} =-\gamma^{02}\gamma^{2}, \quad
\gamma^{1}\gamma^{02} = \gamma^{02}\gamma^{1},
\end{eqnarray}
we obtain
\begin{eqnarray}
&&-q_{0}{\Upsilon_{\gamma^2}} +P_1\Upsilon_{\gamma^{012}}
-{q_2}{\Upsilon_{\gamma^0}} \nonumber \\
&& =G^{-1}(k)\gamma^{02} + \gamma^{02}G^{-1}(p) = \mathcal{A}_2.
\end{eqnarray}

(4) Choose $\Theta = \sigma^{12} = i\gamma^{12} = i\gamma^1
\gamma^2$. The definition of $\sigma^{12}$ can be found in Appendix
\ref{App:gammamatrices}. This $\Theta$ satisfies constraints I and
III simultaneously, thus Eq.~(\ref{Eq:WTIS1}) should be adopted.
Notice that
\begin{eqnarray}
\sigma^{12}\gamma^{0}=\gamma^{0}\sigma^{12}=i \gamma^{012}, \quad
\sigma^{12}\gamma^{1} = - \gamma^{1}\sigma^{12} = i\gamma^{2}, \quad
\sigma^{12}\gamma^{2} = -\gamma^{2}\sigma^{12} =-i\gamma^{1}.
\end{eqnarray}
For this choice we get
\begin{eqnarray}
&&q_0 \Upsilon_{\gamma^{012}} - P_1{\Upsilon_{\gamma^2}}
+ P_2{\Upsilon_{\gamma^1}} \nonumber \\
&& = i{G}^{-1}(k){\sigma^{12}} - i{\sigma^{12}}G^{-1}(p) =
\mathcal{A}_3.
\end{eqnarray}

Now we see that the four unknown current vertex functions
$\Upsilon_{\gamma^0}$, $\Upsilon_{\gamma^1}$, $\Upsilon_{\gamma^2}$,
and $\Upsilon_{\gamma^{012}}$ satisfy four coupled WTIs, which can
be expressed in the following compact form
\begin{eqnarray}
M_{\mathcal{A}}\left({\begin{array}{*{20}{c}}
{{\Upsilon_{\gamma^{0}}}}\\
{{\Upsilon_{\gamma^{1}}}}\\
{{\Upsilon_{\gamma^{2}}}}\\
{\Upsilon_{\gamma^{012}}}
\end{array}}\right) \equiv {\begin{pmatrix}
{q_0}&{q_1}&{q_2}&0\\
-q_1&-q_0&0&{-P_2}\\
-q_2&0&-q_0&{P_1}\\
0&P_2&{-P_1}&{q_0}
\end{pmatrix}}\left({\begin{array}{*{20}{c}}
{{\Upsilon_{\gamma^{0}}}}\\
{{\Upsilon_{\gamma^{1}}}}\\
{{\Upsilon_{\gamma^{2}}}}\\
{\Upsilon_{\gamma^{012}}}
\end{array}}\right) = \left({\begin{array}{*{20}{c}}
{{\mathcal{A}_0}}\\
{{\mathcal{A}_1}}\\
{{\mathcal{A}_2}}\\
{{\mathcal{A}_3}}
\end{array}}\right).
\label{Eq:Matrix12}
\end{eqnarray}
%where the matrix $M_{\mathcal{A}}$ is given by
%\begin{eqnarray}
%M_{\mathcal{A}} = {\begin{pmatrix}
%{q_0}&{q_1}&{q_2}&0\\
%-q_1&-q_0&0&{-P_2}\\
%-q_2&0&-q_0&{P_1}\\
%0&P_2&{-P_1}&{q_0}
%\end{pmatrix}}.
%\end{eqnarray}
From Eq.~(\ref{Eq:Matrix12}), we obtain
\begin{eqnarray}
\left({\begin{array}{*{20}{c}}
{{\Upsilon_{\gamma^{0}}}}\\
{{\Upsilon_{\gamma^{1}}}}\\
{{\Upsilon_{\gamma^{2}}}}\\
{\Upsilon_{\gamma^{012}}}
\end{array}} \right) =
{M_{\mathcal{A}}^{-1}}\left({\begin{array}{*{20}{c}}
{{\mathcal{A}_0}}\\
{{\mathcal{A}_1}}\\
{{\mathcal{A}_2}}\\
{{\mathcal{A}_3}}
\end{array}} \right).
\end{eqnarray}
We are only interested in $\Upsilon_{\gamma_{0}}$. It is easy to
find that $\Upsilon_{\gamma_{0}}$ has the form
\begin{eqnarray}
\Upsilon_{\gamma_{0}}(k,p) &=& \frac{1}{\det(M_{\mathcal{A}})}
\big[q_0\left(q_0^2-P_1^2-P_2^2\right){\mathcal{A}_0} +
\left(q_1P_1^2 + q_2P_1P_2-q_0^2q_1\right){\mathcal{A}_1} \nonumber
\\
&& +\left(q_1P_1P_2+q_2P_2^2-q_0^2q_2\right){\mathcal{A}_{2}} -
q_0\left(q_2P_1-q_1P_2\right){\mathcal{A}_{3}}\big],
\end{eqnarray}
where the determinant of matrix $M_{\mathcal{A}}$ is
\begin{eqnarray}
\det(M_{\mathcal{A}}) = q_0^2\left(q_0^2 - q_1^2 - q_2^2\right) -
P_1\left(P_1q_0^2 - P_1q_1^2 - P_2q_1q_2\right) - P_2\left(P_2q_0^2
- P_2q_2^2 - P_1 q_1 q_2\right).
\end{eqnarray}
The above $\Upsilon_{\gamma_{0}}(k,p)$ will be utilized to study the
Coulomb interaction in graphene in Sec.~\ref{Sec:graphene}. Let us
take a closer look at its expression. The matrix $M_{\mathcal{A}}$
is not invertible if $\det(M_{\mathcal{A}})=0$. It is therefore
necessary to examine under what conditions
$\det(M_{\mathcal{A}})=0$. Since $\det(M_{\mathcal{A}})$ is the
denominator of $\Upsilon_{\gamma_{0}}(k,p)$, this is equivalent to
examining under what conditions $\Upsilon_{\gamma_{0}}(k,p)$
diverges. For this purpose, we re-write $\det(M_{\mathcal{A}})$ as
\begin{eqnarray}
\det(M_{\mathcal{A}}) = q_0^4 - 2q_0^2\left(\mathbf{k}^2 +
\mathbf{p}^2\right) + \left(\mathbf{q}\cdot \mathbf{P}\right)^2.
\end{eqnarray}

If we work within the Matsubara formalism of finite-temperature
quantum field theory, we should take the boson energy as $q_{0} =
i\omega_{n} = i2nk_{B}T$, which leads to
\begin{eqnarray}
\det(M_{\mathcal{A}}) = \omega_{n}^4 + 2\omega_{n}^{2}(\mathbf{k}^2
+ \mathbf{p}^2) + \left(\mathbf{q}\cdot \mathbf{P}\right)^2.
\end{eqnarray}
For any nonzero $\omega_{n}$, $\det(M_{\mathcal{A}})$ is always
nonzero, irrespective of the value of $\mathbf{q}\cdot \mathbf{P}$.
Apparently, $\det(M_{\mathcal{A}})$ vanishes only when
$\omega_{n}=0$ and $\mathbf{q}\cdot \mathbf{P}=0$ simultaneously.
After substituting $\omega_{n}=0$ and $\mathbf{q}\cdot \mathbf{P}=0$
into $\Upsilon_{\gamma_{0}}(k,p)$, we verify that the numerator and
denominator of $\Upsilon_{\gamma_{0}}(k,p)$ both vanish but
$\Upsilon_{\gamma_{0}}(k,p)$ itself remains finite. Indeed, the
zeroes and the poles of $\Upsilon_{\gamma_{0}}(k,p)$ cancel exactly.
Thus, $\Upsilon_{\gamma_{0}}(k,p)$ is free of singularity and can be
safely inserted into the DS equation of $G(p)$.

Alternatively, we can use real energies at zero temperature. To make
integrals converge, we should introduce an infinitesimal factor
$i\delta$ to the energies of fermion and boson, namely, $k_{0}
\rightarrow k_{0}+i\delta$, $p_{0} \rightarrow p_{0}+i\delta$, and
$q_{0} \rightarrow q_{0}+i\delta$. The factor $i\delta$ enters into
the fermion propagator $G(p)$ and boson propagator $F_{0}(q)$, and
also into the vertex function $\Upsilon_{\gamma_{0}}(k,p)$. Then
$G(p)$, $F_{0}(q)$, and $\Upsilon_{\gamma_{0}}(k,p)$ becomes complex
functions and have poles on the complex plane for certain values of
$k$ and $p$. Such functions should be treated by standard
manipulations of quantum many-body theory \cite{AGD}: divide complex
functions into real and imaginary parts, and employ principal value
integral to define DS equations. The retarded fermion propagator,
denoted by $G_{\mathrm{ret}}(p_{0}+i\delta,\mathbf{p})$, could be
computed by numerically solving its self-consistent DS integral
equation. However, this framework is rather complicated and less
convenient than the Matsubara formalism. In Sec.~\ref{Sec:graphene},
we will adopt the Matsubara formalism to study the DS equation of
$G(p)$.

The above analysis of the zeros of $\det(M_{\mathcal{A}})$ is
applicable to the two matrices $M_{\mathcal{B}}$ and
$M_{\mathcal{D}}$ obtained in the last section and also to the
matrix $M_{\mathcal{C}}$ to be derived in the next subsection.

\subsection{$(1+3)$ dimensions \label{Sec:gamma013}}

The same calculational procedure adopted in the case of $(1+2)$
dimensions can be directly applied to $(1+3)$ dimensions. There are
eight mutually related WTIs.

(1) If we choose $\Theta=I$, the constraints I and III are satisfied
simultaneously. Thus Eq.~(\ref{Eq:WTIS1}) is reduced to the ordinary
WTI:
\begin{eqnarray}
&&q_0 \Upsilon_{\gamma^0} + q_1 \Upsilon_{\gamma^1}
+ q_2\Upsilon_{\gamma^2} + q_3 \Upsilon_{\gamma^3} \nonumber \\
&& = -G^{-1}(k) + G^{-1}(p) = {\mathcal{C}_0}.
\end{eqnarray}
This identity contains four unknown current vertex functions
${\Upsilon_{\gamma^0}}$, ${\Upsilon_{\gamma^1}}$,
${\Upsilon_{\gamma^2}}$, and ${\Upsilon_{\gamma^3}}$.

(2) Choose $\Theta=\gamma^{01} = \gamma^{0}\gamma^{1}$. This matrix
satisfies constraints II and IV. Notice the following relations
hold:
\begin{eqnarray}
&&\gamma^0 \gamma^{01} = -\gamma^{01}{\gamma^0} = \gamma^1, \quad
\gamma^1\gamma^{01} = -\gamma^{01}\gamma^1 = \gamma^0, \\
&&{\gamma^2}{\gamma^{01}}={\gamma ^{01}} {\gamma^2} =
-i\tau^3\otimes I, \quad \gamma^3\gamma^{01} = \gamma^{01} \gamma^3
= -i \tau^1\otimes\tau^1.
\end{eqnarray}
From Eq.~(\ref{Eq:WTIS2}), one finds that
\begin{eqnarray}
&&-q_0{\Upsilon_{\gamma^1}} - {q_1} \Upsilon_{\gamma^0} +
iP_2\Upsilon_{\tau^3 \otimes I} +
i{P_3}\Upsilon_{\tau^1\otimes\tau^1} \nonumber \\
&& = G^{-1}(k)\gamma^{01}+\gamma^{01}G^{-1}(p) = \mathcal{C}_{1}.
\end{eqnarray}
It is clear that $\Upsilon_{\gamma^0}$, $\Upsilon_{\gamma^1}$,
$\Upsilon_{\gamma^2}$, and $\Upsilon_{\gamma^3}$ do not form a
closed set of self-consistently coupled functions, because
${\Upsilon_{\gamma^0}}$ and ${\Upsilon_{\gamma^1}}$ are related to
two new functions $\Upsilon_{\tau^3 \otimes I}$ and
$\Upsilon_{\tau_1\otimes\tau^1}$. Four WTIs are not sufficient and
we need more WTIs.

(3) Choose $\Theta = \gamma^{02} = \gamma^{0}\gamma^{2}$. This
$\Theta$ satisfies constraints II and IV. One can verify that
\begin{eqnarray}
&&\gamma^{0}\gamma^{02} = -\gamma^{02}\gamma^{0}=\gamma^2 ,\quad
\gamma^{1}\gamma^{02} = \gamma^{02}\gamma^1 = i\tau^3\otimes I, \\
&&\gamma^{2}\gamma^{02} = -\gamma^{02}\gamma^{2}=\gamma^0, \quad
\gamma^{3}\gamma^{02} = \gamma^{02} \gamma^{3} = -i
\tau^{1}\otimes\tau^{2}.
\end{eqnarray}
From Eq.~(\ref{Eq:WTIS2}), we obtain
\begin{eqnarray}
&&-q_0 \Upsilon_{\gamma^2} - i{P_1}\Upsilon_{\tau^3 \otimes I}
-q_2\Upsilon_{\gamma^0} + iP_3\Upsilon_{\tau^1\otimes\tau^2} \nonumber \\
&& =-iG^{-1}(k)\sigma^{02} - i\sigma^{02}G^{-1}(p) = \mathcal{C}_2.
\end{eqnarray}
$\Upsilon_{\tau^1\otimes\tau^2}$ is the seventh relevant unknown
current vertex function.

(4) Choose $\Theta=\sigma^{12}$. This $\Theta$ satisfies constraints
I and III. Notice that
\begin{eqnarray}
&&\gamma^{0}\sigma^{12} = \sigma^{12}{\gamma^0}=\tau^3\otimes I,
\quad \gamma^{1}\sigma^{12} = -\sigma^{12}{\gamma^1}=-i
\gamma^{2}, \\
&&\gamma^{2}\sigma^{12} = -\sigma^{12}{\gamma^2} =i \gamma^{1},
\quad \gamma^{3}\sigma^{12} = { \sigma}^{12}\gamma^{3}= -\tau^{1}
\otimes \tau^{3}.
\end{eqnarray}
From Eq.~(\ref{Eq:WTIS1}), we obtain
\begin{eqnarray}
&& -iq_0 \Upsilon_{\tau^3\otimes I} - P_1\Upsilon_{\gamma^2}
+ P_2\Upsilon_{\gamma^1}- q_3 \Upsilon_{\tau^1\otimes\tau^3} \nonumber \\
&& =iG^{-1}(k)\sigma^{12} - i\sigma^{12}G^{-1}(p) = \mathcal{C}_3.
\end{eqnarray}
Here we encounter the eighth unknown current vertex function
$\Upsilon_{\tau^1\otimes\tau^3}$.

(5) Choose $\Theta = \gamma^{03} = \gamma^{0}{\gamma^3}$. This
$\Theta$ satisfies constraints II and IV. Notice that
\begin{eqnarray}
&&\gamma^0 \gamma^{03} = -\gamma^{03}\gamma^0 = \gamma^3, \quad
\gamma^1 \gamma^{03} = \gamma^{03}\gamma^1 = -\tau^1 \otimes \tau^1,
\\
&&\gamma^2\gamma^{03} = \gamma^{03}\gamma^2 = -\tau^1 \otimes
\tau^2, \quad {\gamma^3}{\gamma^{03}} =
-{\gamma^{03}}{\gamma^3}={\gamma^0}.
\end{eqnarray}
From Eq.~(\ref{Eq:WTIS1}), we obtain
\begin{eqnarray}
&& -q_0\Upsilon_{\gamma^3} + {P_1}\Upsilon_{\tau^1\otimes\tau^1} +
{P_2} \Upsilon_{\tau^1\otimes\tau^2} - {q_3}\Upsilon_{\gamma^0} \nonumber \\
&& =G^{-1}(k){\gamma^{03}} + {\gamma^{03}}G^{-1}(p) =
\mathcal{C}_{4}.
\end{eqnarray}

(6) Choose $\Theta=\sigma^{13}$. This $\Theta$ satisfies constraints
I and III. Notice that
\begin{eqnarray}
&&{\gamma^0}\sigma^{13} = \sigma^{13}\gamma^{0}=\tau^1 \otimes
\tau^1, \quad \gamma^{1}\sigma^{13} =
-\sigma^{13}\gamma^{1}=-i\gamma^{3},\\
&&\gamma^{2}\sigma^{13} = \sigma^{13}\gamma^{2} = -\tau^1 \otimes
\tau^3, \quad \gamma^3 \sigma^{13}=-\sigma^{13}\gamma^{3} = i
\gamma^{1}.
\end{eqnarray}
From Eq.~(\ref{Eq:WTIS1}), we obtain
\begin{eqnarray}
&& -iq_0\Upsilon_{\tau^1\otimes\tau^1} - P_1 \Upsilon_{\gamma^3} +
iq_2 \Upsilon_{\tau^1\otimes\tau^3} + P_3 \Upsilon_{\gamma^1} \nonumber \\
&& =iG^{-1}(k)\sigma^{13}-i\sigma^{13}G^{-1}(p) = \mathcal{C}_{5}.
\end{eqnarray}

(7) Choose $\Theta=\sigma^{23}$. This $\Theta$ satisfies constraints
I and III. Notice that
\begin{eqnarray}
&&\gamma^0\sigma^{23} = \sigma^{23}\gamma^{0}=\tau^1 \otimes \tau^2,
\quad \gamma^1\sigma^{23} = \sigma^{23}\gamma^{1} =
-i\tau^1 \otimes \tau^3, \\
&&\gamma^2\sigma^{23} = - \sigma^{23}\gamma^{2} = -i\gamma^{3},
\quad \gamma^3 \sigma ^{23}=-\sigma^{23}\gamma^{3}= i\gamma^{2}.
\end{eqnarray}
From Eq.~(\ref{Eq:WTIS1}), we obtain
\begin{eqnarray}
&& -iq_0 \Upsilon_{\tau^1 \otimes \tau^2} - q_1
\Upsilon_{\tau^1\otimes\tau^3} - P_2 \Upsilon_{\gamma^3} + P_3
\Upsilon_{\gamma^2} \nonumber \\
&& =iG^{-1}(k)\sigma^{23} - i\sigma^{23}G^{-1}(p) = \mathcal{C}_{6}.
\end{eqnarray}

(8) Choose $\Theta = i\gamma^{0123} = i\gamma^0\gamma^1
\gamma^2\gamma^3$. This $\Theta$ satisfies constraints II and IV.
Notice that
\begin{eqnarray}
&&{\gamma^0}\gamma^{0123} = -\gamma^{0123}\gamma^0 =
-\tau^1\otimes\tau^3, \quad \gamma^1 \gamma^{0123} = -\gamma^{0123}
\gamma^1 = -i\tau^1\otimes\tau^2,\\
&& \gamma^2\gamma^{0123} =-\gamma^{0123}\gamma^2 =
i\tau^1\otimes\tau^1, \quad \gamma^3\gamma^{0123} = -\gamma^{0123}
\gamma^3 = -i\tau^3\otimes I.
\end{eqnarray}
From Eq.~(\ref{Eq:WTIS2}), we obtain
\begin{eqnarray}
&& iq_0\Upsilon_{\tau^1\otimes\tau^3} -  q_1
\Upsilon_{\tau^1\otimes\tau^2} + q_2 \Upsilon_{\tau^1\otimes\tau^1}
- q_3\Upsilon_{\tau^3\otimes I} \nonumber \\
&& =iG^{-1}(k)\gamma^{0123} + i\gamma^{0123}G^{-1}(p) =
\mathcal{C}_{7}.
\end{eqnarray}

It turns out that eight unknown functions ${\Upsilon_{\gamma_0}}$,
${\Upsilon_{\gamma_1}}$, ${\Upsilon_{\gamma_2}}$,
${\Upsilon_{\gamma_3}}$, $\Upsilon_{\tau^3 \otimes I}$,
$\Upsilon_{\tau^1\otimes\tau^1}$, $\Upsilon_{\tau^1 \otimes
\tau^2}$, and $\Upsilon_{\tau^1 \otimes \tau^3}$ are mutually
related via eight WTIs. The eight coupled WTIs can be written as
follows
\begin{eqnarray}
M_{\mathcal{C}}
\begin{pmatrix}
\Upsilon_{\gamma_0}\\
\Upsilon_{\gamma_1}\\
\Upsilon_{\gamma_2}\\
\Upsilon_{\gamma_3}\\
\Upsilon_{\tau^3\otimes I}\\
\Upsilon_{\tau^1\otimes\tau^1}\\
\Upsilon_{\tau^1\otimes\tau^2}\\
\Upsilon_{\tau^1\otimes \tau^3}
\end{pmatrix} \equiv \begin{pmatrix}
q_0 &q_1 &q_2 &-q_3& 0   &0    & 0   & 0\\
-q_1&q_0 &0   &0   & iP_2&iP_3 &0    & 0 \\
-q_2&0   &-q_0&0   &-iP_1& 0   &iP_3 & 0\\
0   &P_2 &-P_1&0   &-iq_0& 0   &0    &-q_3 \\
-q_3& 0  & 0  &-q_0&0    &P_1  &P_2  & 0\\
0   &P_3 & 0  &P_1 &0    &-iq_0&0    &-q_2\\
0   & 0  &P_3 &P_2 &0    & 0   &-iq_0&-q_1 \\
0   & 0  &0   &0   &-q_3 &q_2  & -q_1&iq_0\\
\end{pmatrix}
\begin{pmatrix}
\Upsilon_{\gamma_0}\\
\Upsilon_{\gamma_1}\\
\Upsilon_{\gamma_2}\\
\Upsilon_{\gamma_3}\\
\Upsilon_{\tau^3\otimes I}\\
\Upsilon_{\tau^1\otimes\tau^1}\\
\Upsilon_{\tau^1\otimes\tau^2}\\
\Upsilon_{\tau^1\otimes \tau^3}
\end{pmatrix}=\begin{pmatrix}
{\mathcal{C}}_0\\
{\mathcal{C}}_1\\
{\mathcal{C}}_2\\
{\mathcal{C}}_3\\
{\mathcal{C}}_4\\
{\mathcal{C}}_5\\
{\mathcal{C}}_6\\
{\mathcal{C}}_7
\end{pmatrix}.
\end{eqnarray}
%where the $8\times 8$ matrix has the form
%\begin{eqnarray}
%M_{\mathcal{C}}=\begin{pmatrix}
%q_0 &q_1 &q_2 &-q_3& 0   &0    & 0   & 0\\
%-q_1&q_0 &0   &0   & iP_2&iP_3 &0    & 0 \\
%-q_2&0   &-q_0&0   &-iP_1& 0   &iP_3 & 0\\
%0   &P_2 &-P_1&0   &-iq_0& 0   &0    &-q_3 \\
%-q_3& 0  & 0  &-q_0&0    &P_1  &P_2  & 0\\
%0   &P_3 & 0  &P_1 &0    &-iq_0&0    &-q_2\\
%0   & 0  &P_3 &P_2 &0    & 0   &-iq_0&-q_1 \\
%0   & 0  &0   &0   &-q_3 &q_2  & -q_1&iq_0\\
%\end{pmatrix}.
%\end{eqnarray}
Using the inverse of $M_{\mathcal{C}}$, one can express
$\Upsilon_{\gamma^0}$ in terms of full fermion propagator. This
$\Upsilon_{\gamma^0}$ can be used to study the Coulomb interaction
in $(1+3)$-dimensional Dirac semimetals.

\section{Relation between interaction and current vertex functions \label{Sec:D0D}}

All the current vertex functions $\Gamma_{M}^{\mu}(k,p)$ obtained in
the last two sections are defined via a number of generalized currents $j_{M}^{\mu}={\bar \psi}M^{\mu}\psi$, which may or may not
be conserved. They are closely related, but certainly not identical,
to the fermion-boson interaction vertex function
$\Gamma_{\mathrm{int}}(k,p)$ that enters into the DS equation of
fermion and boson propagators. In this section, we demonstrate how
to determine $\Gamma_{\mathrm{int}}(k,p)$ from its corresponding
$\Gamma_{M}^{\mu}$ function, using the strategy developed in
Ref.~\cite{Liu19}. We know from Eq.~(\ref{eq:Gammaint}) that
$\Gamma_{\mathrm{int}}$ is defined via the correlation function
$\langle \phi \psi {\bar\psi}\rangle$. In order to derive the
relation between $\Gamma_{M}^{\mu}$ and $\Gamma_{\mathrm{int}}$, we
need first to study the relation between $\langle{\bar
\psi}M^{\mu}\psi\psi{\bar \psi}\rangle$ and $\langle \phi \psi
{\bar\psi}\rangle$.

In Sec.~\ref{Sec:WTIs}, we have derived the WTIs by using the
equations $\delta \mathcal{Z} = 0$ under arbitrary infinitesimal
variations $\delta \psi$ and $\delta {\bar \psi}$. Here, in order to
unveil the relation between $\langle{\bar \psi}M^{\mu}\psi\psi{\bar
\psi}\rangle$ and $\langle \phi \psi {\bar\psi}\rangle$, we make use
of the fact that $\delta \mathcal{Z} = 0$ under an arbitrary
infinitesimal variation $\delta \phi$, which leads to the mean value
of the EOM of boson field $\phi(x)$:
\begin{eqnarray}
g\sum^N_{\sigma=1}\langle{\bar \psi}_{\sigma}(x)\gamma^{m}
\psi_{\sigma}(x)\rangle_J = \langle-\mathbb{D}\phi(x) -J(x)\rangle_J
= -\mathbb{D}\frac{\delta W}{\delta J(x)}-\langle J\rangle_J.
\label{Eq:EOMphi}
\end{eqnarray}
One might compare this equation to Eq.~(\ref{Eq:EOMpsi}) for
${\bar\psi}(x)$ and Eq.~(\ref{Eq:EOMbarpsi}) for $\psi(x)$. These
three equations have the same physical origin. The symbol $W = -i
\ln \mathcal{Z}$ is the generating functional of connected
correlation functions \cite{Itzykson}. As shown by Eq.~(B11), the
mean value of $\phi(x)$ is identical to $\delta W/\delta J(x)$,
which is used in the derivation of Eq.~(\ref{Eq:EOMphi}). Starting
from Eq.~(\ref{Eq:EOMphi}), we carry out functional derivatives
$\frac{\delta}{i\delta{\bar \eta}_\alpha(y)}$ and
$\frac{\delta}{-i\delta\eta_\beta(z)}$ in order on both sides and
then obtain
\begin{eqnarray}
g\langle{\bar \psi}_{\sigma}(x)\gamma^{m}\psi_{\sigma}(x)
\psi_\alpha(y){\bar \psi}_\beta(z)\rangle_c = -\mathbb{D}\langle
\phi(x)\psi_\alpha(y){\bar\psi}_\beta(z) \rangle_c =
-\mathbb{D}\frac{\delta^3 W}{\delta
J(x)\delta{\bar\eta}_\alpha(y)\delta \eta_\beta(z)}.\label{Eq:gmD}
\end{eqnarray}
This equation will then be used to derive the relation between the
current and interaction vertex functions.

We learn from the generic rules of function integral (see the
standard textbook \cite{Itzykson} for more details) that for each
fermion flavor $\sigma$
\begin{eqnarray}
\frac{\delta^3 W}{\delta J(x)\delta{\bar\eta}_\sigma(y)\delta
\eta_\sigma(z)} = \int
dx'D(x,x')\frac{\delta}{\delta\phi(x')}\left[\frac{\delta
^{2}\Xi}{\delta{\bar\psi}_\sigma(y)\delta\psi_\sigma(z)}\right]^{-1},
\end{eqnarray}
where $\Xi$ is the generating functional of proper vertices and is
connected to $W$ by the Legendre transformation given by
Eq.~(\ref{Eq:Legendre}). Here, for notational simplicity we drop the
indices $\alpha$ and $\beta$ but retain the flavor index $\sigma$.
Making use of the following identity for an arbitrary matrix
$\mathcal{M}$
\begin{eqnarray}
\frac{\delta}{\delta \phi(x')}\mathcal{M}^{-1}(y,z)=-\int dy'dz'
\mathcal{M}^{-1}(y,y')\frac{\delta \mathcal{M}
(y',z')}{\delta\phi(x')}\mathcal{M}^{-1}(z',z),
\end{eqnarray}
one obtains
\begin{eqnarray}
\frac{\delta^3 W}{\delta J(x)\delta{\bar\eta}_\sigma(y)
\delta\eta_\sigma(z)}= -\int dx'dy'dz' D(x,x')G(y,y')
\frac{\delta^{3}\Xi}{\delta\phi(x')\delta{\bar\psi}_\sigma(y')
\delta\psi_\sigma(z')}G(z',z).
\end{eqnarray}
According to the elementary rules of functional integral, one can
verify that
\begin{eqnarray}
\frac{\delta^{3}\Xi}{\delta \phi(x')\delta{\bar\psi}_{\sigma}(y')
\delta\psi_{\sigma}(z')}\Big|_{\phi,{\bar\psi},\psi=0} = g
\Gamma_{\mathrm{int}}(y'-x',x'-z').
\end{eqnarray}
This then implies that
\begin{eqnarray}
\frac{\delta^3 W}{\delta J(x)\delta{\bar\eta}_\sigma(y)
\delta\eta_\sigma(z)} = -g\int dx'dy'dz'D(x,x')G(y,y')
\Gamma_{\mathrm{int}}(y'-x',x'-z') G(z',z).\label{Eq:WJbaretaeta}
\end{eqnarray}
Combining Eq.~(\ref{Eq:gmD}) and Eq.~(\ref{Eq:WJbaretaeta}) gives
rise to
\begin{eqnarray}
\langle{\bar \psi}_{\sigma}(x)\gamma^{m} \psi_{\sigma}(x)
\psi_\alpha(y){\bar \psi}_\beta(z)\rangle_c = \mathbb{D}\int
dx'dy'dz'D(x,x')(G(y,y') \Gamma_{\mathrm{int}}(y'-x',x'-z')
G(z',z))_{\alpha\beta}.\nonumber
\\ \label{Eq:Gammaint}
\end{eqnarray}

In the above expressions, the product ${\bar\psi}_{\sigma}(x)
\gamma^{m}\psi_{\sigma}(x)$ comes from the fermion-boson interaction
term $\mathcal{L}_{fb} = g\phi(x){\bar\psi}_{\sigma}(x)
\gamma^{m}\psi_{\sigma}(x)$. However, one may also regard ${\bar
\psi}_{\sigma}(x)\gamma^{m}\psi_{\sigma}(x)$ as one component of a
generalized (flavor-independent) current $j_{M}^{\mu}(x)$,
which is previously defined by Eq.~(\ref{Eq:vectorcurrent}), with
$\gamma^{m}$ being one component of $M^{\mu}$. According to
Eq.~(\ref{Eq:currentvertex}), one can use current $j_{\gamma^{m}}(x)
= {\bar \psi}_{\sigma}(x)\gamma^{m}\psi_{\sigma}(x)$ to define a
current vertex function $\Gamma_{\gamma^{m}}$ as follows
\begin{eqnarray}
\langle j_{\gamma_{m}}(x)\psi_\alpha(y){\bar\psi}_\beta(z)\rangle_c
&\equiv& \langle{\bar \psi}_{\sigma}(x)\gamma^{m}
\psi_{\sigma}(x)\psi_\alpha(y){\bar \psi}_\beta(z)\rangle_c
\nonumber \\
&=& \int dy'dz'(G(y,y')\Gamma_{\gamma^{m}}(y'-x,x-z)
G(z',z))_{\alpha\beta}. \label{Eq:Gammagammam}
\end{eqnarray}

Comparing Eq.~(\ref{Eq:Gammaint}) and Eq.~(\ref{Eq:Gammagammam}), it
is easy to find that
\begin{eqnarray}
\mathbb{D}\int dx' D(x,x')\Gamma_{\mathrm{int}}(y'-x',x'-z') =
\Gamma_{\gamma^{m}}(y'-x,x-z').
\end{eqnarray}
After performing the following Fourier transformations
\begin{eqnarray}
\Gamma_{\mathrm{int}}(y'-x',x'-z')&=&\int\frac{dk dp}{(2
\pi)^{2(1+d)}}\Gamma_{\mathrm{int}}(k,p)e^{-ik(y'-x')-ip(x'-z')},\\
D(x-x') &=& \int\frac{dq}{(2\pi)^{1+d}}D(q)e^{-iq(x-x')}, \\
\Gamma_{\gamma^{m}}(y'-x',x'-z') &=& \int \frac{dk
dp}{(2\pi)^{2(1+d)}}\Gamma_{\gamma^{m}}(k,p)
e^{-ik(y'-x')-ip(x'-z')},
\end{eqnarray}
we immediately obtain an identity relating current vertex function
to interaction vertex function
\begin{eqnarray}
\Gamma_{\gamma^{m}}(k,p) = D_{0}^{-1}(k-p)
D(k-p)\Gamma_{\mathrm{int}}(k,p), \label{Eq:gammaD0Dgamma}
\end{eqnarray}
where the free boson propagator $D_{0}^{-1}(q)$ is the Fourier
transformation of $\mathbb{D}$. This identity is derived by
performing rigorous functional analysis, and thus is strictly valid.

Recall that the DS equation of Dirac fermion propagator is
\begin{eqnarray}
G^{-1}(p) = G^{-1}_{0}(p) + ig^2 \int \frac{dk}{(2
\pi)^{1+d}}\gamma^{m}G(k)D(k-p)\Gamma_{\mathrm{int}}(k,p). \nonumber
\end{eqnarray}
At first glance, this DS equation is not closed since it couples to
an infinite number of DS equations of $D(k-p)$,
$\Gamma_{\mathrm{int}}(k,p)$, and other higher-point correlation
functions. Luckily, this equation can be made self-closed by
properly employing several identities. A key point is that, one does
not need to separately determine $D(k-p)$ and
$\Gamma_{\mathrm{int}}(k,p)$. It is only necessary to determine
their product. According to the identity given by
Eq.~(\ref{Eq:gammaD0Dgamma}), the replacement
$$D(k-p)\Gamma_{\mathrm{int}}(k,p) \rightarrow
D_{0}(k-p)\Gamma_{\gamma^{m}}(k,p)$$ can be made, which then turns
the DS equation of $G(p)$ into a new form
\begin{eqnarray}
G^{-1}(p) = G^{-1}_{0}(p) + ig^2 \int \frac{dk}{(2\pi)^{1+d}}
\gamma^{m}G(k)D_{0}(k-p)\Gamma_{\gamma^{m}}(k,p).
\end{eqnarray}
In this new DS equation, the free boson propagator $D_{0}(k-p)$ can
be easily obtained and is supposed to be known, whereas the current
vertex function $\Gamma_{\gamma^{m}}(k,p)$ can be completely
determined by the full fermion propagator. In the last two sections,
we have shown how to obtain $\Gamma_{I}(k,p)$ and
$\Gamma_{\gamma^{0}}(k,p)$ by solving several coupled WTIs in (1+2)-
and (1+3)-dimensional Dirac semimetals. The generalization to other
cases, such as $\Gamma_{\gamma^{1}}(k,p)$ and
$\Gamma_{\gamma^{2}}(k,p)$, is straightforward. Now we can see that
the DS equation of fermion propagator $G(p)$ is indeed completely
self-closed and can be numerically solved once the free fermion
propagator $G_{0}(p)$ and the free boson propagator $D_{0}(q)$ are
known. Based on the numerical solutions, one can analyze various
interaction-induced effects. Since no small expansion parameter is
adopted, all the results are reliable no matter whether the
fermion-boson interaction is in the weak-coupling or strong-coupling
regime.

The identity given by Eq.~(\ref{Eq:gammaD0Dgamma}) is strictly valid
in the case of Coulomb interaction, and also in the case of
fermion-boson interaction under the harmonic oscillation
approximation. If the boson field $\phi$ represents the quantum
fluctuation of an order parameter, the identity
Eq.~(\ref{Eq:gammaD0Dgamma}) becomes invalid. The reason is that,
the action of bosonic order parameter always has self-coupling
terms, such as $u\phi^{4}$. When such a quartic term is present, an
additional $4u \phi^{3}$ term should be added to the mean value of
the EOM of $\phi$ field given by Eq.~(\ref{Eq:EOMphi}), namely
\begin{eqnarray}
g\sum^N_{\sigma=1}\langle{\bar \psi}_{\sigma}(x)\gamma^{m}
\psi_{\sigma}(x) \rangle_J= \langle-\mathbb{D}\phi(x) -
4u\phi^{3}(x) -J(x)\rangle_J.
\end{eqnarray}
Performing functional derivatives $\frac{\delta}{i\delta{\bar
\eta}_\alpha(y)}$ and $\frac{\delta}{-i\delta\eta_\beta(z)}$ yields
\begin{eqnarray}
g\langle{\bar \psi}_{\sigma}(x)\gamma^{m}\psi_{\sigma}(x)
\psi_\alpha(y){\bar \psi}_\beta(z)\rangle_c = -\mathbb{D}\langle
\phi(x)\psi_\alpha(y){\bar\psi}_\beta(z) \rangle_c - 4u\langle
\phi^{3}(x)\psi_\alpha(y){\bar\psi}_\beta(z)\rangle_c.
\end{eqnarray}
The $u\phi^{4}$ terms gives rise to a complicated five-point correlation function $\langle \phi^{3}\psi {\bar\psi} \rangle_c$. This extra term spoils the identity given by Eq.~(\ref{Eq:gammaD0Dgamma}). As a consequence, the DS equation of fermion propagator $G(p)$ is no longer self-closed. The same problem is encountered as one goes beyond the harmonic oscillation of lattice vibration and includes a self-interaction of phonons. If the coupling term $u\phi^{4}$ is sufficiently weak, one might take into account its contribution to $D_{0}(q)$ by performing weak perturbative expansion in powers of small $u$ and then substitute the modified boson propagator into the DS equation of $G(p)$. However, for strong $u\phi^{4}$, this approximation breaks down. We will investigate the impact of $u\phi^{4}$ term in the future.

\section{An example: Coulomb interaction in graphene \label{Sec:graphene}}

In this section we apply our generic approach to a concrete example.
We will investigate the quantum many-body effects of massless Dirac
fermions produced by the long-range Coulomb interaction in intrinsic
(undoped) graphene, which is the most prototypical
$(1+2)$-dimensional Dirac semimetal. This problem has been
theoretically investigated for over twenty-five years. However, due
to the absence of a reliable non-perturbative tool, there are still
some open questions regarding the impact of Coulomb interaction on
the low-energy behaviors of Dirac fermions. Taking advantage of our
approach, we will be able to conclusively answer these open
questions.

The Lagrangian of $(1+2)$-dimensional Dirac fermion system is
already given in Sec.~\ref{Sec:Model}. But for readers' convenience
we wish to make this section self-contained and re-write the
Lagrangian density as follows
\begin{eqnarray}
\mathcal{L}_{\mathrm{DF}} =
\sum^N_{\sigma=1}{\bar\psi}_\sigma(i\partial_{t} \gamma^{0} - v
\partial_{i}\gamma^{i}){\psi_\sigma} + a_{0}
\frac{|\nabla|}{8\pi v\alpha}a_{0} - \sum^N_{\sigma=1}a_{0}{\bar
\psi}_{\sigma} \gamma^{0}\psi_{\sigma}.
\end{eqnarray}
The fermion flavor is fixed at its physical value $N=2$ throughout
this section, The strength of Coulomb interaction is characterized
by a dimensionless parameter $\alpha = e^{2}/v \epsilon$, where $v$
is a uniform Fermi velocity and $\varepsilon$ is dielectric
constant, which can be regarded as an effective fine structure
constant. Notice that the velocity $v$ is explicitly written down
throughout this section. For simplicity, we consider the isotropic
graphene with the fermion velocity being a constant in all
directions. The above Lagrangian density respects a continuous
chiral symmetry
\begin{eqnarray}
\psi \rightarrow e^{i\theta \gamma^{5}}.
\end{eqnarray}
If the originally massless Dirac fermions acquire a finite mass due
to the formation of excitonic pairs, this symmetry would be
dynamically broken. The order parameter of the excitonic insulating
phase is $m(x)=\langle {\bar \psi}(x)\psi(x)\rangle$.

The free boson propagator is
\begin{eqnarray}
D_{0}(\mathbf{q}) = \frac{2\pi \alpha}{|\mathbf{q}|}.
\end{eqnarray}
The free fermion propagator is
\begin{eqnarray}
G_{0}(p) \equiv G_{0}(p_{0},\mathbf{p}) = \frac{1}{\gamma^{0}p_{0} -
v\mathbf{\gamma}\cdot \mathbf{p}},
\end{eqnarray}
where $\mathbf{\gamma}\cdot \mathbf{p} = \gamma^{i}p^{i}$. After
including the interaction-induced corrections, it is significantly
renormalized and becomes
\begin{eqnarray}
G(p) \equiv G(p_{0},\mathbf{p}) = \frac{1}{A_{0}(p)\gamma^{0}p_{0} -
A_{1}(p)\mathbf{\gamma}\cdot \mathbf{p} + m(p)}, \label{Eq:fullGp}
\end{eqnarray}
where we have introduced three functions: $A_0(p) \equiv
A_{0}(p_{0},\mathbf{p})$ embodies the (Landau-type) fermion damping,
$A_1(p)\equiv A_{1}(p_{0},\mathbf{p})$ reflects the fermion velocity
renormalization, and $m(p)\equiv m(p_{0},\mathbf{p})$ represents the
excitonic mass gap.

Before performing non-perturbative analysis, below we would first
review some previous perturbative studies on the problem. It will
become clear why it is necessary to abandon perturbative approaches
and develop a non-perturbative approach.

\subsection{Weak-coupling perturbation theory}

From the perspective of quantum field theory, the long-range Coulomb interaction between Dirac fermions in graphene can be described by a variant of the well-studied $(1+3)$-dimensional QED, dubbed QED$_{4}$. The graphene-version of QED is defined in $(1+2)$ dimensions and Dirac fermions couple to a real scalar boson $a_{0}$, rather than a vector field $a_{\mu}$. Unlike QED$_{4}$, the graphene-version of QED does not really suffer from ultraviolet (UV) divergences, since, being an effective low-energy theory, it has an explicit UV cutoff $\Lambda$, which can be determined by the inverse of lattice spacing. Despite such
differences, these two models basically have the same field-theoretical structure and thus are expected to be analyzed in an analogous way. It is well known that weak perturbation theory \cite{Itzykson} is the standard method of treating QED$_{4}$. To compute a physical quantity, one always expands it into a power series in the fine structure constant $\alpha$. The UV divergence of each coefficient is eliminated by the renormalization procedure. The combination of perturbation theory and renormalization \cite{Itzykson}, developed by Tomanaga, Schwinger, Feynman, and Dyson, is incredibly successful. In particular, the anomalous magnetic moment of electron has been computed up to the $O(\alpha^{5})$ order \cite{Aoyama12}, and the theoretical results are in extremely good agreement with experiments \cite{Aoyama12}. Given the success of perturbation theory achieved in previous studies of QED$_{4}$ and other weakly interacting quantum field theories, it is natural to employ the techniques of perturbation expansion to theoretically investigate the interaction effects in graphene.

Ten years before monolayer graphene was isolated \cite{Geim05, Geim07}, Gonzalez \emph{et al.} \cite{Gonzalez94} had carried out a perturbative field-theoretical analysis of two-dimensional Dirac fermions subjected to Coulomb interaction. They found that, to the first-order of small-$\alpha$ expansion, i.e., $O(\alpha)$, the
fermion velocity $v_{\mathrm{R}}$ receives a logarithmic renormalization, described by
\begin{eqnarray}
\frac{v_{\mathrm{R}}(\mathbf{p})}{v} \approx 1 - \frac{\alpha}{4}
\ln\left(\frac{|\mathbf{p}|}{\Lambda}\right).\label{Eq:logarithmicv}
\end{eqnarray}
Here, $\mathbf{p}$ is the fermion momentum (relative to Dirac point)
and $\Lambda$ is the UV cutoff. The charge $e$ is not renormalized
by the Coulomb interaction \cite{Ye98, Herbut06}. The flow of
velocity $v$ with varying energy scale drives the parameter $\alpha$
to flow (see \cite{Kotov12} for a review). The influence of
$O(\alpha^{2})$ contributions have been subsequently examined by
several groups of authors \cite{Kotov08, Vozmediano10, Fogler12,
Mishchenko07, Vafek08}. In particular, the polarization function was
computed to $O(\alpha^{2})$ order in Refs.~\cite{Kotov08,
Vozmediano10, Fogler12}, and the fermion self-energy was calculated
to $O(\alpha^{2})$ order in Refs.~\cite{Mishchenko07, Vafek08}. The
results obtained in these theoretical works are not consistent. More
recently, Barnes \emph{et al.} \cite{Barnes14} have performed a
systematic perturbative calculations, and argued that the
first-order result of velocity renormalization can be dramatically
altered by higher-order corrections. In particular, after explicitly
computing the fermion self-energy up to $O(\alpha^{3})$ order and
the polarization function up to $O(\alpha^{2})$ order, Barnes
\emph{et al.} \cite{Barnes14} found that the renormalized velocity
$v_{\mathrm{R}}(\mathbf{p})$ should be expanded as a series that
contains all powers of logarithms, which suggested that
weak-coupling perturbation theory is not an appropriate tool for the
theoretical study of graphene. Sharma and Kopietz \cite{Sharma16}
have applied the functional renormalization group (RG) method to handle the interaction and demonstrated that the multi-logarithmic behavior reported in
Ref.~\cite{Barnes14} can be re-summed by means of functional RG
techniques to yield a simple logarithmic
$v_{\mathrm{R}}(\mathbf{p})$ that is very similar to
Eq.~(\ref{Eq:logarithmicv}). But this conclusion needs to be
verified more carefully since the contributions of three- and
four-point vertices are all neglected in functional RG
calculations.

An apparent fact is that previous perturbative calculations have not
reached a consensus on the behavior of fermion velocity
renormalization. Different results are obtained if different methods
and/or approximations are employed, which manifests the inefficiency
of perturbation theory. The breakdown of perturbation theory is
actually not out of expectation. Within the framework of
perturbation theory, physical quantities are computed as power
series expansions in some small (dimensionless) parameter. The fine
structure constant $\alpha = 1/137$ is small enough in QED$_{4}$,
rendering the applicability of perturbation theory. In contrast, the
effective fine structure constant $\alpha \sim 1$ in undoped
graphene. Specifically, $\alpha \approx 2.2$ for graphene suspended
in vacuum, and $\alpha \approx 0.4$ and $\alpha \approx 0.8$ for
graphene on BN and SiO$_{2}$ substrates, respectively. It is
therefore not surprising that higher-order contributions
substantially alter the first-order result \cite{Barnes14}. We
emphasize that there is actually a fundamental principle that causes
the breakdown of perturbation theory in graphene. In 1952, Dyson
\cite{Dyson52} pointed out that the power series of QED$_{4}$ is not
convergent if all the contributions are included. The series is only
asymptotic in the sense that summing terms up to an optimal
$N_{\mathrm{op}}$-order leads to the best agreement between
theoretical calculations and experiments but adding higher-order
terms would eventually drive the series to diverge. A crude estimate
given by Dyson \cite{Dyson52} indicated that $N_{\mathrm{op}}
\approx 1/\alpha \approx 137$. Migdal and Krainov \cite{Migdal68}
later obtained a different result: $N_{\mathrm{op}} \approx
137^{3/2}$. Recently, Kolomeisky \cite{Kolomeisky15} noticed the
similarity of the collapse of perturbative series to the
gravitational collapse of a star, and proposed that the value of
$N_{\mathrm{op}}$ can be computed by using the method of estimating
the famous Chandrasekhar's limit on the star mass. It was found in
Ref.~\cite{Kolomeisky15} that $N_{\mathrm{op}} \approx 5000$. In
practical theoretical studies on QED$_4$ there is no necessity to
worry about the validity of perturbation theory. But the situation
is sharply different in graphene where $\alpha$ is of the order of
unity. For undoped graphene, the value of $N_{\mathrm{op}}$, beyond which perturbation theory breaks down, should be much smaller than that of QED$_4$.
Kolomeisky \cite{Kolomeisky15} and Barnes \emph{et al.} \cite{Barnes14} have addressed this issue by adopting the analysis leading to the Chandrasekhar's limit. Although the value of $N_{\mathrm{op}}$ obtained in Ref.~\cite{Kolomeisky15} is a little different from that of Ref.~\cite{Barnes14}, the same conclusion is reached that conventional perturbation theory is not applicable in undoped graphene.

Many experimental techniques \cite{Elias11, Lanzara11, Chae12} have
been exploited to measure the momentum dependence of renormalized
fermion velocity in graphene. Surprisingly, the results extracted
from experiments seem to be well consistent with a logarithmic
velocity renormalization \cite{Elias11, Lanzara11, Chae12}. Then a
question arises. Given that weak perturbation theory breaks down,
why do experiments \cite{Elias11, Lanzara11, Chae12} extract a
logarithmic $\mathbf{p}$ dependence of fermion velocity that seems
to agree with the result obtained in first-order perturbative
calculations? Generically, there could be two possibilities. The
first possibility is that, the logarithmic behavior is valid only in
an intermediate range of momentum and is changed by higher-order
corrections in the region of lower momentum, which, nevertheless,
cannot be accessed by measurements due to limited resolution of
experimental techniques. The second possibility is that, the
renormalized fermion velocity $v_{\mathrm{R}}(\mathbf{p})$ still
exhibits a logarithmic $\mathbf{p}$-dependence if one could be able
to compute the contributions of \emph{all} the higher-order
corrections. It is impossible to judge which possibility is correct
within the framework of perturbation theory because nobody is
capable of calculating all the Feynman diagrams.

The DS equation approach developed in this paper provides a powerful
tool to deal with the strong Coulomb interaction and allows us to obtain a conclusive answer of the above question.

\subsection{$1/N$ expansion}

Since the series expansion in $\alpha$ does not work in graphene, we
would like to adopt a more suitable expansion parameter. A natural
alternative is the inverse of fermion flavor, i.e., $1/N$. The $1/N$
expansion \cite{Son07, Vafek07, Son08, Foster08, Hofmann14} provides
a different scheme to organize Feynman diagrams comparing to the
small-$\alpha$ expansion. To implement the $1/N$ expansion, one
needs to first compute the polarization function $\Pi(q)$ at the
level of RPA. The RPA-form of the
polarization \cite{Son07} is given by
\begin{eqnarray}
\Pi_{\mathrm{RPA}}(q) &=& -N\int \frac{d^{3}p}{(2\pi)^{3}}
\mathrm{Tr}[\gamma^{0}G_{0}(p+q)\gamma^{0}G_{0}(p)] \nonumber \\
&=& -\frac{N}{8}\frac{\mathbf{q}^{2}}{\sqrt{q_{0}^{2}+v^{2}\mathbf{q}^{2}}},
\end{eqnarray}
which then leads to the following dressed boson propagator
\begin{eqnarray}
D_{\mathrm{RPA}}(q) = \frac{1}{D_{0}^{-1}(q)-\Pi_{\mathrm{RPA}}(q)}.
\label{Eq:DRPA}
\end{eqnarray}
Each Feynman diagram has a number of boson propagators and fermion
loops. We know that $D_{\mathrm{RPA}}(q) \sim N^{-1}$ and each
fermion loop contributes a factor of $N$. Thus all the Feynman
diagrams can be classified by the powers of $1/N$. It is expected
that most quantum corrections, especially the vertex corrections,
are suppressed in the limit of $N\rightarrow \infty$.

It is technically very difficult to compute Feynman diagrams within
the framework of $1/N$ expansion. The RPA-form of boson propagator,
i.e., $D_{\mathrm{RPA}}(q)$, is more complicated than the bare
propagator $D_{0}(q)$. Hence one is forced to introduce many further
approximations to compute the complicated integrals of multi-loop
diagrams, which inevitably reduces the accuracy of the results. Son
\cite{Son07} has performed an approximate analysis to the
leading-order of $1/N$ expansion and argued that the velocity $v$
acquires a finite anomalous dimension, which, however, has never
been experimentally observed. Hofmann \emph{et al.} \cite{Hofmann14}
have calculated the quasiparticle residue and the renormalized
fermion velocity to next-to-leading order and claimed to obtain
results consistent with experiments. Nevertheless, it is unclear
whether or not such a consistency survives higher-order corrections.
Recall the physical flavor is $N=2$. If Dyson's argument
\cite{Dyson52} and its refined versions \cite{Migdal68,
Kolomeisky15, Barnes14} are applied to analyze the convergence
radius of the formal power series in $1/N$, it is legitimate to
expect that the series would rapidly become out of control as
higher-order corrections are included. In
Sec.~\ref{Sec:nonperturbative}, we will show that the $1/N$
expansion is especially unreliable when it is combined with the DS
equation(s) to treat the non-perturbative effects of Coulomb
interaction.

\subsection{Non-perturbative study on excitonic instability \label{Sec:nonperturbative}}

There is one more reason to distrust perturbation theory: it is not
capable of capturing the non-perturbative effects. One possible
non-perturbative effect of long-range Coulomb interaction is the
occurrence of excitonic pairing instability. As discussed in
Sec.~\ref{Sec:Introdunction}, a finite mass gap could be generated
by the formation of excitonic-type particle-hole pairs when $\alpha$
exceeds a critical value $\alpha_c$. As a consequence, the chiral
(sublattice) symmetry of gapless semimetallic state is dynamically
broken \cite{Khveshchenko01, Gorbar02, Khveshchenko04,
Khveshchenko09, Liu09, Gamayun10, WangLiu12, WangLiu14, Gonzalez12,
Gonzalez12jhep, Carrington16, Carrington18, Drut09A, Armour10,
Buividovich12, Ulybyshev13, Tupitsyn17}, which turns the originally
gapless semimetal into a gapped excitonic insulator. This is an
interaction-driven quantum phase transition that has been studied
for twenty years since the seminal work of Khveshchenko
\cite{Khveshchenko01}. Why is this problem interesting? In 1960,
Pauling \cite{Pauling, Castrophysics} conjectured that the exact
ground state of graphene might be an interaction-induced insulator.
At almost the same time, Nambu and Jona-Lasinio \cite{Nambu61}
proposed a novel scenario in which massless Dirac fermions can
acquire a finite mass via the mechanism of dynamical chiral symmetry
breaking, which plays a fundamental role in the research field of
QCD. Several years later, Keldysh and Kopaev \cite{Kopaev65}
predicted the existence of excitonic insulators driven by
particle-hole pairing. It is remarkable that graphene is a rare
material that might simultaneously realize the above three
theoretical predictions. To judge whether an excitonic gap is opened
in a realistic graphene, it is necessary to determine the accurate
value of $\alpha_c$ and compare it to the physical value of
$\alpha$. The method of weak-coupling perturbation is definitely
failed since dynamical excitonic gap generation is a
non-perturbative effect. No gap is generated at any finite order of
perturbative calculations, no matter whether $\alpha$ or $1/N$ is
adopted to carry out the series expansion.

Two non-perturbative methods are often adopted to compute $\alpha_c$
in the literature. One is the DS equation method combined with $1/N$
expansion. It is now clear that the value of $\alpha_c$ obtained by
this method is strongly approximation dependent
\cite{Khveshchenko01, Gorbar02, Khveshchenko04, Khveshchenko09,
Liu09, Gamayun10, WangLiu12, WangLiu14, Gonzalez12, Gonzalez12jhep,
Carrington16, Carrington18}, ranging from $\alpha_c = 0.9$ to
$7.9$ (see Ref.~\cite{WangLiu12} for a summary). Such
calculations are usually based on the naive assumption that the
corrections to fermion-boson vertex function are suppressed by high
powers of $1/N$. This assumption is apparently problematic because
the physical flavor is $N=2$ if four-component spinor representation
is used (chiral symmetry cannot be defined in terms of two-component
spinor). In the absence of an efficient route to include vertex
corrections, the exact value and even the existence of $\alpha_c$
cannot be convincingly specified. The other non-perturbative method
is the QMC simulation \cite{Drut09A, Armour10, Buividovich12,
Ulybyshev13, Tupitsyn17}. This method suffers from fermion-sign
problem and severe finite-size effects, and also leads to
controversial conclusions \cite{Drut09A, Armour10, Buividovich12,
Ulybyshev13, Tupitsyn17} about the value of $\alpha_{c}$. In a
recent work, Tang \emph{et al.} \cite{Tang18} have proposed an
approach to handle strong interactions in Dirac semimetal by
combining QMC simulation and perturbative RG technique. While their
approach can be applied to treat strong on-site interaction, it
failed to access the regime of strong long-range Coulomb interaction
\cite{Tang18}.

Perturbative RG method is often used to address the possible
existence of a strong-coupling fixed point, which, if exists at all,
is usually expected to signal the happening of some sort of ordering
instability. Vafek and Case \cite{Vafek08} performed a two-loop RG
analysis of the Coulomb interaction and claimed to find an unstable
infrared fixed point $\alpha^{\ast}\approx 0.8$, implying that
$\alpha$ would exhibit a runaway behavior at low energies if its
initial value is greater than $0.8$. However, the existence of such
an fixed point does not necessarily mean that excitonic insulating
transition must occur, because it may indicate the emergence of
other instabilities or the complete breakdown of perturbative RG
method in the strong-coupling regime. To determine under what
circumstance an excitonic instability is triggered by the Coulomb
interaction, the most direct approach is to compute the excitonic
gap $m(p)$ and quantitatively study how it depends on various
parameters, such as $\alpha$ and $T$. Perturbative RG is certainly
incapable of implementing such calculations.

The DS integral equation provides an ideal theoretical framework to
quantitatively compute the excitonic gap $m(p)$. The dependence of
$m(p)$ on $\alpha$ and $T$ can be naturally extracted from the
solutions of its DS equation. The fermion velocity renormalization
and the excitonic gap generation are induced by the same Coulomb
interaction and thus have mutual effects on each other. Using the DS
equation approach, their interplay can be investigated in a
self-consistent manner. Unfortunately, all previous DS equation
studies suffer from the significant uncertainties induced by the
ignorance of the precise form of the vertex function. In this paper,
we can accurately incorporate the exact vertex function into the DS
equation of fermion propagator with the help of several identities,
which makes it possible to obtain reliable and approximation-free
results.

\subsection{Exact Dyson-Schwinger integral equations}

Now we apply our DS equation approach to study the fermion velocity
renormalization and the possibility of excitonic pairing on an equal
footing.

From the analysis presented above, the free and fully renormalized
fermion propagators satisfy the following DS equation
\begin{eqnarray}
G^{-1}(p) = G^{-1}_{0}(p) + i\int\frac{d^3k}{(2\pi)^3}
\gamma^{0}G(k)D(k-p)\Gamma_{\mathrm{int}}(k,p).\nonumber
\end{eqnarray}
Using the identity given by Eq.~(\ref{Eq:gammaD0Dgamma}), we convert
this equation into
\begin{eqnarray}
G^{-1}(p) = G^{-1}_{0}(p) + i \int \frac{d^3k}{(2\pi)^3}
\gamma^{0}G(k)D_{0}(k-p)\Upsilon_{\gamma^{0}}(k,p),
\end{eqnarray}
where $D_{0}(q) = \frac{2\pi \alpha}{|\mathbf{q}|}$ is the bare
Coulomb interaction function. We emphasize that the polarization
function, usually denoted by $\Pi(q)$, should not be included into
$D_{0}(q)$. Otherwise, the influence of the polarization would be
double counted. With the help of Eq.~(\ref{Eq:gammaD0Dgamma}), the
effect of dynamical screening of Coulomb interaction, represented by
full boson propagator $D(q) = \frac{1}{D_{0}^{-1}(q)-\Pi(q)}$, is
included indirectly in the current vertex function
$\Upsilon_{\gamma^{0}}(k,p)$. An advantage of such a manipulation is
that it avoids adopting the so-called RPA, which has been
extensively used in field-theoretic studies \cite{Gonzalez99,
DasSarma07, Polini07, Son07, Vafek07, Son08, Foster08, Hofmann14,
Khveshchenko01, Gorbar02, Khveshchenko04, Khveshchenko09, Liu09,
Gamayun10, WangLiu12, WangLiu14, Gonzalez12, Carrington16,
Carrington18} of the Coulomb interaction but is actually not well
justified for $N=2$. According to Eq.~(77), the current vertex
function $\Upsilon_{\gamma^{0}}(k,p)$ has the form
\begin{eqnarray}
\Upsilon_{\gamma^{0}}(k,p) &=& \frac{1}{\det(M_{\mathcal{A}})}
\big[q_0\left(q_0^2-P_1^2-P_2^2\right){\mathcal{A}_0} +
\left(q_1P_1^2 + q_2P_1P_2-q_0^2q_1\right){\mathcal{A}_1} \nonumber
\\
&& +\left(q_1P_1P_2+q_2P_2^2-q_0^2q_2\right){\mathcal{A}_{2}} -
q_0\left(q_2P_1-q_1P_2\right){\mathcal{A}_{3}}\big],
\end{eqnarray}
where the denominator is
\begin{eqnarray}
\det(M_{\mathcal{A}}) &=& q_0^2\left(q_0^2 - q_1^2 - q_2^2\right) -
P_1\left(P_1q_0^2 - P_1q_1^2 - P_2q_1q_2\right) - P_2\left(P_2q_0^2
- P_2q_2^2 - P_1 q_1 q_2\right) \nonumber \\
&=& q_0^4 - 2q_0^2v^2(\mathbf{k}^2 + \mathbf{p}^2) +
v^4(\mathbf{k}^2 - \mathbf{p}^2)^2
\end{eqnarray}
and $\mathcal{A}_{0,1,2,3}$ are related to the full fermion
propagator as follows
\begin{eqnarray}
\mathcal{A}_{0} &=& -\left[G^{-1}(k)-G^{-1}(p)\right], \\
\mathcal{A}_1 &=& -v\left[G^{-1}(k)\gamma^0\gamma^1 +
\gamma^0\gamma^1G^{-1}(p)\right], \\
\mathcal{A}_2 &=& -v\left[G^{-1}(k)\gamma^0\gamma^2 +
\gamma^0\gamma^2G^{-1}(p)\right], \\
\mathcal{A}_3 &=& -v^2\left[{G}^{-1}(k)\gamma^1\gamma^2
-\gamma^1\gamma^2G^{-1}(p)\right].
\end{eqnarray}
Since $\Upsilon_{\gamma^{0}}(k,p)$ depends only on $G(k)$ and
$G(p)$, the DS equation of $G(p)$ is self-closed, decoupled from
that of the boson propagator and all the other correlation
functions. Now we could substitute the generic form of $G(p)$, given
by Eq.~(\ref{Eq:fullGp}), into its DS equation and then obtain
\begin{eqnarray}
A_{0}(p)\gamma^{0}p_{0} - A_{1}(p)\mathbf{\gamma}\cdot \mathbf{p} +
m(p) = \gamma^{0}p_{0} - \mathbf{\gamma}\cdot \mathbf{p} + i\int
\frac{d^3k}{(2\pi)^3}\gamma^{0} G(k)
D_{0}(k-p){\Upsilon}_{\gamma^{0}}(k,p).
\label{Eq:DSEG3}
\end{eqnarray}
This DS equation can be readily decomposed into three coupled
integral equations of $A_{0}(p)$, $A_{1}(p)$, and $m(p)$.
Calculating the trace of Eq.~(\ref{Eq:DSEG3}) leads to the equation
of $m(p)$. Multiplying matrix $\gamma^{0}$ and $\gamma^{1}$ to both
sides of Eq.~(\ref{Eq:DSEG3}) and then calculating the trace yield
the equations of $A_{0}(p)$ and $A_{1}(p)$, respectively. The
interaction-induced effects of Dirac fermions can be extracted from
the numerical solutions of $A_{0}(p)$, $A_{1}(p)$, and $m(p)$.

The exact integral equations of $A_0(p)$, $A_1(p)$, and $m(p)$ are
\begin{eqnarray}
A_0(p)p_0 - p_0 &=& -i \int\frac{v^2d^3k}{(2\pi)^{3}}\frac{D_{0}
(k-p)}{\left(m^2(k) - A_0^2(k)k_0^2 + A_1^2(k)v^2\mathbf{k}^2\right)
\det(M_{\mathcal{A}})} \nonumber \\
&& \times \Big[A_0(k)k_0\big[q_0\left(v^2P_1^2 + v^2P_2^2 -
q_0^2\right) \left(A_0(k)k_0-A_0(p)p_0\right) \nonumber \\
&& -\left(v^2q_1P_1^2+v^2q_2P_1P_2-q_1q_0^2\right) v
\left(A_1(k)vk_1-A_1(p)vp_1\right) \nonumber \\
&& -\left(v^2q_1P_1P_2+v^2q_2P_2^2-q_2q_0^2\right) v
\left(A_1(k)vk_2-A_1(p)vp_2\right)\big]
\nonumber \\
&& -A_1(k)vk_1\big[q_0\left(v^2P_1^2 + v^2P_2^2 -
q_0^2\right)\left(A_1(k)vk_1 - A_1(p)v p_1\right) \nonumber \\
&& -\left(v^2q_1P_1^2 + v^2q_2P_1P_2 - q_1q_0^2\right) v
\left(A_0(k)k_0-A_0(p)p_0\right) \nonumber \\
&& +q_0\left(q_2P_1-q_1P_2\right) v^2
\left(A_1(k)vk_2+A_1(p)vp_2\right)\big] \nonumber \\
&& -A_1(k)vk_2 \big[q_0\left(v^2P_1^2+v^2P_2^2-q_0^2\right)
\left(A_1(k)vk_2-A_1(p)vp_2\right) \nonumber\\
&& -\left(v^2q_1P_1P_2+v^2q_2P_2^2-q_2q_0^2\right) v
\left(A_0(k)k_0-A_0(p)p_0\right)
\nonumber\\
&&-q_0\left(q_2P_1-q_1P_2\right) v^2
\left(A_1(k)vk_1+A_1(p)vp_1\right)\big]\nonumber \\
&& -m(k)\big[q_0\left(v^2P_1^2+v^2P_2^2+q_0^2\right)
\left(m(k)-m(p)\right)\big]\Big],
\label{Eq:eqa0} \\
A_1(p)vp_1 - vp_1 &=& -i \int \frac{v^2d^3k}{(2\pi)^{3}} \frac{D_{0}
(k-p)}{\left(m^2(k) - A_0^2(k)k_0^2 + A_1^2(k)v^2\mathbf{k}^2\right)
\det(M_{\mathcal{A}})} \nonumber \\
&& \times \Big[A_0(k)k_0\big[q_0 \left(v^2P_1^2 + v^2P_2^2 -
q_0^2\right)\left(A_1(k) k_1 - A_1(p)vp_1\right) \nonumber \\
&& -\left(v^2q_1P_1^2+v^2q_2P_1P_2-q_1q_0^2\right) v
\left(A_0(k)k_0-A_0(p)p_0\right) \nonumber \\
&&+q_0\left(q_2P_1-q_1P_2\right)v^2\left(A_1(k)vk_2+A_1(p)vp_2\right)\big]
\nonumber\\
&& - A_1(k)vk_1\big[q_0\left(v^2P_1^2 + v^2P_2^2 -
q_0^2\right)\left(A_0(k)k_0-A_0(p)p_0\right) \nonumber \\
&& -\left(v^2q_1P_1^2 + v^2q_2P_1P_2 - q_1q_0^2\right) v
\left(A_1(k)vk_1-A_1(p)vp_1\right)
\nonumber \\
&&-\left(v^2q_1P_1P_2 + v^2q_2P_2^2 - q_2q_0^2\right) v
\left(A_1(k)vk_2 - A_1(p)vp_2\right)\big] \nonumber \\
&& +A_1(k)vk_2\big[\left(v^2q_1P_1^2 + v^2q_2P_1P_2 -
q_1q_0^2\right) v \left(A_1(k)v k_2 + A_1(p)vp_2\right) \nonumber \\
&& -\left(v^2q_1P_1P_2 + v^2q_2P_2^2 - q_2q_0^2\right) v
\left(A_1(k)vk_1+A_1(p)vp_1\right)
\nonumber \\
&& -q_0\left(q_2P_1-q_1P_2\right) v^2\left(A_0(k)k_0 -
A_0(p)p_0\right)\big] \nonumber \\
&& +m(k)\big[\left(v^2q_1P_1^2 + v^2q_2P_1P_2 - q_1q_0^2\right) v
\left(m(k) + m(p)\right)\big]\Big],
\label{Eq:eqa1}\\
m(p) &=& -i \int \frac{v^2d^3k}{(2\pi)^{3}}\frac{D_{0}
(k-p)}{\left(m^2(k) - A_0^2(k)k_0^2 + A_1^2(k)v^2\mathbf{k}^2\right)
\det(M_{\mathcal{A}})} \nonumber \\
&& \times\Big[A_0(k)k_0 q_0\left(v^2P_1^2 + v^2P_2^2 - q_0^2\right)
\left(m(k)-m(p)\right) \nonumber \\
&& -A_1(k)vk_1 \left(v^2q_1P_1^2 + v^2q_2P_1 P_2 - q_1 q_0^2\right)
v \left(m(k)+m(p)\right) \nonumber \\
&& -A_1(k)vk_2 \left(v^2q_1P_1P_2 + v^2q_2P_2^2 - q_2q_0^2\right)
v\left(m(k)+m(p)\right) \nonumber \\
&& -m(k)\big[q_0(v^2P_1^2 + v^2P_2^2- q_0^2)(A_0(k)k_0-A_0(p)p_0)
\nonumber \\
&& -(v^2q_1P_1^2 + v^2q_2P_1P_2 - q_1q_0^2)v(A_1(k)vk_1-A_1(p)vp_1)
\nonumber \\
&&-\left(v^2q_1P_1P_2 + v^2q_2P_2^2 - q_2q_0^2\right)
v\left(A_1(k)vk_2-A_1(p)vp_2\right)\big]\Big]. \label{Eq:eqm}
\end{eqnarray}
As discussed in Sec.~\ref{Sec:gamma012}, it is most convenient to
work in the Matsubara formalism and set $p_0 = i(2n+1)k_{B}T$. The
zero-temperature results can be obtained by taking the $T
\rightarrow 0$ limit. The integration range is initially
$[0,\Lambda]$, where $\Lambda$ is a UV cutoff for $k$. For
calculational convenience, we rescale all momenta by defining
dimensionless variables $p_{\mu}\rightarrow p_{\mu}/\Lambda$ and
$k_{\mu}\rightarrow k_{\mu}/\Lambda$, which changes the integration
range to $[0,1]$. In practical numerical computations, it is also
necessary to introduce a small IR cutoff. The influence of different
IR cutoffs will be discussed later.

These three equations are self-consistently coupled, implying that
the fermion damping, velocity renormalization, and excitonic pairing
are treated on an equal footing. It is unlikely that these equations
have analytical solutions. We will numerically solve them by using
the iteration method. This method involves several steps. We first
choose some initial values of $A_{0}(p)$, $A_{1}(p)$, and $m(p)$,
and substitute the chosen initial values into the coupled integral
equations to obtain a set of new values. Then we substitute this set
of new values into the same equations to obtain another set of new
values. Repeat the same operation over and over again until
convergence is achieved. Here the criterion of convergence is that
solutions do not change after carrying out further iterations. The
final results should not depend on the initial values of $A_{0}(p)$,
$A_{1}(p)$, and $m(p)$. For a detailed elaboration of the iteration
method, please refer to Ref.~\cite{Liu19}.

Over the last 20 years, a variety of approximations have been
employed to solve the DS equation of the fermion propagator. Before
solving the above exact equations, we first review some of the
results obtained under various approximations. To the leading order
of the $1/N$ expansion, the vertex function takes its bare form,
namely
\begin{eqnarray}
\Gamma_{\mathrm{int}}(k,p) = \gamma^{0},
\end{eqnarray}
and all the corrections to the renormalization functions are
ignored, implying that
\begin{eqnarray}
A_{0}(p) = A_{1}(p) = 1.
\end{eqnarray}
Under such approximations, the equation of fermion mass gap
\cite{Khveshchenko01, Gorbar02, Khveshchenko04, Khveshchenko09,
Liu09, Gamayun10} has a simple expression
\begin{eqnarray}
m(p) = \int \frac{d^3k}{(2\pi)^{3}}\frac{m(k)}{m^2(k) + k_0^2 +
\mathbf{k}^2}D_{\mathrm{RPA}}(k-p),
\end{eqnarray}
where the boson propagator $D_{\mathrm{RPA}}(k-p)$ is given by
Eq.~(\ref{Eq:DRPA}). Khveshchenko \cite{Khveshchenko01} solved this
equation in the instantaneous approximation, which amounts to
omitting the energy-dependence of $D_{\mathrm{RPA}}(k-p)$, and
argued that an excitonic gap is generated if $N < N_{c} \approx 2.5$
in the strong coupling limit $\alpha \gg 1$. Gorbar \emph{et al.}
\cite{Gorbar02} also analyzed this equation under the same
approximation, showing that $\alpha_{c} \approx 2.33$ for the
physical fermion flavor $N=2$. Khveshchenko \cite{Khveshchenko09}
studied the influence of fermion velocity renormalization on the gap
generation, but still ignoring all the vertex corrections, and
revealed that excitonic transition occurs at $\alpha_c \approx 1.13$
for $N=2$. Liu \emph{et al.} \cite{Liu09} numerically solved this
gap equation by using the energy-dependent propagator
$D_{\mathrm{RPA}}(k-p)$ and found $\alpha_c \approx 1.2$ for $N=2$.
Gamayun \emph{et al.} \cite{Gamayun10} discovered that $\alpha_c
\approx 0.92$ after analytically solving nearly the same gap
equation. The above gap equation is apparently oversimplified,
because it neglects all the contributions due to $A_{0}(p)$,
$A_{1}(p)$, and $\Gamma_{\mathrm{int}}(k,p)$. Their contributions
must be taken into account simultaneously. Otherwise, the
U(1)-symmetry induced WTI, given by Eq.~(\ref{Eq:ordinaryWTI12}),
would be violated. Including the impact of $A_{0}(p)$, $A_{1}(p)$,
and $\Gamma_{\mathrm{int}}(k,p)$ is extremely difficult because the
vertex function $\Gamma_{\mathrm{int}}(k,p)$ seems too complicated
to tackle. In 2012, Wang and Liu \cite{WangLiu12} considered a
simple \emph{Ansatz} for the vertex function that respects the
ordinary WTI, and revealed that such a vertex function significantly
increases the critical coupling to $\alpha_{c} \approx 3.2$, which
implies the absence of excitonic gap generation in suspended
graphene. Subsequently, Carrington \emph{et al.} \cite{Carrington16,
Carrington18} made a more detailed analysis of the impact of several
different \emph{Ans\"{a}tze} of the vertex function. The value of
$\alpha_{c}$ obtained in \cite{Carrington16} ranges from $2.89$ to
$7.80$ under several different approximations. Gonzalez
\cite{Gonzalez12, Gonzalez12jhep} studied the zero energy/momentum
($q = k-p = 0$) limit of the vertex function
$\Gamma_{\mathrm{int}}(k,p)$ in the so-called ladder approximation
(without crossing of boson lines). The free fermion propagator
$G_{0}(p)$ and free boson propagator $D_{0}(q)$ were used in
Refs.~\cite{Gonzalez12, Gonzalez12jhep} to analyze the behavior of
$\Gamma_{\mathrm{int}}(k=p)$, which simplifies analytical
calculations but neglects the contributions from the fermion
self-energy and the dynamical screening effect. To summarize,
although the possibility of excitonic gap generation has been
investigated by the DS equation approach for 20 years, it is
still far from clear whether an excitonic insulating state can
emerge in any realistic graphene material.

All the previous DS equation studies \cite{Khveshchenko01, Gorbar02,
Khveshchenko04, Khveshchenko09, Liu09, Gamayun10, WangLiu12,
Gonzalez12, Gonzalez12jhep, Carrington16, Carrington18} have
introduced a certain number of unjustified approximations, and the
value of $\alpha_c$ obtained in these works is strongly dependent of
the adopted approximations. To compute the precise value of
$\alpha_c$, it is necessary not to use any approximation. In this paper, the vertex function is completely determined by solving a number of strictly valid identities. The three self-consistent integral equations of $A_{0}(p)$, $A_{1}(p)$, and
$m(p)$ given by Eqs.(\ref{Eq:eqa0})-(\ref{Eq:eqm}) are exact, which
allows us to unambiguously determine whether an excitonic gap is
opened by Coulomb interaction, and, if the answer is yes, the
accurate value of $\alpha_{c}$.

Below we present our numerical solutions and analyze their physical
implications.

\begin{figure}[htbp]
\centering
\includegraphics[width=2.6in]{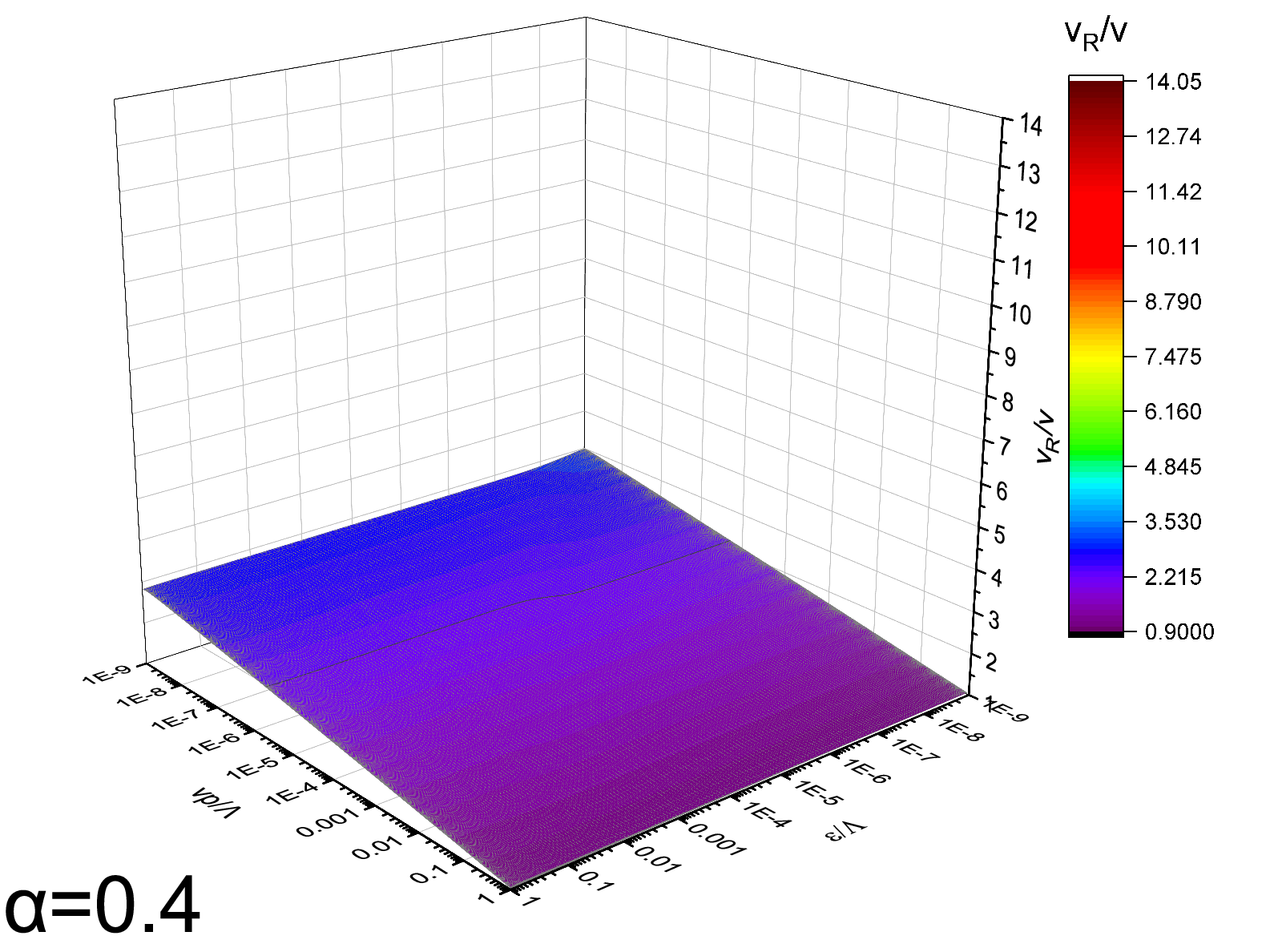}
\includegraphics[width=2.6in]{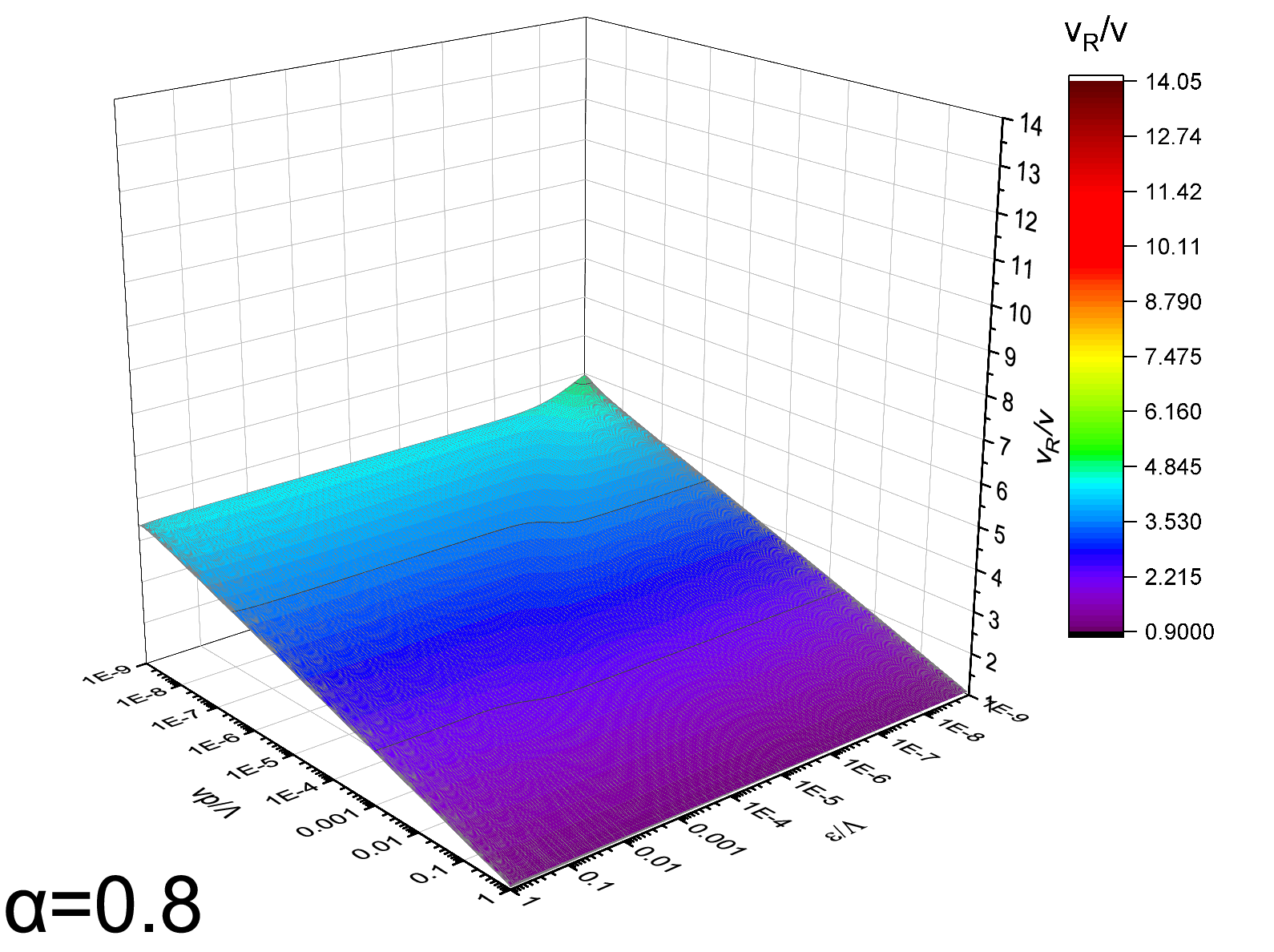}
\includegraphics[width=2.6in]{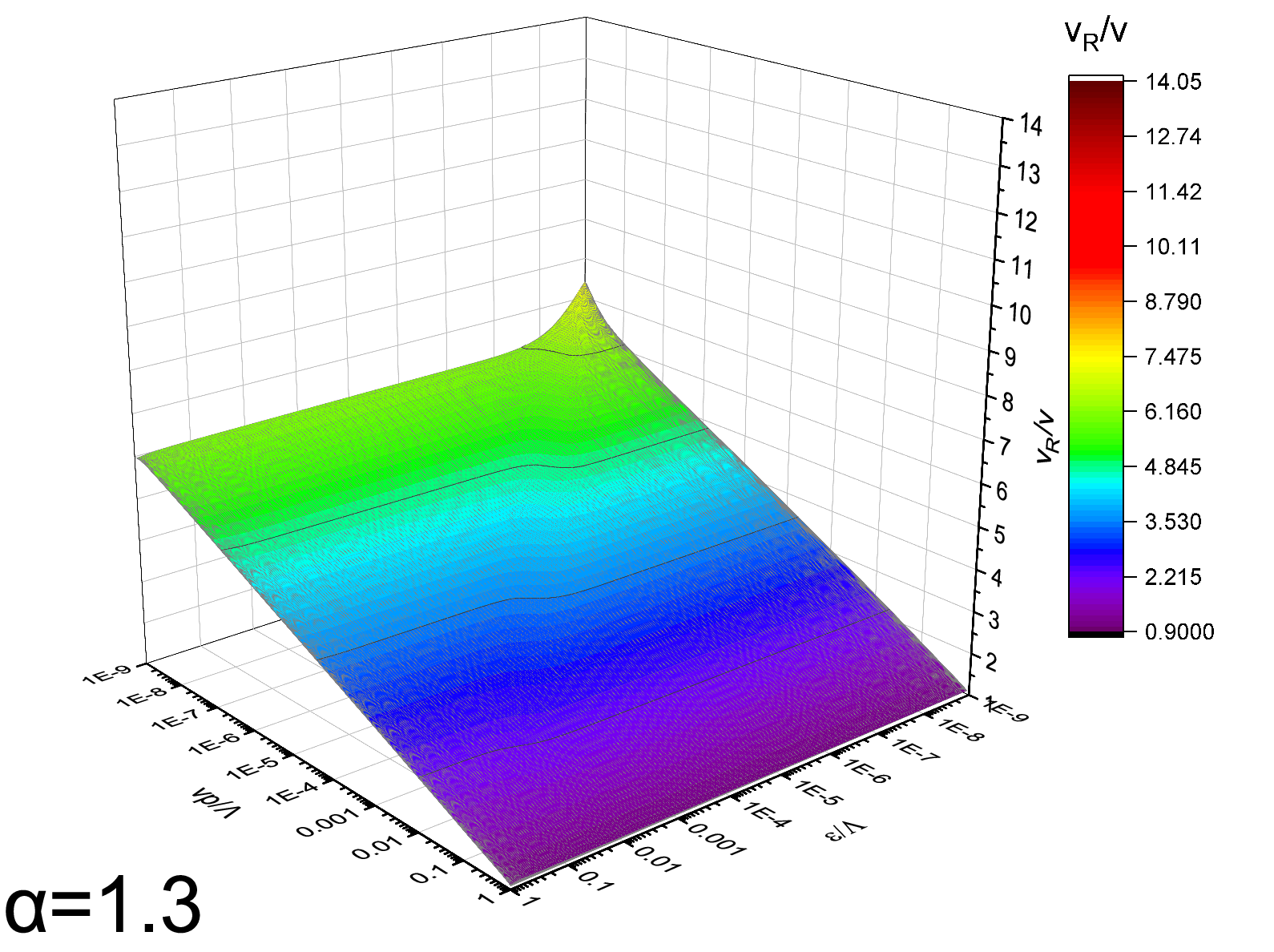}
\includegraphics[width=2.6in]{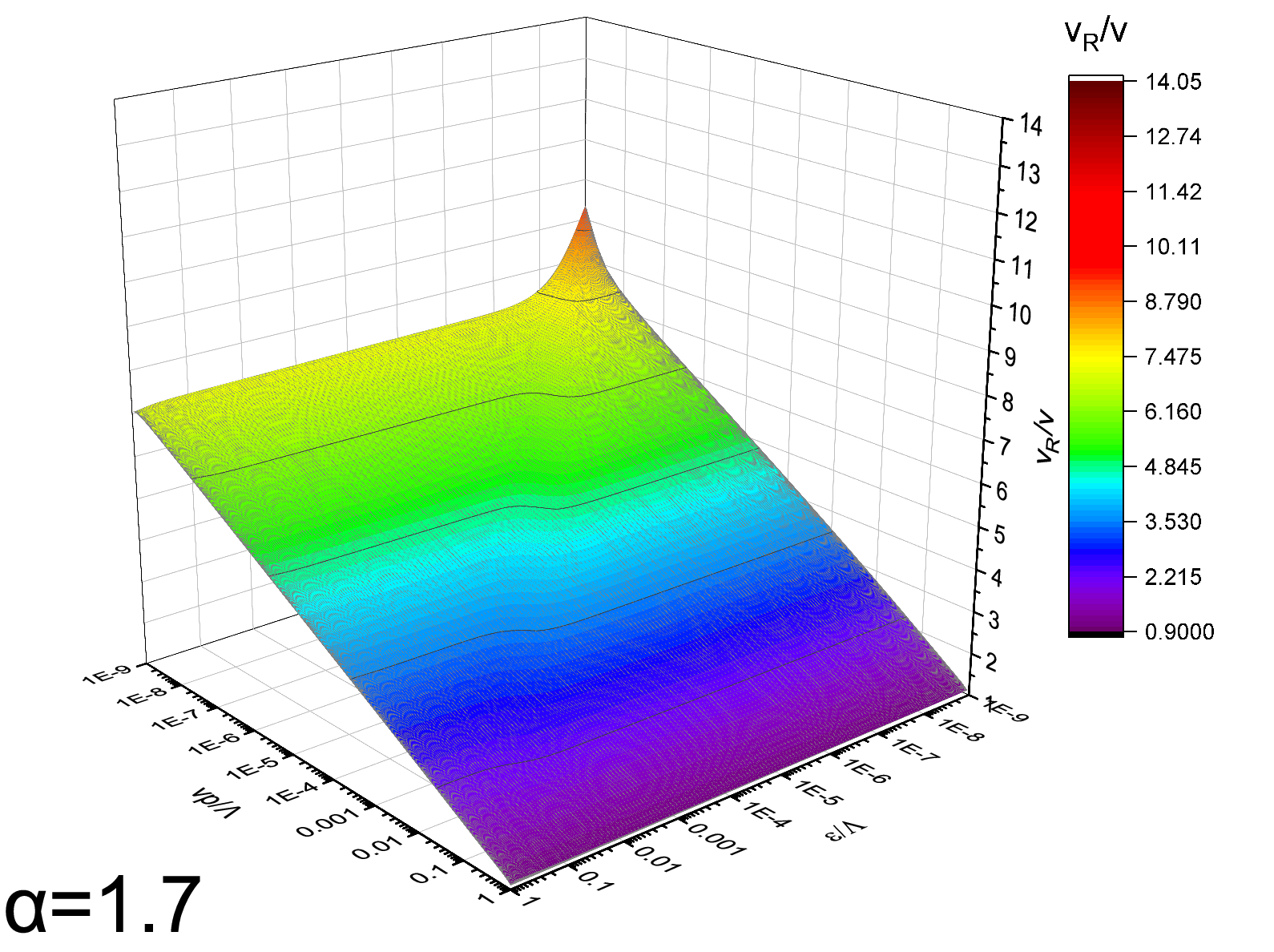}
\includegraphics[width=2.6in]{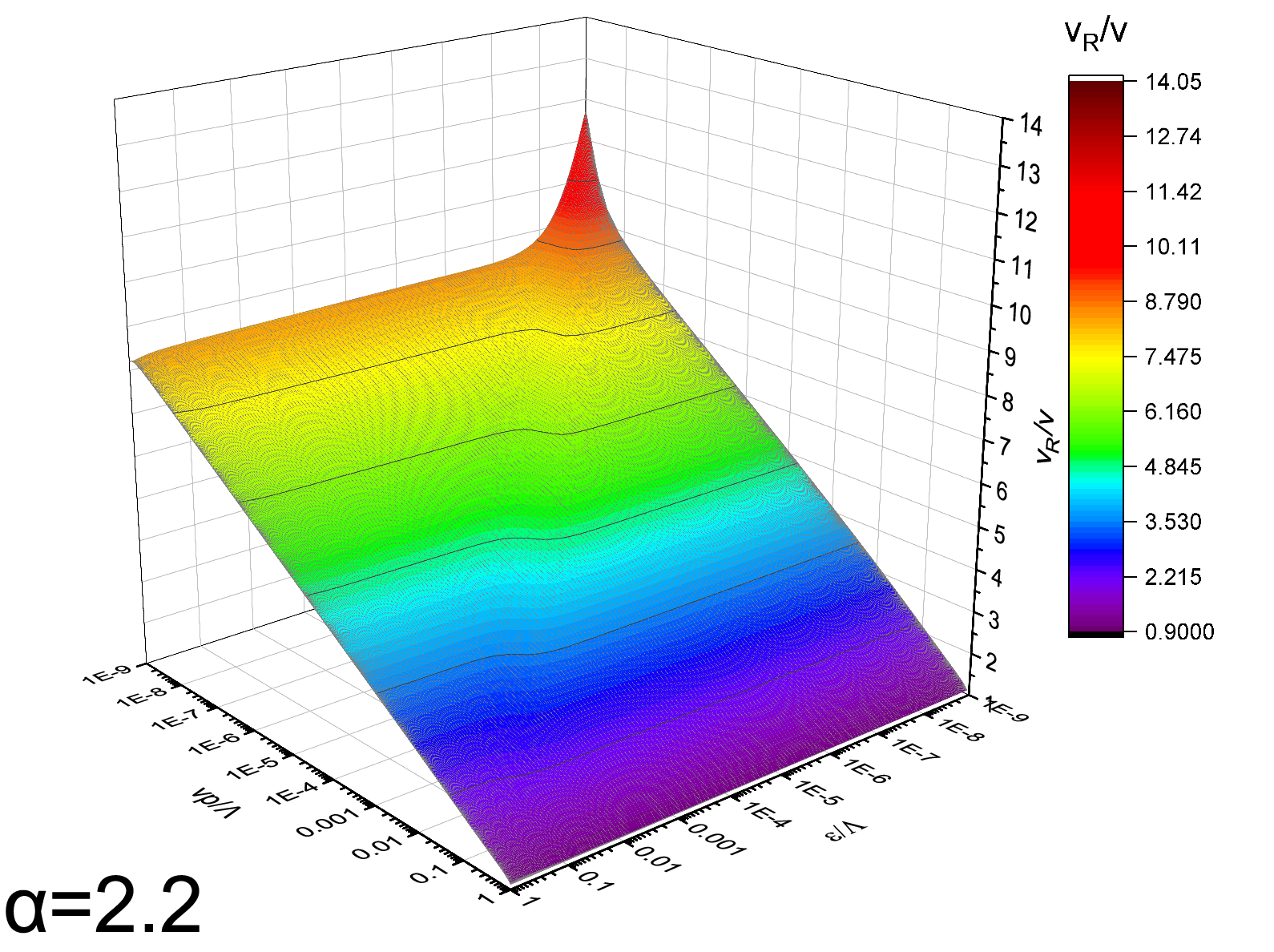}
\includegraphics[width=2.6in]{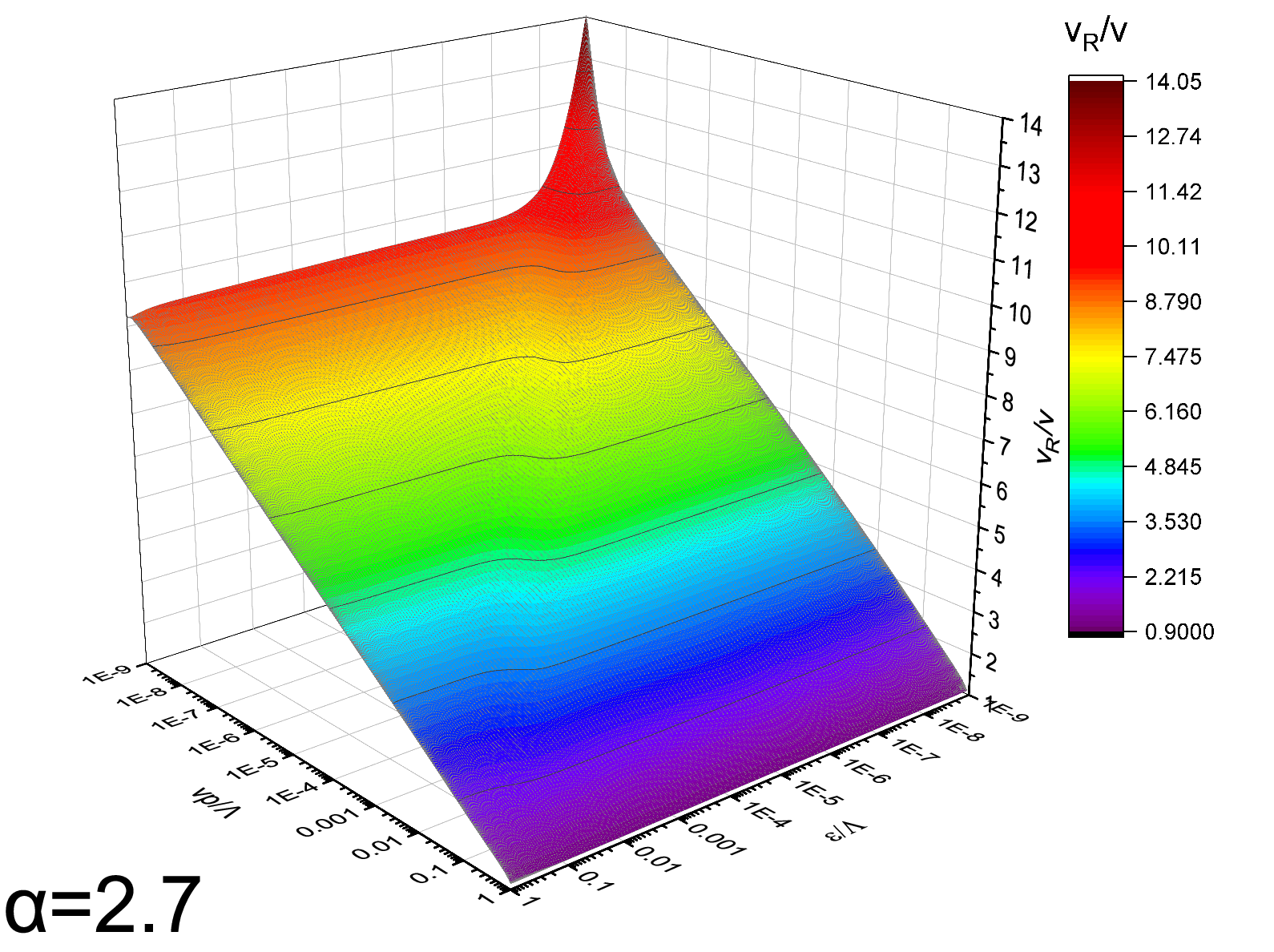}
\caption{The energy-momentum dependence of renormalized velocity
$v_{\mathrm{R}}(\varepsilon,\mathbf{p})$ obtained by using the full
fermion-boson vertex function for $\alpha=0.4$, $\alpha=0.8$,
$\alpha=1.3$, $\alpha=1.7$, $\alpha=2.2$, and $\alpha=2.7$. Over a
wide range of $\varepsilon$ and $\mathbf{p}$,
$v_{\mathrm{R}}(\varepsilon,\mathbf{p})$ exhibits a logarithmic
dependence on $|\mathbf{p}|$ but is nearly independent of
$\varepsilon$. Close to the IR cutoffs of $\varepsilon$ and
$\mathbf{p}$, $v_{\mathrm{R}}(\varepsilon,\mathbf{p})$ appears to
deviate from the normal behavior and rises abruptly. The origin of
such an abrupt rise is explained in the main text.}
\label{renorvelocity}
\end{figure}

We first analyze the behavior of fermion velocity renormalization.
For concreteness, here we take the UV cutoff \cite{Elias11} as
$\Lambda = 2.0$ eV. It is important to emphasize that the solutions
are independent of the value of $\Lambda$. Here, we choose six
different values of $\alpha$: $\alpha=0.4$ (graphene on BN
substrate), $\alpha=0.8$ (graphene on SiO$_{2}$ substrate),
$\alpha=1.3$, $\alpha=1.7$, $\alpha=2.2$ (suspended graphene), and
$\alpha=2.7$. After solving the most generic equations given by
Eqs.~(\ref{Eq:eqa0}-\ref{Eq:eqm}) without making any approximation,
we extract the full energy-momentum dependence of the renormalized
velocity
\begin{eqnarray}
\frac{v_{\mathrm{R}}(p)}{v} = \frac{A_{1}(p)}{A_{0}(p)}.
\end{eqnarray}
from the numerical solutions of $A_{0}(p)$ and $A_{1}(p)$ and show
the results in Fig.~\ref{renorvelocity}. $m(p)$ has only a zero
solution. To the best of our knowledge, the accurate energy-momentum
dependence of $v_{\mathrm{R}}(p)$ has never been obtained
previously. Here it is convenient to introduce the symbol
$\varepsilon$ to denote $-ip_{0}$. At a fixed $\varepsilon$,
$v_{\mathrm{R}}(\mathbf{p})$ exhibits a logarithmic dependence on
$|\mathbf{p}|$ within a wide range of $|\mathbf{p}|$. These results
are qualitatively well consistent with experimental observations of
renormalized velocity \cite{Elias11, Lanzara11, Chae12}. It seems
incredible that the function $v_{\mathrm{R}}(\mathbf{p})$ obtained
by solving the exact DS equation of $G(p)$ displays the same
logarithmic behavior obtained in first-order ($O(\alpha)$)
perturbative calculations. This perfectly explains why existing
experimental data fit well with the $O(\alpha)$ result in graphene
materials that actually have a relatively large $\alpha$ (comparing
to $\alpha=1/137$ in QED$_4$).

\begin{figure}[htbp]
\centering
\includegraphics[width=2.48in]{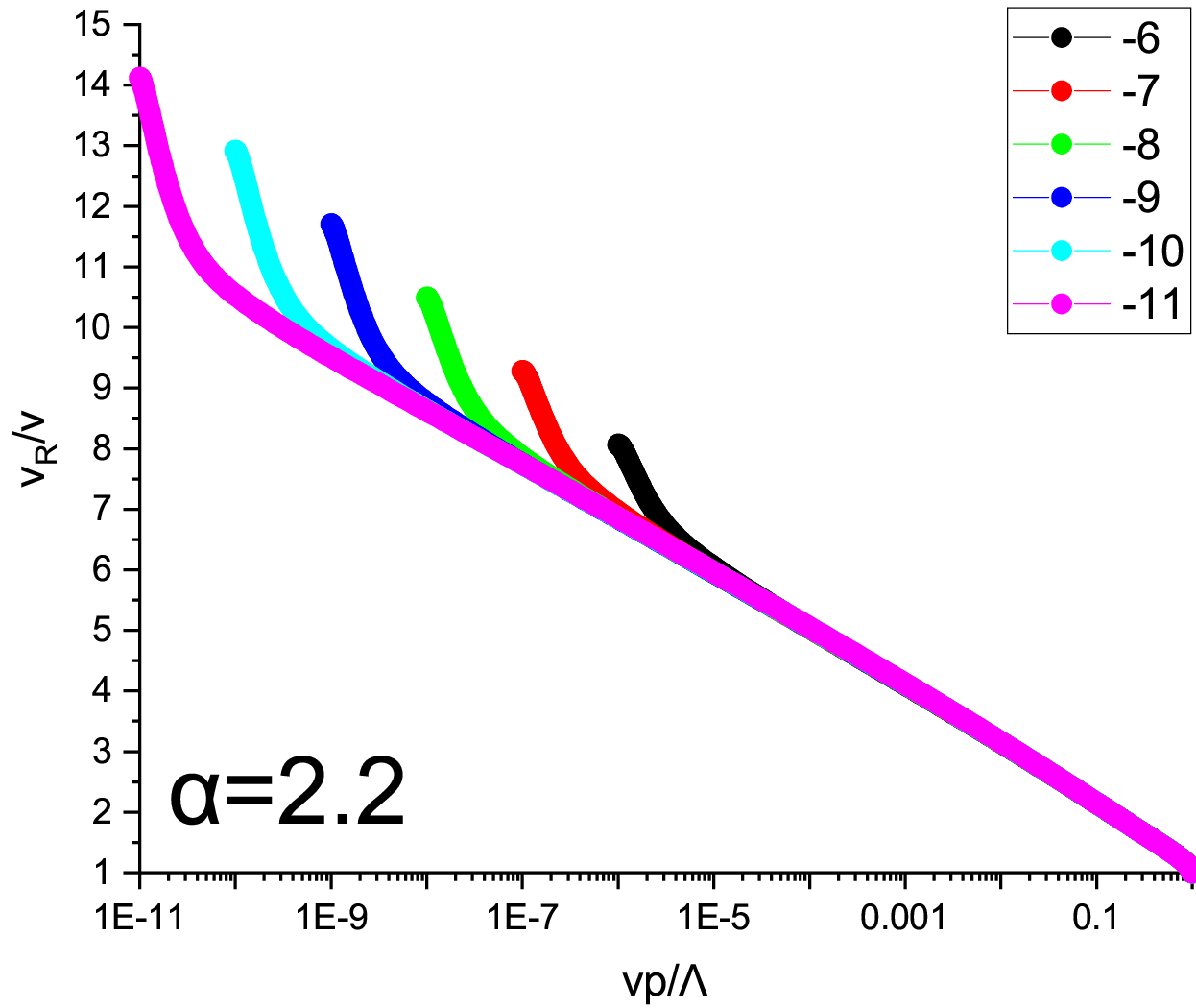}
\includegraphics[width=2.48in]{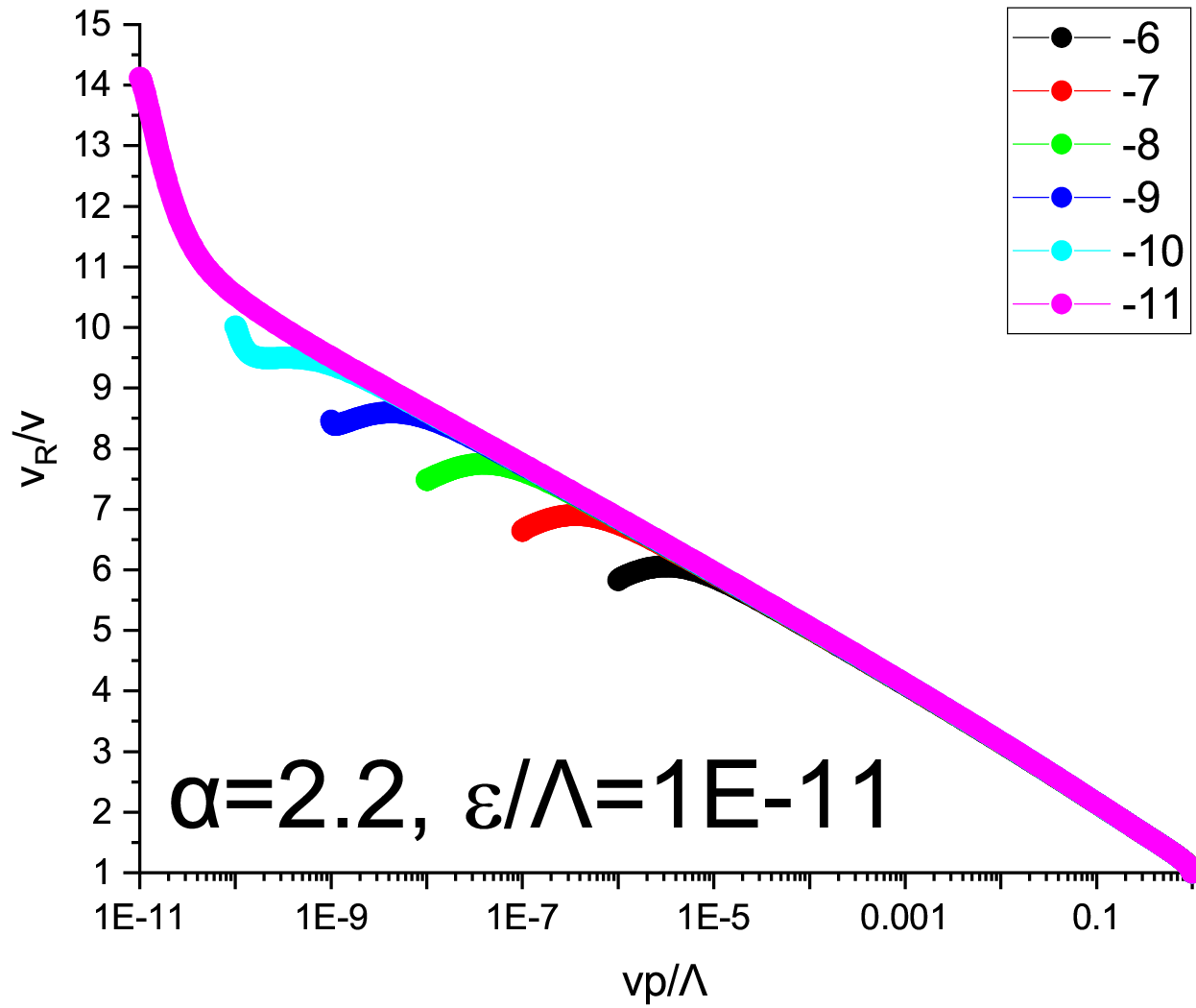}
\caption{Renormalized velocity obtained by using different IR
cutoffs at $\alpha=2.2$. Left panel: The IR cutoff of $\varepsilon$
is equal to that of $v|\mathbf{p}|$. Here, $\varepsilon$ is assumed
to take the value of its IR cutoff. Six different IR cutoffs
(relative to UV cutoff) are considered: $10^{-6}$, $10^{-7}$,
$10^{-8}$, $10^{-9}$, $10^{-10}$, and $10^{-11}$. The logarithmic
$|\mathbf{p}|$-dependence of $v_{\mathrm{R}}(\mathbf{p})$ extends
for several orders of magnitude of scaled momentum. Close to IR
cutoffs, $v_{\mathrm{R}}(\mathbf{p})$ seems to deviate from the
standard logarithmic behavior. However, such a seeming deviation is an
artifact and the logarithmic behavior is always extended to lower
energy and momentum region as IR cutoff is decreasing. Right panel: the
energy $\varepsilon$ is fixed at $\varepsilon/\Lambda = 10^{-11}$,
which also sets its IR cutoff, and the IR cutoff of $v|\mathbf{p}|$
takes six different values. The logarithmic behavior continues going
leftwards with lowering IR cutoff of $v|\mathbf{p}|$.}
\label{alpha22cutoffs}
\end{figure}

According to Fig.~\ref{renorvelocity}, it turns out that
$v_{\mathrm{R}}(\varepsilon,\mathbf{p})$ deviates from logarithmic
$|\mathbf{p}|$-dependence and $\varepsilon$-independence in the
region of small $\varepsilon$ and small $|\mathbf{p}|$ and appears
to be considerably increased as $\varepsilon$ and $|\mathbf{p}|$
decrease. We emphasize that such an abrupt deviation is unphysical
and stems from the infrared (IR) cutoffs that inevitably exist in
practical numerical calculations. This can be understood as follows.
In solid state physics, the metallic state is usually described by
the jellium model, which assumes that the positive charges are
uniformly distributed in space so as to maintain the global
neutrality of the system. Aside from the free part (kinetic term)
$H_{0}$, the total Hamiltonian contains three interaction terms:
$H_{C}$ for Coulomb interaction between electrons, $H_{B}$ for the
electrostatic energy of the uniform positive background, and
$H_{EB}$ for the interaction energy between the electrons and the
background. The term $H_{C}$, which sums over all the possible
values of transferred momentum $\mathbf{q}$, is further divided into
two parts: $H_{C}(\mathbf{q}) =
H_{C}(\mathbf{q}=0)+\sum_{\mathbf{q}\neq 0}H_{C}(\mathbf{q})$. It is
easy to check \cite{Mahan} that
$H_{C}(\mathbf{q}=0)+H_{B}+H_{EB}=0$. As a result, one can omit all
the contributions from positive background and at the same time
remove the $\mathbf{q} = 0$ contribution from the effective
Lagrangian density. That means, $\mathbf{q}$ appearing in the boson
propagator $D_{0}(q)\equiv D_{0}(q_0,\mathbf{q})$ can be made
arbitrarily small but cannot be set to zero. In the process of doing
numerical calculations, it is always necessary to choose an IR
cutoff $\Lambda_{\mathrm{IR}}^{\mathbf{q}}$ for $\mathbf{q}$. The
contributions from the range of $|\mathbf{q}|\in
(0,\Lambda_{\mathrm{IR}}^{\mathbf{q}})$ are always neglected. Since
$D_{0}(\mathbf{q})$ is inversely proportional to $|\mathbf{q}|$,
smaller $|\mathbf{q}|$ gives rise to a larger contribution to the
fermion self-energy. This is a salient feature of long-range
interaction. On the one hand, it indicates that large-$|\mathbf{q}|$
processes are unimportant and ensures that the results are
independent of the specific value of UV cutoff. On the other hand,
it implies that the neglected contributions from the range
$(0,\Lambda_{\mathrm{IR}}^{\mathbf{q}})$ are indeed not small, which
explains why an abrupt deviation from the standard logarithmic
behavior emerges as $\varepsilon$ and $|\mathbf{p}|$ are close to
their IR cutoffs. We choose six different values of
$\Lambda_{\mathrm{IR}}$ for $|\mathbf{p}|$. We see from
Fig.~\ref{alpha22cutoffs} that decreasing the IR cutoffs of
$\varepsilon$ and $|\mathbf{p}|$ always extends the logarithmic
behavior into the region of lower momenta. If we fix the fermion
energy at $\epsilon/\Lambda = 10^{-11}$ and choose UV cutoff
$\Lambda = 2.0$ eV, the logarithmic velocity renormalization holds
over a wide momentum range $v|\mathbf{p}|\in [2.0\times
10^{-10}\mathrm{eV},2.0\mathrm{eV}]$, as shown in the right panel of
Fig.~\ref{alpha22cutoffs}. Of course we can further decrease the
value of $\Lambda_{\mathrm{IR}}$, which would extend the logarithmic
behavior into lower momenta.

The logarithmic velocity renormalization would be eventually altered
as $|\mathbf{p}|$ becomes very small. This is because the
renormalized velocity $v_{\mathrm{R}}$ cannot be greater than the
speed of light $c$. When $v_{\mathrm{R}}$ is increased to a
magnitude close to $c$, the electromagnetic radiation effect becomes
significant and the non-relativistic model of Coulomb interaction
between Dirac fermions should be replaced with the fully
relativistic $(1+2)$-dimensional QED. As $v_{\mathrm{R}}\rightarrow
c$, the corrections to fermion velocity due to the longitudinal
(Coulomb-type) and transverse components of gauge field cancel each
other, leaving the fermion velocity unrenormalized \cite{Vafek07}.
But $v_{\mathrm{R}}\rightarrow c$ only at extremely low energies,
which can never be realized in graphene materials. Thus the
logarithmic velocity renormalization is robust at energy scales
accessible to experiments.

\begin{figure}[htbp]
\centering
\includegraphics[width=2.6in]{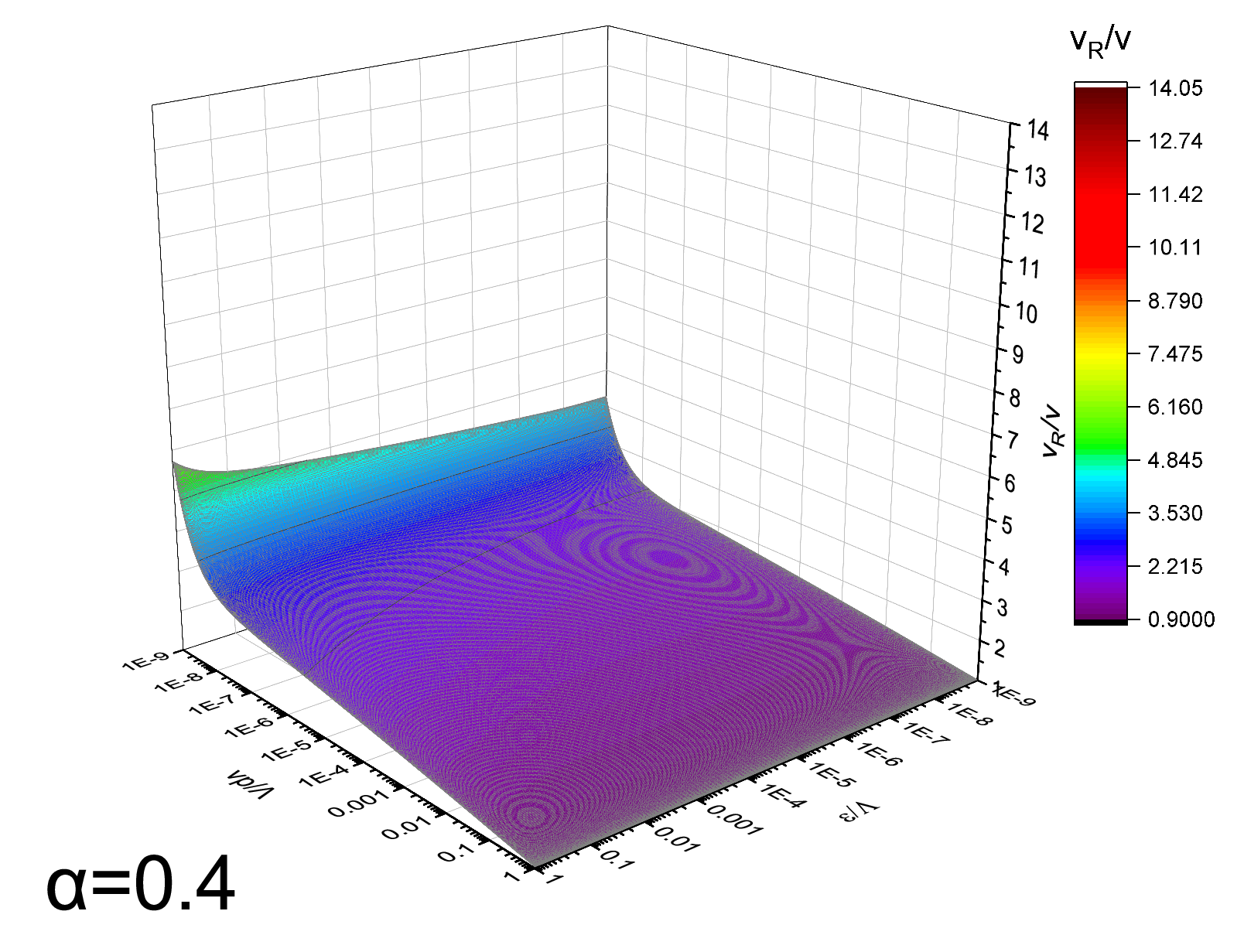}
\includegraphics[width=2.6in]{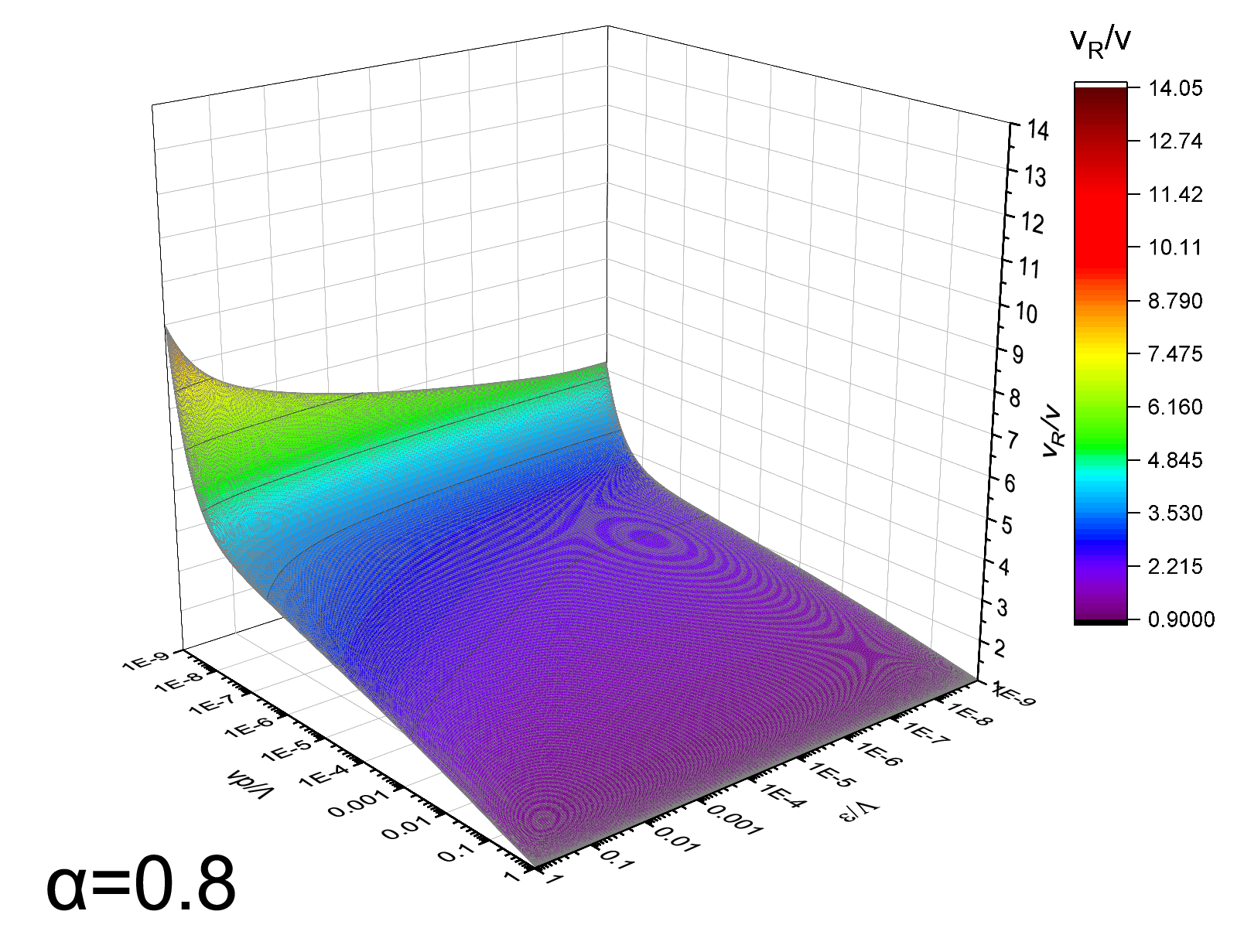}
\includegraphics[width=2.6in]{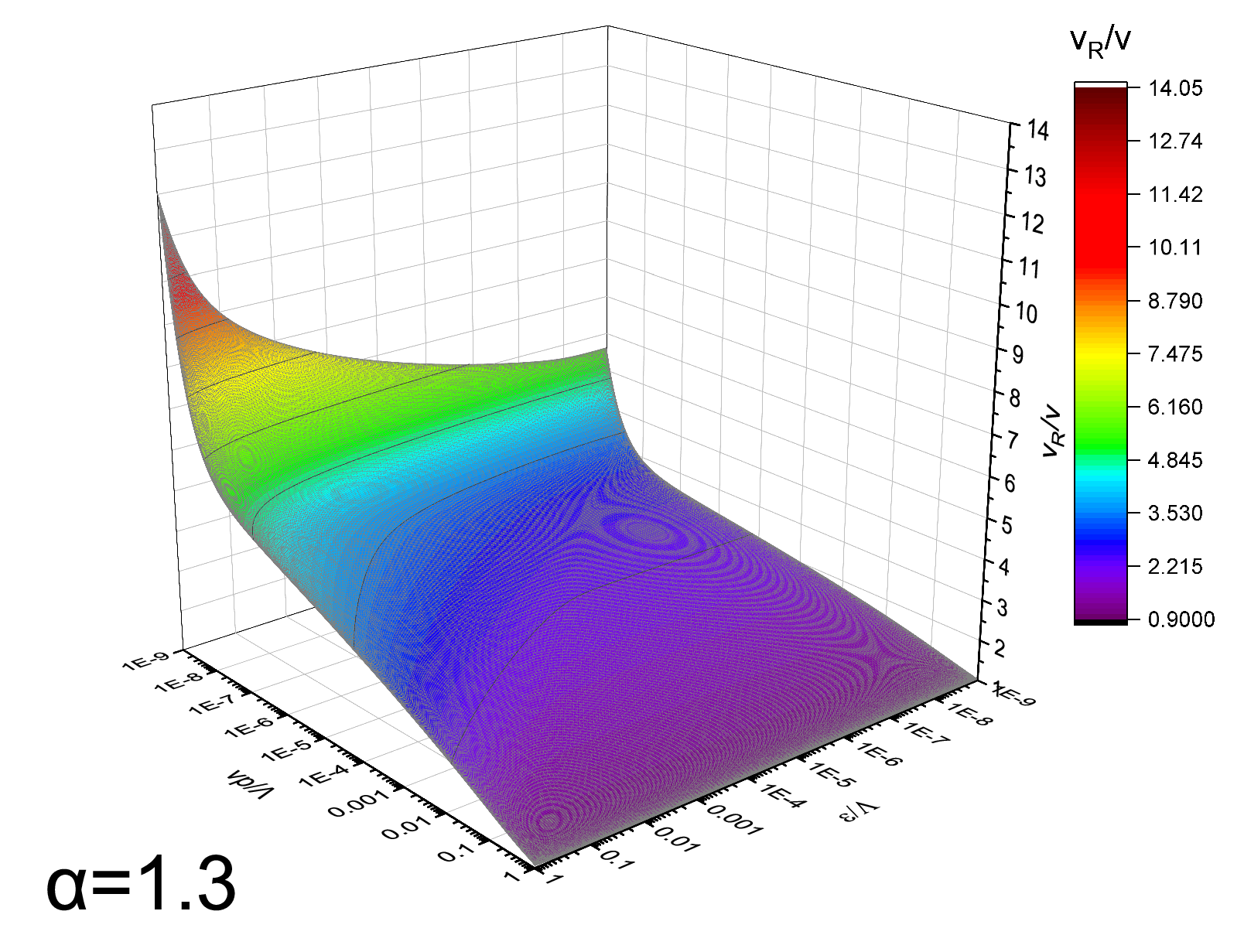}
\includegraphics[width=2.6in]{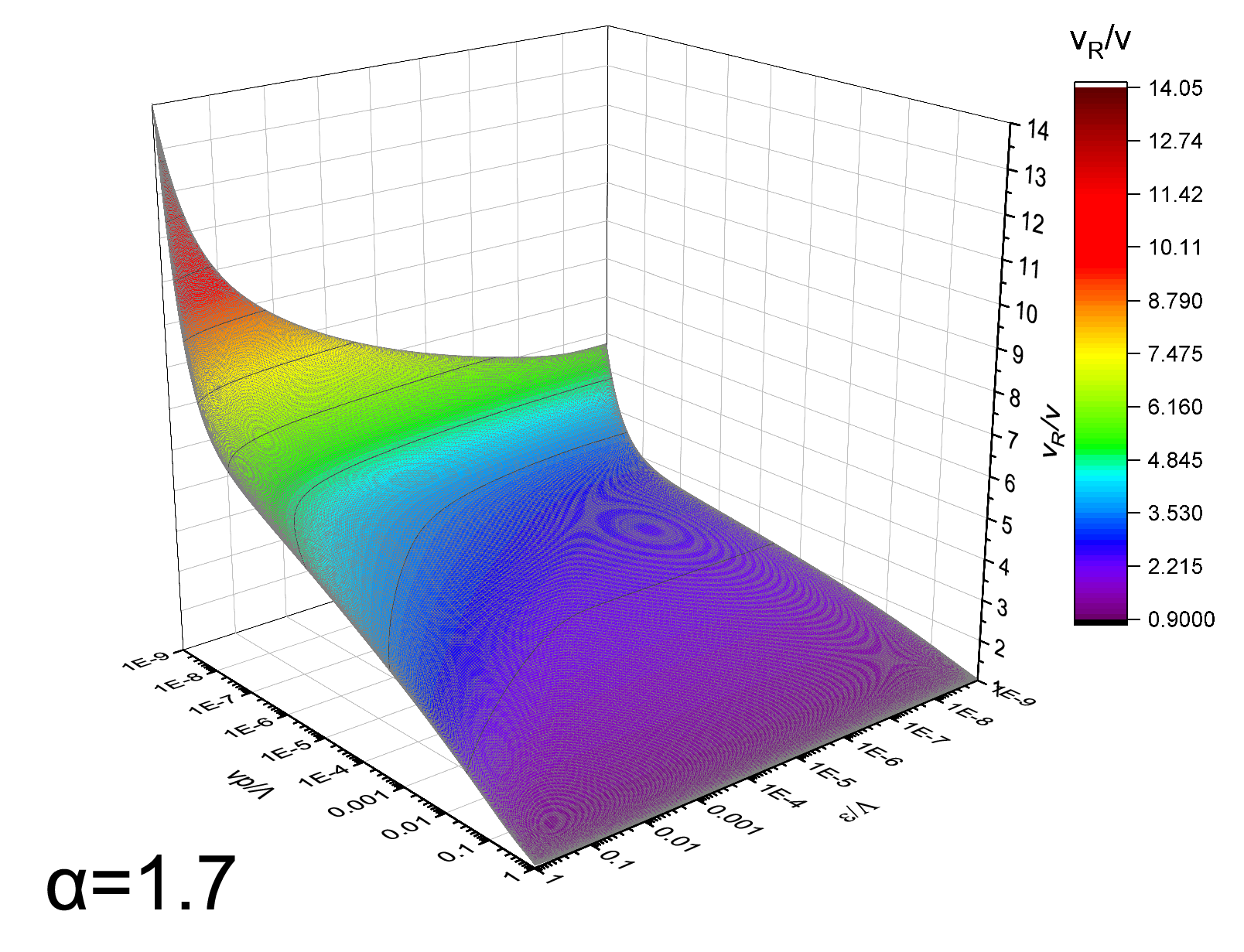}
\includegraphics[width=2.6in]{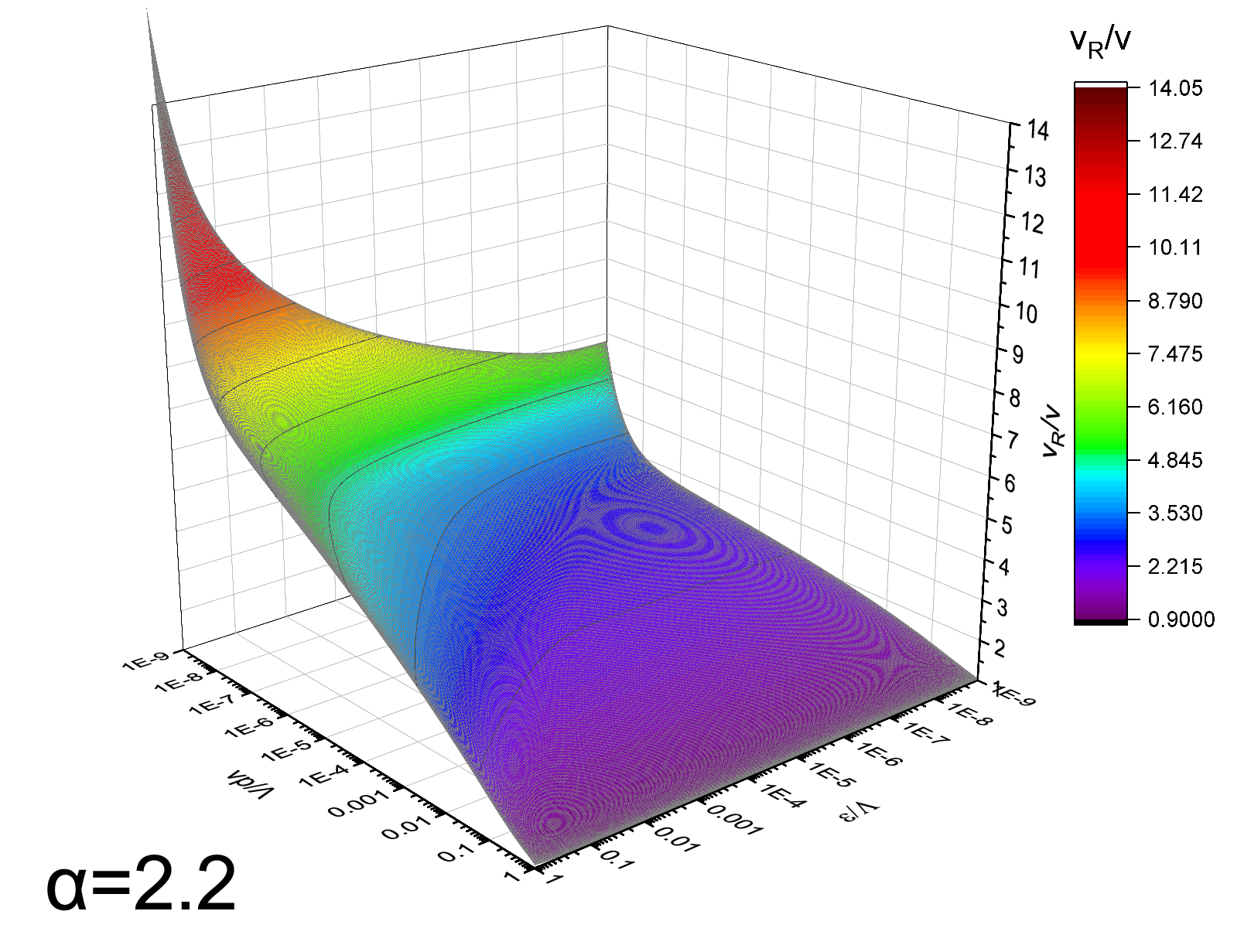}
\includegraphics[width=2.6In]{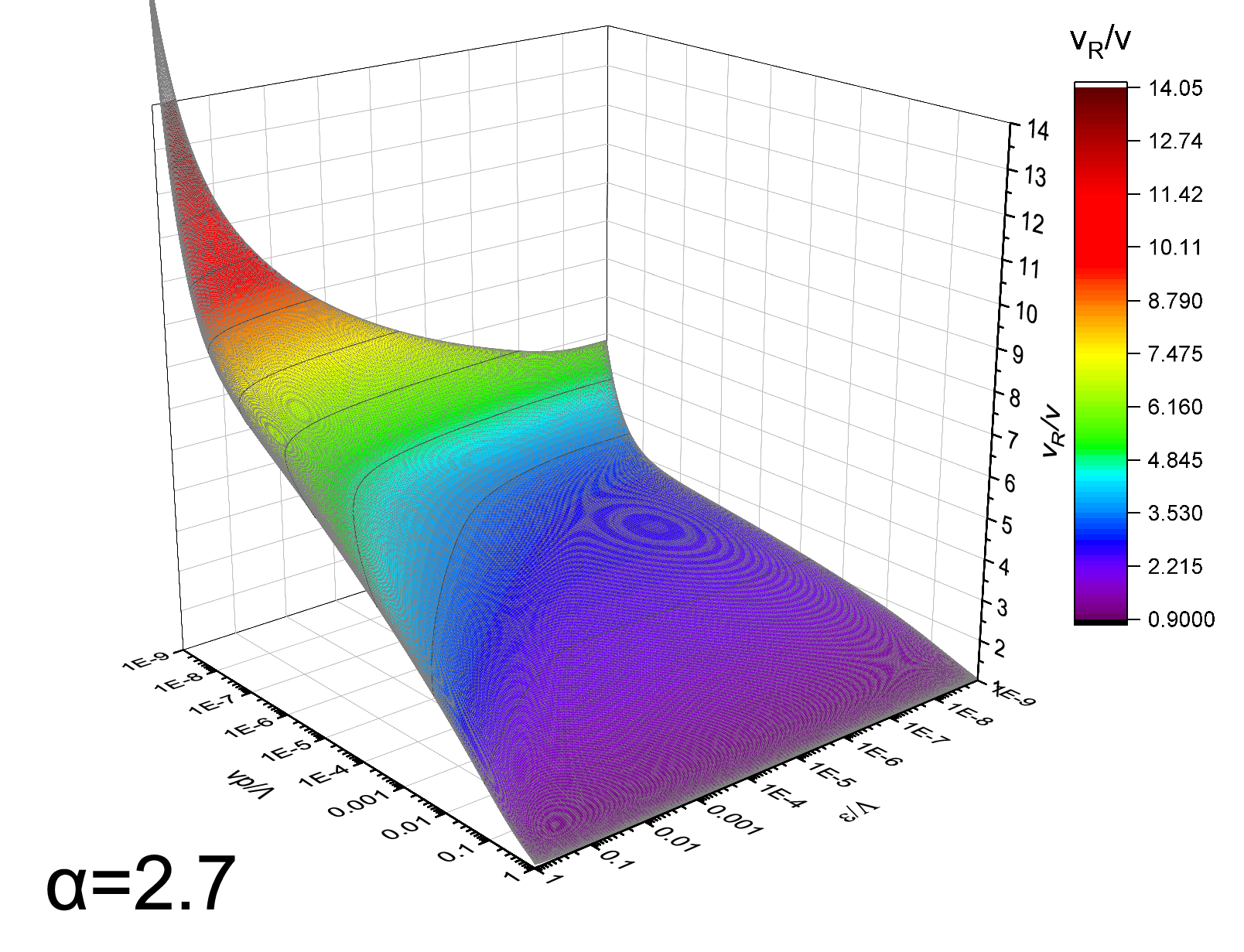}
\caption{The energy-momentum dependence of
$v_{\mathrm{R}}(\varepsilon,\mathbf{p})$ obtained by using the bare
vertex function and the RPA-form of boson propagator
$D_{\mathrm{RPA}}(k-p)$ for $\alpha=0.4$, $\alpha=0.8$,
$\alpha=1.3$, $\alpha=1.7$, $\alpha=2.2$, and $\alpha=2.7$.
$v_{\mathrm{R}}(\varepsilon,\mathbf{p})$ shows a strong dependence
on $\varepsilon$, which, however, is an artifact of incorrect
approximation.} \label{barevertexv}
\end{figure}

Although the inclusion of exact vertex function leads to the same
logarithmic $\mathbf{p}$-dependence of $v_{\mathrm{R}}(\mathbf{p})$
as $O(\alpha)$ order calculations, it would be false to say that
vertex corrections are not important. To demonstrate the impact of
vertex corrections, we also have solved the equations of $A_{0}(p)$
and $A_{1}(p)$ by using the bare vertex, with results being
presented in Fig.~\ref{barevertexv}. Comparing
Fig.~\ref{barevertexv} to Fig.~\ref{renorvelocity}, we find that
$v_{\mathrm{R}}(\mathbf{p})$ exhibits a logarithmic
$\mathbf{p}$-dependence at a fixed $\varepsilon$ no matter whether
bare vertex or full vertex is utilized. However, the magnitude of
$v_{\mathrm{R}}(\varepsilon,\mathbf{p})$ at any given point
$(\varepsilon,|\mathbf{p}|)$ is significantly increased due to the
inclusion of vertex corrections. In addition, we observe from
Fig.~\ref{barevertexv} that $v_{\mathrm{R}}(\varepsilon,\mathbf{p})$
is nearly energy independent if the exact vertex function is
adopted. In contrast, ignoring the vertex corrections would lead to
an incorrect result that $v_{\mathrm{R}}(\varepsilon,\mathbf{p})$ is
strongly energy dependent. All these results point to conclusions
that the vertex corrections do play a vital role and should be
seriously taken into account.

Next we discuss the possibility of excitonic gap generation. To
elaborate how $\alpha_c$ is influenced by various ingredients, we
have solved the equations of $A_{0}(p)$, $A_{1}(p)$, and $m(p)$
under several different approximations. For instance, we found
$\alpha_{c} \approx 1.0$ if the bare vertex $\gamma^{0}$ and the
free boson propagator $D_{0}(q)$ are employed. If we use bare vertex
$\gamma^{0}$ but promote $D_{0}(q)$ to RPA propagator
$D_{\mathrm{RPA}}(q)$, then $\alpha_c \approx 3.9$. If we use
$D_{\mathrm{RPA}}(q)$ and the leading term of the so-called
Ball-Chiu \emph{Ansatz} of vertex function (see \cite{WangLiu12,
Carrington16} for an explanation), we found $\alpha_c \approx 2.9$.
Apparently, the value of $\alpha_c$ is very sensitive to the chosen
approximation. In order to eliminate the unpleasant ambiguity in
results of $\alpha_c$, it is important to go beyond all
approximations and adopt the exact vertex function derived from
coupled WTIs. We have solved the most generic equations
(\ref{Eq:eqa0})-(\ref{Eq:eqm}) and found that no excitonic gap is
generated for $\alpha < 5$. An immediate indication is that the
semimetallic ground state of graphene is surprisingly robust against
Coulomb interaction.

Resistivity measurements \cite{Elias11, Mayorov12} have been
performed to detect the possible existence of excitonic insulating
transition in clean graphene. No sign of insulating state was found
\cite{Elias11, Mayorov12} down to roughly $1$ K. Indeed, thus far
there is no experimental signature of the excitonic-type pairing
instability in graphene. Our theoretical results are consistent with
the experimental situation.

When $\alpha > 5$, anomalous behaviors emerge. While the two
functions $A_{0}(p)$ and $A_{1}(p)$ exhibit regular behaviors
(without singularities) and lead to logarithmic velocity
renormalization for $\alpha < 5$, they no longer have stable
solutions once $\alpha$ exceeds $5$. It turns out that the system
undergoes an instability as $\alpha$ is increased across $5$. But
the nature of such an instability remains elusive. The transition
into an excitonic insulator can be directly precluded since the
equation of excitonic gap always has a vanishing solution (i.e.,
$m=0$) for all values of $\alpha$. Further investigations are called
for to uncover the nature of such an instability.

If two-component spinor and $2\times 2$ gamma matrices are utilized
to describe Dirac fermions, the integral equations of $A_{0}(p)$ and
$A_{1}(p)$ would still be given by Eqs.~(\ref{Eq:eqa0}) and (\ref{Eq:eqa1}). All the results about the velocity renormalization would not be changed. The only difference is that chiral symmetry cannot be explicitly defined.

As shown in Ref.~\cite{Liu19}, one can make proper use of the
solutions of $A_{0}(p)$ and $A_{1}(p)$ to explore the properties of
the boson. Substituting the full fermion propagator $G(p)$ and the
full vertex function $\Gamma_{\mathrm{int}}(k,p) =
D_{0}(q)\Upsilon_{\gamma^{0}}(k,p) D^{-1}(q)$ into the DS equation
of boson propagator $D(q)$, we find
\begin{eqnarray}
D^{-1}(q) = D^{-1}_{0}(q) - i N D_{0}(q)D^{-1}(q)
\int\frac{d^{3}k}{(2\pi)^{3}} {\mathrm{Tr}}\left[\gamma^{0}G(k+q)
\Upsilon_{\gamma^{0}}(k,p)G(k)\right],
\end{eqnarray}
which can be further written as
\begin{eqnarray}
D(q) = D_{0}(q) + i N D_{0}^{2}(q)\int\frac{d^{3}k}{(2\pi)^{3}}
{\mathrm{Tr}}\left[\gamma^{0}G(k+q)\Upsilon_{\gamma^{0}}(k,p)
G(k)\right].
\end{eqnarray}
Then the full polarization function $\Pi(q)$ can be calculated from
$D(q)$, based on the relation
\begin{eqnarray}
\Pi(q) = D_{0}^{-1}(q) - D^{-1}(q).
\end{eqnarray}
This $\Pi(q)$ is exact and can be used to investigate such effects
as plasmon and Friedel oscillation, which is out of the scope of this paper.

In this paper we consider only undoped graphene. However, the
graphene samples prepared in laboratory are always doped. Thus the
Fermi level is not exactly located at the neutral Dirac point. It
was found \cite{Elias11, Lanzara11, Chae12} that the renormalized
velocity displays a logarithmic dependence on carrier density. As
elaborated by Barnes \emph{et al.} \cite{Barnes14}, although the
density actually becomes unimportant as the temperature scale
$k_{B}T$ is greater than the Fermi energy $E_{F}$, the results
obtained at Dirac point cannot be directly used to account for the
density dependence of renormalized velocity. To explain the observed
density dependence at a quantitative level, it is necessary to
extend the analysis of undoped graphene to the case of doped
graphene \cite{DasSarma13}. The impact of finite carrier density can
be taken into account by adding a finite chemical potential $\mu$ to
the free fermion Hamiltonian via the replacement
\begin{eqnarray}
H_{f} = \sum_{\sigma=1}^{N}\bar{\psi}_{\sigma}\gamma\cdot
\partial \psi_{\sigma} \rightarrow H_{f}-\mu
\sum_{\sigma=1}^{N}\bar{\psi}_{\sigma}\gamma^{0}\psi_{\sigma}.
\end{eqnarray}
Then the free fermion propagator becomes
\begin{eqnarray}
G_{0}(p,\mu) \equiv G_{0}(p_{0},\mathbf{p},\mu) =
\frac{1}{\gamma^{0}\left(p_{0}-\mu\right) - v\mathbf{\gamma}\cdot
\mathbf{p}}.
\end{eqnarray}
The generalized WTIs and the equations of $A_{0}(p,\mu)$,
$A_{1}(p,\mu)$, and $m(p,\mu)$ can be similarly derived and solved,
following the calculational procedure developed in the case of
$\mu=0$, which will allow for a more quantitative comparison between
field-theoretic results and experiments. This issue is out of the
scope of this paper and will be addressed systematically in a
forthcoming work.

If graphene is made anisotropic, the free and full fermion
propagators would take the forms
\begin{eqnarray}
G_{0}(p) &=& \frac{1}{\gamma^{0}p_{0} - v_1 \gamma^{1}p_{1} - v_2
\gamma^{2}p_{2}}, \\
G(p) &=& \frac{1}{A_{0}(p)\gamma^{0}p_{0} - A_{1}(p)\gamma^{1}p_{1}
- A_{2}(p) \gamma^{2}p_{2} + m(p)}.
\end{eqnarray}
The interaction effects are now embodied in the three functions
$A_{0,1,2}(p)$ and $m(p)$. The renormalization of velocities $v_{1}$
and $v_{2}$ can be analyzed based on $A_{1}(p)/A_{0}(p)$ and
$A_{2}(p)/A_{0}(p)$, respectively.

The same calculational procedure can be applied to study the
fermion-phonon coupling in graphene by replacing the bare Coulomb
interaction function $D_{0}(q) = \frac{2\pi
e^{2}}{v\epsilon|\mathbf{q}|}$ with the free phonon propagator
$D_{0}(q) = -\frac{\Omega_{\mathbf{q}}}{q_{0}^{2} -
\Omega_{\mathbf{q}}^{2}}$. Application of the approach to
$(1+3)$-dimensional Dirac semimetal is straightforward. In this
case, the current vertex function should be computed based on the
expressions shown in Sec.~\ref{Sec:gamma013}.

\section{Summary and Discussion \label{Sec:Summary}}

In this paper we have developed a powerful non-perturbative DS
equation approach to study the strong coupling of massless Dirac
fermions to a scalar boson. The full vertex function of
fermion-boson coupling is incorporated into the DS equation of full
fermion propagator by solving a number of coupled WTIs that are
derived rigorously from several symmetric and asymmetric global U(1)
transformations. Based on this result, we prove that the DS equation
of full fermion propagator is entirely self-closed and can be
numerically solved. After solving this DS equation, the fermion
damping, the fermion velocity renormalization, and the possible
excitonic pairing can be investigated in a self-consistent way. In
using our approach, there is no need to expand physical quantities
into powers of small parameter. All the interaction-induced effects
on Dirac fermions are extracted from the solutions of exact DS
equation(s). Therefore, the results are reliable no matter whether
the fermion-boson coupling is weak or strong.

We have applied our approach to revisit the strong Coulomb
interaction in undoped graphene and solved the exact self-consistent
integral equations of wave function renormalizations $A_{0,1}(p)$
and excitonic gap $m(p)$. Our numerical results indicate that the
renormalized fermion velocity displays a logarithmic momentum
dependence over a wide range of momentum at a fixed energy, and that
the Coulomb interaction cannot open an excitonic gap. These results
are qualitatively in agreement with experiments. More in-depth
theoretical analysis is required to carry out a more quantitative
explanation of relevant experiments. Potential directions of future
research include analyzing the carrier-density dependence of renormalized fermion velocity and computing a number of observable quantities, such as specific heat and optical conductivity.

Our approach is applicable to long-range Coulomb interaction and
fermion-phonon interaction in both $(1+2)$ and $(1+3)$ dimensions.
But, the approach is no longer exact if the boson action has a
self-coupling term, such as $\phi^{4}$. We emphasize that the
coupled WTIs derived in Sec.~\ref{Sec:WTIs} and the current vertex
functions obtained in Secs.~\ref{Sec:unitymatrix} and \ref{Sec:gamma0} are always valid, irrespective of whether there is a self-interaction of scalar boson. This is because the WTIs originate from the variation of the action under infinitesimal
transformations of the fermion field. The real difficulty brought by
the boson self-interaction is that the identity given by
Eq.~(\ref{Eq:gammaD0Dgamma}) would have a complicated additional
term. In order to adopt our approach to investigate the fertile
quantum critical phenomena of Dirac fermion systems \cite{Kim08,
Wangmit17, Pan18, Liunematic12, Wang19, Xiao19, Lang19, LeeSS07,
Grover14, Jian15, Liu19npj}, we need to find a controllable method
to either exactly or approximately treat such an additional term.
This problem will be studied in a subsequent project.

We believe that the DS equation approach can also be applied to
study the superconducting instability of Dirac fermion systems,
mediated by phonons or other bosonic modes, and the interplay
between superconductivity and CDW. The Nambu spinor of Dirac
fermions usually has eight components, thus the structure of WTIs
would be very complicated. One might have to solve eight or even
sixteen coupled WTIs to obtain one specific current vertex function.

\section*{ACKNOWLEDGEMENTS}

We thank Jing-Rong Wang, Ying-Hai Wu, and Hai-Xiao Xiao for helpful
discussions.

G.-Z.L. motivated and designed the project and wrote the manuscript.
X.-Y.P. carried out the analytical calculations. Z.-K.Y. developed the
numerical program and performed the numerical computations. X.-Y.P.,
Z.-K.Y., and G.-Z.L. analyzed and interpreted the results. X.L.
contributed to the generic analysis of functional integral.

\appendix

\section{Definitions of some matrices \label{App:gammamatrices}}

Here we present the conventions and define all the matrices used in
the paper.

The metric tensor in $(1+2)$ and $(1+3)$ dimensions are
\begin{eqnarray}
g_{\mu\nu} = \left( {\begin{array}{*{20}{c}}
1&0&0\\
0& -1&0 \\
0&0&-1
\end{array}}\right), \quad
g_{\mu\nu} = \left( {\begin{array}{*{20}{c}}
1&0&0&0\\
0& -1&0&0 \\
0&0&-1&0 \\
0&0&0&-1
\end{array}}\right).
\end{eqnarray}
Three- and four-vectors for coordinate and momentum are written as
$x^{\mu}=(x^{0},x^{i})=(x^{0},\mathbf{x})$ and
$p^{\mu}=(p^{0},p^{i})=(p^{0},\mathbf{p})$. The following relations
are frequently used:
\begin{eqnarray}
x_\mu = g_{\mu\nu}x^{\nu}, \quad p_\mu = g_{\mu\nu}p^{\nu}, \quad
\gamma_\mu=g_{\mu\nu}\gamma^{\nu}.
\end{eqnarray}
Standard Pauli matrices are $$\tau^1 = {\begin{pmatrix}
0&1\\
1&0
\end{pmatrix}}, \quad \tau^2 = {\begin{pmatrix}
0& - i\\
i&0
\end{pmatrix}}, \quad \tau^3 = {\begin{pmatrix}
1&0\\
0& - 1
\end{pmatrix}}.$$

In both $(1+2)$ and $(1+3)$ dimensions, we will use the following
five $4\times 4$ gamma matrices:
\begin{eqnarray}
&&\gamma^0 = \gamma_0 = {\begin{pmatrix}
\tau^3&0\\
0& - \tau^3
\end{pmatrix}},\quad \gamma^1 = -\gamma_1 = {\begin{pmatrix}
i\tau^2&0\\
0& -i\tau^2
\end{pmatrix}},\quad\gamma^2 = -\gamma_2 = {\begin{pmatrix}
-i\tau^1&0\\
0& i\tau^1
\end{pmatrix}},
\end{eqnarray}
and
\begin{eqnarray}
\gamma^3 = -\gamma_{3} = -i{\begin{pmatrix}
0& 1 \\
1 & 0
\end{pmatrix}}, \quad \gamma^5 \equiv i\gamma^{0123} =
i\gamma^{0}\gamma^{1} \gamma^{2}\gamma^{3} = i{\begin{pmatrix}
0& 1 \\
-1 & 0
\end{pmatrix}}.
\end{eqnarray}

To derive the coupled WTIs in Sec.~\ref{Sec:unitymatrix} and
Sec.~\ref{Sec:gamma0}, we need to construct several $4 \times 4$
matrices:
\begin{eqnarray}
&&\sigma^{01}=\frac{i}{2}[\gamma^0,\gamma^1]=i\gamma^0 \gamma^1 =
{\begin{pmatrix}
i\tau^1&0 \\
0& i\tau^1
\end{pmatrix}},\\
&&\sigma^{02}=\frac{i}{2}[\gamma^0,\gamma^2]=i\gamma^0 \gamma^2 =
{\begin{pmatrix}
i\tau^2&0\\
0& i\tau^2
\end{pmatrix}},\\
&&\sigma^{12} = \frac{i}{2}[\gamma^1,\gamma^2] = i\gamma^1 \gamma^2
= {\begin{pmatrix}
\tau^3&0 \\
0&\tau^3
\end{pmatrix}},\\
&&\{{\sigma^{01},\gamma^2}\} = 2\tau^3\otimes I= 2{\begin{pmatrix}
I&0\\
0& -I
\end{pmatrix}},\\
&&\{{\sigma^{02},\gamma^1} \}=-2 \tau^3\otimes I=2{\begin{pmatrix}
-I&0 \\
0&I \\
\end{pmatrix}},\\
&&\{{\sigma^{12},\gamma^0} \} =2 \tau^3\otimes I= 2{\begin{pmatrix}
I&0\\
0&{-I}
\end{pmatrix}}.
\end{eqnarray}
In $(1+3)$ dimensions, we also need three additional matrices:
\begin{eqnarray}
\sigma^{03} &=& \frac{i}{2}[\gamma^0,\gamma^3]=i\gamma^0\gamma^3
=\begin{pmatrix}
0 & \tau^3 \\
-\tau^3 & 0
\end{pmatrix}, \\
\sigma^{13} &=& \frac{i}{2}[\gamma^1,\gamma^3]=i\gamma^1\gamma^3
=\begin{pmatrix}
0 & i\tau^2 \\
-i\tau^2 & 0
\end{pmatrix} , \\
\sigma^{23} &=& \frac{i}{2}[\gamma^2,\gamma^3]=i\gamma^2\gamma^3 =
\begin{pmatrix}
0 & -i\tau^1 \\
i\tau^1 & 0
\end{pmatrix}.
\end{eqnarray}

As mentioned in Sec.~\ref{Sec:Model}, one can alternatively use
$2\times 2$ matrices to describe two-component spinor in $(1+2)$
dimensions. This representation would lead to the same results as
four-component spinor representation, if we are not intended to
consider chiral symmetry (breaking). Although we adopt
four-component spinor throughout the main text of the paper, here
for completeness we also show how our approach works if
two-component spinor is adopted. One can choose
\begin{eqnarray}
\gamma^0=\tau^3, \quad \gamma^1=i\tau^1, \quad \gamma^2 = i\tau^2.
\end{eqnarray}
These three matrices also satisfy $\{\gamma^\mu,\gamma^\nu \} =
2g^{\mu\nu}$. The following three matrices are needed:
\begin{eqnarray}
\sigma^{01} = -i\gamma^0 \gamma^1 = -i\tau^2, \quad \sigma^{02} =
-i\gamma^0 \gamma^2 = -i\tau^1, \quad \sigma^{12} = i\gamma^1
\gamma^2 = \tau^3.
\end{eqnarray}
The corresponding WTIs can be readily obtained by substituting the
above expressions of $\gamma^0$, $\gamma^1$, $\gamma^2$,
$\sigma^{01}$, $\sigma^{02}$, and $\sigma^{12}$ into the general
expressions of Eqs.~(\ref{Eq:WTIS1}) and (\ref{Eq:WTIS2}).

\section{Derivation of Dyson-Schwinger equations \label{App:DSEs}}

In this Appendix we derive the DS equations of fermion and boson
propagators within the functional-integral formalism of quantum
field theory. Similar derivations have previously be presented in
Ref.~\cite{Liu19}. However, we feel it helpful to provide some
crucial calculational details here.

The starting point is the partition function
\begin{eqnarray}
\mathcal{Z}[J,{\bar \eta},\eta] &=& \int \mathcal{D}\phi
\mathcal{D}\psi\mathcal{D}{\bar \psi} e^{i\int dx [{\mathcal
L}+J\phi+{\bar \eta}\psi + {\bar \psi}\eta)} \nonumber \\
&=& e^{iW[J,{\bar\eta},\eta]}.
\end{eqnarray}
The Lagrange density is given by
\begin{eqnarray}
\mathcal{L} = \sum^N_{\sigma=1}\left[{\bar \psi}_{\sigma}(x)
i\gamma^{\mu}\partial_{\mu}\psi_{\sigma}(x) + g\phi(x){\bar
\psi}_{\sigma}(x)\gamma^m\psi_{\sigma}(x)\right] +
\frac{1}{2}\phi(x) \mathbb{D}\phi(x).
\end{eqnarray}
The average of an arbitrary operator $\mathcal{O}$ is defined as
\begin{eqnarray}
\langle \mathcal{O}(x)\rangle_J = \frac{[[O(x)]]_J}{[[1]]_J},
\end{eqnarray}
where $[[1]]_J$ is just the partition function $\mathcal{Z}$ and
\begin{eqnarray}
[[O(x)]]_J=\int \mathcal{D}\phi \mathcal{D}\psi \mathcal{D}{\bar
\psi} e^{i\int dx [\mathcal{L}+J\phi+ {\bar \eta}\psi + {\bar
\psi}\eta]}O(x).
\end{eqnarray}
Here we use one single subscript $J$ to stand for all the possible
external sources, i.e., $\langle \mathcal{O}\rangle_{J} \equiv
\langle \mathcal{O} \rangle_{J,{\bar \eta},\eta}$.

\subsection{Dyson-Schwinger equation of boson propagator}

Since $\delta \mathcal{Z} = 0$ under an arbitrary infinitesimal
variation $\delta \phi$, we have
\begin{eqnarray}
0 &=& \int \mathcal{D}\phi\mathcal{D}\psi\mathcal{D}{\bar\psi}
\left[\frac{\delta \mathcal{L}}{\delta\phi(x)}+J(x)\right] e^{i\int
dx [\mathcal{L} + J\phi + {\bar \eta}\psi + {\bar \psi} \eta]}
\nonumber \\
&=& \left[\frac{\delta \mathcal{L}}{\delta \phi(x)}
\left(\frac{\delta}{i\delta J},\frac{\partial}{i{\bar
\eta}_{\sigma}},-\frac{\delta}{i\delta \eta_{\sigma}}\right) +
J\right] \mathcal{Z}[J,{\bar \eta},\eta].
\end{eqnarray}
Since
\begin{eqnarray}
\frac{\delta \mathcal{L}}{\delta\phi(x)} = g\sum^N_{\sigma=1}
{\bar\psi}_{\sigma}(x) \gamma^m \psi_{\sigma}(x) +
\mathbb{D}\phi(x),
\end{eqnarray}
one can verify that
\begin{eqnarray}
J(x)\mathcal{Z} + \mathbb{D}\frac{\delta \mathcal{Z}}{i\delta J(x)}
+ g\sum^N_{\sigma=1}\frac{\delta}{-i\delta \eta_{\sigma}(x)}\gamma^m
\frac{\delta}{i\delta{\bar \eta}_{\sigma}(x)}\mathcal{Z} = 0.
\end{eqnarray}
Dividing this equation by $\mathcal{Z}$ yields
\begin{eqnarray}
J(x)+\mathbb{D}\frac{\delta W}{\delta J(x)}+\frac{g}{\mathcal{Z}}
\sum^N_{\sigma=1} \frac{\delta}{\delta \eta_{\sigma}(x)}\gamma^m
\frac{\delta}{\delta{\bar\eta}_\sigma(x)}e^{iW} = 0.
\end{eqnarray}
The last term of the l.h.s. of the above equation is
\begin{eqnarray}
\frac{g}{\mathcal{Z}}\frac{\delta}{\delta\eta_{\sigma}(x)}\gamma^m
\frac{\delta}{\delta{\bar \eta}_\sigma(x)}e^{iW} =- ig{\mathrm{Tr}}
[\gamma^m \frac{\delta^2 W}{\delta{\bar \eta}_\sigma(x)\delta
\eta_\sigma(y)} ]-g\frac{\delta W}{\delta \eta_{\sigma}(x)} \gamma^m
\frac{\delta W}{\delta{\bar\eta}_{\sigma}(x)}.
\end{eqnarray}
The second term of the r.h.s vanishes as the fields are set to be
zero.

To proceed, we define the following Legendre transformation
\begin{eqnarray}
\Xi(\phi,\psi,{\bar\psi}) = W(J,{\bar\eta},\eta) -
\sum^N_{\sigma=1}\int dx \left[J\phi + {\bar\psi}_{\sigma}
\eta_{\sigma}+{\bar\eta}_{\sigma}\psi_\sigma\right].
\label{Eq:Legendre}
\end{eqnarray}
It is known \cite{Itzykson} that the following identities hold
\begin{eqnarray}
&&\phi(x) = \frac{\delta W}{\delta J(x)}, \quad \psi_{\sigma}(x) =
\frac{\delta W}{\delta{\bar \eta}_{\sigma}(x)}, \quad {\bar
\psi}_{\sigma}(x)=-\frac{\delta W}{\delta \eta_{\sigma}(x)}
\nonumber
\\
&&J(x)=-\frac{\delta \Xi}{\delta\phi(x)},\quad \eta_{\sigma}(x) =
-\frac{\delta\Xi}{\delta{\bar \psi}_{\sigma}(x)}, \quad
{\bar\eta}_{\sigma}(x)=\frac{\delta \Xi}{\delta\psi(x)}.
\end{eqnarray}
The boson propagator and its inverse are defined as
\begin{eqnarray}
D(x,y)&=&-\frac{\delta^2 W}{\delta J(x)\delta J(y)}=-\frac{\delta
\phi (y)}{\delta J(x)} = -i\langle\phi(x)\phi(y)\rangle_c,\\
D^{-1}(x,y)&=&\frac{\delta^2\Xi}{\delta \phi(x)\delta\phi(y)} =
-\frac{\delta J(x)}{\delta\phi(y)}.
\end{eqnarray}
It is easy to check that
\begin{eqnarray}
\int dy D(x,y)D^{-1}(y,z) = \int dy\frac{-\delta^2 W}{\delta
J(x)\delta J(y)}\frac{\delta^2 \Xi}{\delta \phi (y)\delta \phi
(z)}=\int dy\frac{\delta \phi(x)}{\delta J(y)}\frac{\delta
J(y)}{\delta \phi(z)} = \delta(x-z).
\end{eqnarray}
Similarly, for each flavor $\sigma$ of the fermion propagator and
its inverse we have
\begin{eqnarray}
&&G_{\alpha \beta}(x,y)=\frac{\delta^2 W}{\delta{\bar
\eta}_{\alpha}(x)\delta \eta_{\beta }(y)} = -\frac{\delta
\psi_{\alpha}(x)}{\delta \eta_{\beta}(y)} = -\frac{\delta{\bar
\psi}_{\beta}(y)}{\delta{\bar\eta}_{\alpha}(x)} =
-i\langle\psi_{\alpha}(x){\bar \psi}_{\beta}(y)\rangle_c, \\
&& G_{\beta \rho}^{-1}(y,z)=-\frac{\delta^{2}\Xi}{\delta{\bar
\psi}_{\beta}(y)\delta \psi_{\rho}(z)} = -\frac{\delta
\eta_{\beta}(y)}{\delta\psi_{\rho}(z)} = -\frac{\delta{\bar
\eta}_{\rho}(z)}{\delta{\bar\psi}_{\beta}(y)}.
\end{eqnarray}
%where $\alpha,\beta,\rho$ stands for the short nation
%$(\sigma,\alpha),(\sigma,\beta),(\sigma,\rho)$. We have dropped the
%flavor index $\sigma$ for the fermion propagator since it is assumed
%to be flavor-independent.
Then they fulfill the relation
\begin{eqnarray}
\int G_{\alpha \beta}(x,y)G_{\beta \rho}^{-1}(y,z)dy = \delta
(x-z)\delta_{\alpha\rho}.
\end{eqnarray}

Eq.~(B8) can be re-written as
\begin{eqnarray}
J(x) = -\mathbb{D}\frac{\delta W}{\delta J(x)} +
ig\sum^N_{\sigma=1}{\mathrm{Tr}}\left[\gamma^m \frac{\delta^2
W}{\delta{\bar \eta}_\sigma(x)\delta \eta_\sigma(x)}\right],
\end{eqnarray}
Making the variation $\frac{\delta}{\delta J(y)}$ on both sides
of Eq.~(B18)
we obtain
\begin{eqnarray}
\delta(x-y) = \mathbb{D}D(x-y) + ig\sum^N_{\sigma=1}\mathrm{Tr}
\left[\gamma^m \frac{\delta^3 W}{\delta J(y)\delta{\bar
\eta}_\sigma(x)\delta \eta_\sigma(x)}\right].
\end{eqnarray}
Using the relation of Eq.~(\ref{Eq:WJbaretaeta}), now we can write
the DS equation of boson propagator in the form
\begin{eqnarray}
\delta(x-y) = \mathbb{D}D(x-y) - ig^2 N\int dx' dy' dz'
{\mathrm{Tr}}\left[\gamma^m D(y,x') G(x,y')
\Gamma_{\mathrm{int}}(y'-x',x'-z')G(z',x)\right], \nonumber \\
\end{eqnarray}
which in the momentum space becomes
\begin{eqnarray}
D^{-1}(q) = D_{0}^{-1}(q) - ig^2 N\int
\frac{dk}{(2\pi)^{(1+d)}}\mathrm{Tr}\left[\gamma^m
G(k+q)\Gamma_{\mathrm{int}}(k+q,k)G(k)\right].
\end{eqnarray}

\subsection{Dyson-Schwinger equation of fermion propagator}

The DS equation of fermion propagator can be similarly derived.

Since $\delta \mathcal{Z} = 0$ under an arbitrary infinitesimal
variation $\delta \psi$, we obtain an equation
\begin{eqnarray}
0 = \int \mathcal{D}\phi \mathcal{D}\psi \mathcal{D}{\bar\psi}
\left[\frac{\delta \mathcal{L}}{\delta{\bar
\psi}(x)}\left(\frac{\delta}{i\delta J},\frac{\delta}{i\delta{\bar
\eta}_{\sigma}},\frac{\delta}{-i\delta
\eta_\sigma}\right)+\eta_\sigma(x)\right]\mathcal{Z}(J,{\bar
\eta},\eta),
\end{eqnarray}
which implies that
\begin{eqnarray}
\eta_\sigma(x)\mathcal{Z} + i\gamma^{\mu}{\partial_{\mu}}\mathcal{Z}
\frac{\delta W}{\delta{\bar\eta}_{\sigma}(x)} +
g\frac{\delta}{i\delta J(x)}\gamma^m \left(\mathcal{Z} \frac{\delta
W}{\delta{\bar\eta}_{\sigma}(x)}\right) = 0.
\end{eqnarray}
Operating the functional derivative $\frac{\delta}{\delta
\eta_\sigma(y)}$ on both sides of the above equation and then
setting $\psi = {\bar\psi}=0$, one finds
\begin{eqnarray}
\delta(x-y)\mathcal{Z} + i\gamma^{\mu}{\partial_\mu}
\mathcal{Z}\frac{\delta^{2}W}{\delta\eta_\sigma(y)\delta{\bar
\eta}_\sigma(x)} + g\frac{\delta}{i\delta J(x)}\gamma^m \mathcal{Z}
\frac{\delta^2W}{\delta \eta_\sigma(y)\delta{\bar
\eta_\sigma}(x)}=0,
\end{eqnarray}
which in turn leads to for each flavor $\sigma$
\begin{eqnarray}
i \gamma^{\mu}{\partial_\mu}G(x,y) - ig \gamma^m \frac{\delta^3
W}{\delta J(x)\delta{\bar\eta}_\sigma(x) \delta \eta_\sigma(y)} =
\delta(x-y).
\end{eqnarray}
The second term of the l.h.s of above equation can be calculated
with the help of Eq.~(\ref{Eq:WJbaretaeta}). Fourier transformation
of the above equation yields the following equation
\begin{eqnarray}
\gamma^{\mu}p_{\mu}G(p) + ig^2\int \frac{dk}{(2\pi)^{(1+d)}}
\gamma^m G(k)D(k-p)\Gamma_{\mathrm{int}}(k,p)G(p)=1,
\end{eqnarray}
which can be turned into the DS equation of fermion propagator
\begin{eqnarray}
G^{-1}(p) = G^{-1}_{0}(p) + ig^{2}\int \frac{dk}{(2\pi)^{(1+d)}}
\gamma^mG(k)D(k-p)\Gamma_{\mathrm{int}}(k,p).
\end{eqnarray}

\end{document}